\pdfoutput=1
\documentclass[twoside,twocolumn,9pt]{article}
\usepackage{extsizes}
\usepackage[super,sort&compress,comma]{natbib} 
\usepackage[version=3]{mhchem}
\usepackage[left=1.5cm, right=1.5cm, top=1.785cm, bottom=2.0cm]{geometry}
\usepackage{balance}
\usepackage{times,mathptmx}
 
\usepackage{sectsty}
\usepackage{graphicx} 
\usepackage{lastpage}
\usepackage[format=plain,justification=justified,singlelinecheck=false,font={stretch=1.125,small,sf},labelfont=bf,labelsep=space]{caption}
\usepackage{float}
\usepackage{color}
\usepackage{fancyhdr}
\usepackage{fnpos}
\usepackage{comment}
\usepackage{bm}
\usepackage{subcaption}
\usepackage[english]{babel}
\addto{\captionsenglish}{%
  
}
\usepackage{array}
\usepackage{droidsans}
\usepackage{charter}
\usepackage[T1]{fontenc}
\usepackage[usenames,dvipsnames]{xcolor}
\usepackage{setspace}
\usepackage[compact]{titlesec}
\usepackage{hyperref}
\DeclareSymbolFont{epsilon}{OML}{ntxmi}{m}{it}
\DeclareMathSymbol{\epsilon}{\mathord}{epsilon}{"0F}

\allowdisplaybreaks

\usepackage{epstopdf}%This line makes .eps figures into .pdf - please comment out if not required.

\definecolor{cream}{RGB}{222,217,201}

\begin{document}

\pagestyle{fancy}
\thispagestyle{plain}
\fancypagestyle{plain}{

%%%HEADER%%%
%\fancyhead[C]{\includegraphics[width=18.5cm]{head_foot/header_bar}}
%\fancyhead[L]{\hspace{0cm}\vspace{1.5cm}\includegraphics[height=30pt]{head_foot/journal_name}}
%\fancyhead[R]{\hspace{0cm}\vspace{1.7cm}\includegraphics[height=55pt]{head_foot/RSC_LOGO_CMYK}}
\renewcommand{\headrulewidth}{0pt}
}
%%%END OF HEADER%%%

%%%PAGE SETUP - Please do not change any commands within this section%%%
\makeFNbottom
\makeatletter
\renewcommand\LARGE{\@setfontsize\LARGE{15pt}{17}}
\renewcommand\Large{\@setfontsize\Large{12pt}{14}}
\renewcommand\large{\@setfontsize\large{10pt}{12}}
\renewcommand\footnotesize{\@setfontsize\footnotesize{7pt}{10}}
\makeatother

\renewcommand{\thefootnote}{\fnsymbol{footnote}}
\renewcommand\footnoterule{\vspace*{1pt}% 
\color{cream}\hrule width 3.5in height 0.4pt \color{black}\vspace*{5pt}} 
\setcounter{secnumdepth}{5}

\makeatletter 
\renewcommand\@biblabel[1]{#1}            
\renewcommand\@makefntext[1]% 
{\noindent\makebox[0pt][r]{\@thefnmark\,}#1}
\makeatother 
\renewcommand{\figurename}{\small{Fig.}~}
\sectionfont{\sffamily\Large}
\subsectionfont{\normalsize}
\subsubsectionfont{\bf}
\setstretch{1.125} %In particular, please do not alter this line.
\setlength{\skip\footins}{0.8cm}
\setlength{\footnotesep}{0.25cm}
\setlength{\jot}{10pt}
\titlespacing*{\section}{0pt}{4pt}{4pt}
\titlespacing*{\subsection}{0pt}{15pt}{1pt}
%%%END OF PAGE SETUP%%%

%%%FOOTER%%%
\fancyfoot{}
%\fancyfoot[LO,RE]{\vspace{-7.1pt}\includegraphics[height=9pt]{head_foot/LF}}
%\fancyfoot[CO]{\vspace{-7.1pt}\hspace{13.2cm}\includegraphics{head_foot/RF}}
%\fancyfoot[CE]{\vspace{-7.2pt}\hspace{-14.2cm}\includegraphics{head_foot/RF}}
%\fancyfoot[RO]{\footnotesize{\sffamily{1--\pageref{LastPage} ~\textbar  \hspace{2pt}\thepage}}}
%\fancyfoot[LE]{\footnotesize{\sffamily{\thepage~\textbar\hspace{3.45cm} 1--\pageref{LastPage}}}}
\fancyhead{}
\renewcommand{\headrulewidth}{0pt} 
\renewcommand{\footrulewidth}{0pt}
\setlength{\arrayrulewidth}{1pt}
\setlength{\columnsep}{6.5mm}
\setlength\bibsep{1pt}
%%%END OF FOOTER%%%

%%%FIGURE SETUP - please do not change any commands within this section%%%
\makeatletter 
\newlength{\figrulesep} 
\setlength{\figrulesep}{0.5\textfloatsep} 

\newcommand{\topfigrule}{\vspace*{-1pt}% 
\noindent{\color{cream}\rule[-\figrulesep]{\columnwidth}{1.5pt}} }

\newcommand{\botfigrule}{\vspace*{-2pt}% 
\noindent{\color{cream}\rule[\figrulesep]{\columnwidth}{1.5pt}} }

\newcommand{\dblfigrule}{\vspace*{-1pt}% 
\noindent{\color{cream}\rule[-\figrulesep]{\textwidth}{1.5pt}} }

\makeatother
%%%END OF FIGURE SETUP%%%

%%%TITLE, AUTHORS AND ABSTRACT%%%
\twocolumn[
  \begin{@twocolumnfalse}
\vspace{3cm}
\sffamily
\begin{tabular}{m{4.5cm} p{13.5cm} }

 & \noindent\LARGE{\textbf{Simulation of dense non-Brownian suspensions with the lattice Boltzmann method: Shear jammed and fragile states}} \\ \indent%Article title goes here instead of the text "This is the title"
\vspace{0.3cm} & \vspace{0.3cm} \\ \indent

 & \noindent\large{Pradipto$^{\ast}$and Hisao Hayakawa} \\ \indent%Author names go here instead of "Full name", etc.

& \noindent\normalsize{Dense non-Brownian suspensions including both the hydrodynamic interactions and the frictional contacts between particles are numerically studied under simple and oscillatory shears in terms of the lattice Boltzmann method.
We successfully reproduce the discontinuous shear thickening (DST) under a simple shear for bulk three-dimensional systems.
For our simulation of an oscillatory shear in a quasi-two-dimensional system, we measure the mechanical response when we reduce the strain amplitude after the initial oscillations with a larger strain amplitude.
Here, we find the existence of the shear-jammed state under this protocol in which the storage modulus $G^{\prime}$ is only finite for high initial strain amplitude $\gamma_0^{I}$.
We also find the existence of the fragile state in which both fluid-like and solid-like responses can be detected for an identical area fraction and an initial strain amplitude $\gamma_0^{I}$ depending on the initial phase $\Theta$ (or the asymmetricity of the applied strain) of the oscillatory shear.
We also observe the DST-like behavior under the oscillatory shear in the fragile state.
Moreover, we find that the stress anisotropy becomes large in the fragile state. 
Finally, we confirm that the stress formula based on the angular distribution of the contact force recovers the contact contributions to the stress tensors for both simple and oscillatory shears with large strains.} 

\end{tabular}

 \end{@twocolumnfalse} \vspace{0.6cm}

  ]
%%%END OF TITLE, AUTHORS AND ABSTRACT%%%

%%%FONT SETUP - please do not change any commands within this section
\renewcommand*\rmdefault{bch}\normalfont\upshape
\rmfamily
\section*{}
\vspace{-1cm}

%%%FOOTNOTES%%%

\footnotetext{\textit{Yukawa Institute for Theoretical Physics, Kyoto University, Kitashirakawaoiwake-cho, Sakyo-ku, Kyoto 606-8502, Japan; E-mail: pradipto@yukawa.kyoto-u.ac.jp }}
%\footnotetext{\textit{$^{b}$~Address, Address, Town, Country. }}

%Please use \dag to cite the ESI in the main text of the article.
%If you article does not have ESI please remove the the \dag symbol from the title and the footnotetext below.
%\footnotetext{\dag~Electronic Supplementary Information (ESI) available: [details of any supplementary information available should be included here]. See DOI: 10.1039/b000000x/}
%additional addresses can be cited as above using the lower-case letters, c, d, e... If all authors are from the same address, no letter is required

%\footnotetext{\ddag~Additional footnotes to the title and authors can be included \textit{e.g.}\ `Present address:' or `These authors contributed equally to this work' as above using the symbols: \ddag, \textsection, and \P. Please place the appropriate symbol next to the author's name and include a \texttt{\textbackslash footnotetext} entry in the the correct place in the list.}

%%%END OF FOOTNOTES%%%

%%%MAIN TEXT%%%%

\section{Introduction}
The behavior of suspended particles in solvents (suspensions) has been studied since Einstein published his seminal work in 1905\cite{einstein1905}. 
However, our understanding of rheological properties of suspensions is still limited. One of the interesting phenomena in the rheology of dense suspensions is the discontinuous shear thickening (DST) which is an abrupt jump of the viscosity at a critical shear rate $\dot{\gamma}$ if the volume fraction $\phi$ is larger enough. 
The DST can be observed easily in a mixture of cornstarch and water.
The DST has also industrial applications such as traction controls and protective vests.
Since the first observation of it about 90 years ago\cite{williamson1931}, the DST has been studied extensively\cite{barnes1989,egres2005,brown2012}. 
Some experiments clarified the roles of shape of particles\cite{brown2011,cwalina2016}, and the boundary effects on the DST\cite{allen2018,maharjan2018}. 

We have substantial progresses on simulations of dense suspensions.
By using the Lubrication-Friction Discrete Element Method (LF-DEM) \cite{seto2013,mari2014,mari2015} which is a simplified version of the Stokesian dynamics \cite{brady1988}, Seto and his coworkers confirmed an important role of frictional contact force between particles for the DSTs in non-Brownian\cite{seto2013,mari2014} and Brownian suspensions\cite{mari2015}. 
They successfully reproduced the DST which quantitatively agrees with experimental results. 
A good review along this line\cite{denn2018} has been published recently.
Despite these efforts, the theoretical understanding is still limited. 
By using a phenomenology that taking into account frictional contacts between grains\cite{wyart2014}, Thomas et al. suggested that the anisotropy of the radial distribution function plays an important role in the DST\cite{thomas2018}.
Furthermore, the DST is also found in dry frictional granular particles\cite{otsuki2011}.

The isotropic jammed state\cite{liu1998,liu2010} in which the system cannot flow above a critical volume fraction is well-defined, at least, for dry granular particles in the zero shear limit.
Bi et al. introduced the shear-jammed state below the critical fraction of the isotropic jamming and the fragile state in an experiment for two-dimensional dry granular materials under a pure shear\cite{bi2011}. 
However, the definitions of these states are not clear if we discuss different setups.
Recently, Otsuki and Hayakawa\cite{otsuki2018} have proposed one definition of the shear jamming in oscillatory shear as a memory effect of the initial shear after reducing the strain amplitude.
They have also adopted a new definition of the fragile state in which the system has or loses the rigidity depending on the initial phase (or the type of oscillatory shear) of the oscillatory shear.
Moreover, they demonstrated that a DST-like phenomenon as a discontinuous jump of the dynamic viscosity can be observed in the fragile state\cite{otsuki2018}.

Until recently, there are only a few studies in suspensions under oscillatory shear, numerically \cite{lee2015,marenne2017,moghimi2017} and experimentally\cite{park2015}.
Ness et al. also studied suspensions under an oscillatory shear below the jamming point where they observed the strain hardening and its frequency dependence in 3D systems in their experiment and numerical simulation\cite{ness2017}.
One drawback of the oscillatory shear is that the storage and loss moduli \cite{doiedwards} beyond the linear response regime are not well-defined \cite{ganeriwala1987}.
There are several proposals to handle the nonlinear response using so called \textit{the Fourier transform rheology}\cite{brader2010,marenne2017} or Chebyshev polynomials decomposition\cite{ewoldt2008,marenne2017}.
Instead of taking into account the nonlinearities, Ref. \cite{otsuki2018} adopts a protocol in which they initially use a finite or a large strain amplitude and then reduce the amplitude to be in the linear response regime.

The situations are more controversial if we discuss the shear-jammed and the fragile states in suspensions.
The definitions of the shear-jammed state, and the fragile state, as well as their relations with the DST in suspensions are still unclear. 
It should be noted that the DST takes place below the shear jamming density in suspensions\cite{mari2014}.
The distinction between shear jamming and the DST also exists in the experimental results by Peters et al.\cite{peters2016}, which suggested that the DST might correspond to the fragile state\cite{peters2016}.
Meanwhile, another experiment suggests that DST takes place at the lower onset of the shear-jammed states\cite{fall2015}.
In this paper, we demonstrate that the definitions of the shear jammed and the fragile states proposed by Ref. \cite{otsuki2018} can be used even in suspensions.

The LF-DEM only considers the lubrication interaction with ignoring the long range hydrodynamic interaction between particles.
Although many people believe that the LF-DEM is sufficient in describing the rheology of dense suspensions near the DST and the shear jamming, the existence of the long-range interaction might affect the critical behavior slightly lower densities or the critical densities of such rheological transitions.
Therefore, we adopt the lattice Boltzmann method (LBM) for suspensions\cite{ladd1994a,ladd1994b} with its lubrication correction\cite{nguyen2002} in order to calculate the hydrodynamic interaction between the particles.
There are also several studies on suspensions using the LBM\cite{shakib-manesh2002,raiskinmaki2003,kulkarni2008,lee2014,lee2015}, where this method has been confirmed to be efficient and accurate for sedimentation simulations\cite{nguyen2004}.
Although we can apply the LBM to suspensions embedded in a fluid with finite Reynolds number, we restrict our analysis to suspensions embedded in a fluid with low Reynolds number limit in this paper.

The outline of this paper is as follows.
In the next section, we explain our simulation method briefly.
In section 3, we present the results of our simulations under the simple shear to verify the relevance of our LBM as well as some new aspects of our analysis.
Then the results for the oscillatory shear are presented as the main result in section 4.
In the last section, we summarize our results and discuss some future perspective.
In Appendices, we describe some details of our analysis and some supplemental information to the main text.
\section{Simulation Method}
	\begin{figure}[htbp]
	 \centering
	 \includegraphics[height=0.5\linewidth]{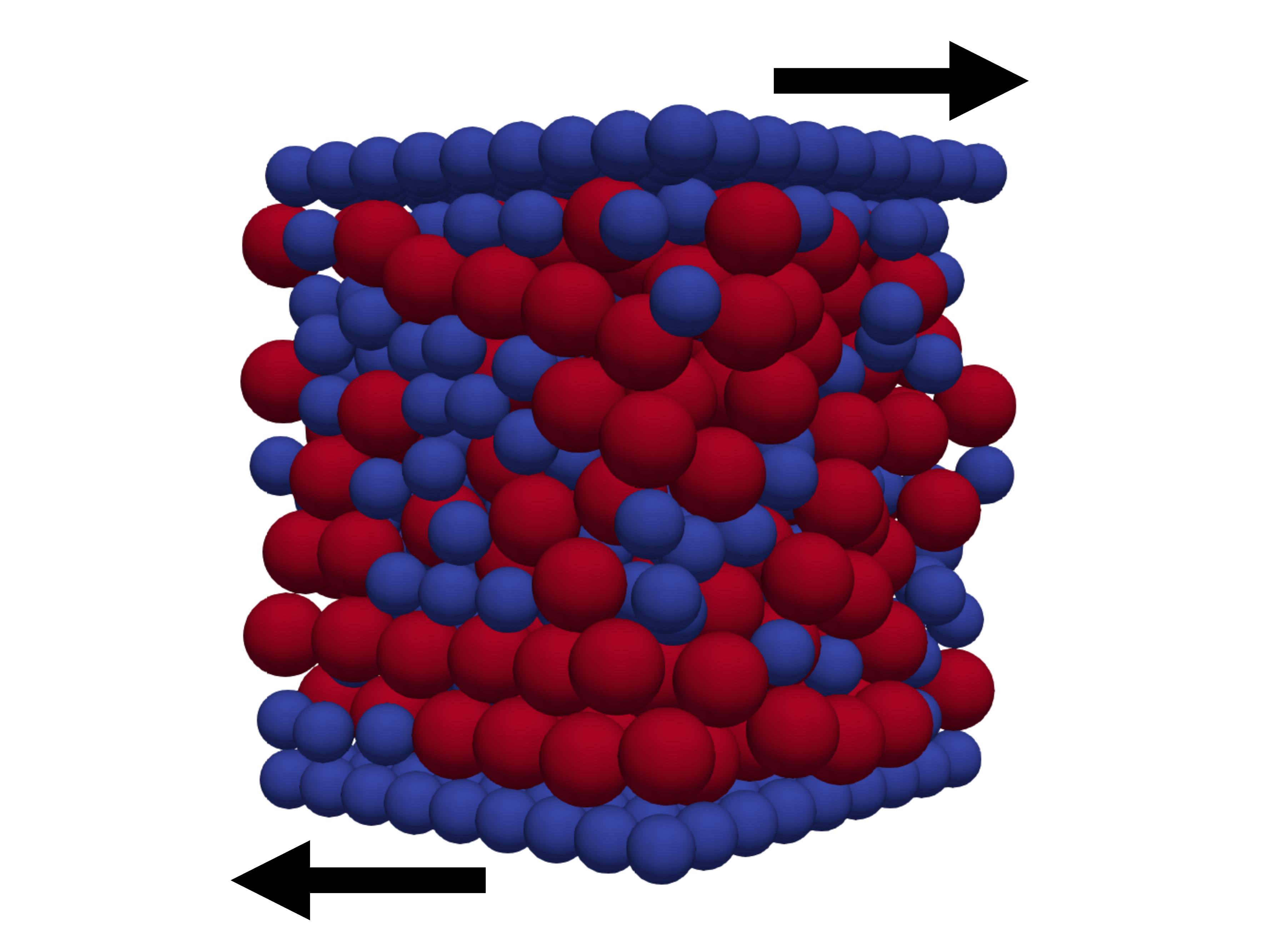}
	 \caption{A snapshot of our simulation for a simple shear flow, where the arrows indicate the motion of the wall.}
	 \label{fgr:snapshot_std}
	\end{figure}
	\begin{figure}[htbp]
	 \centering
	 \includegraphics[height=0.5\linewidth]{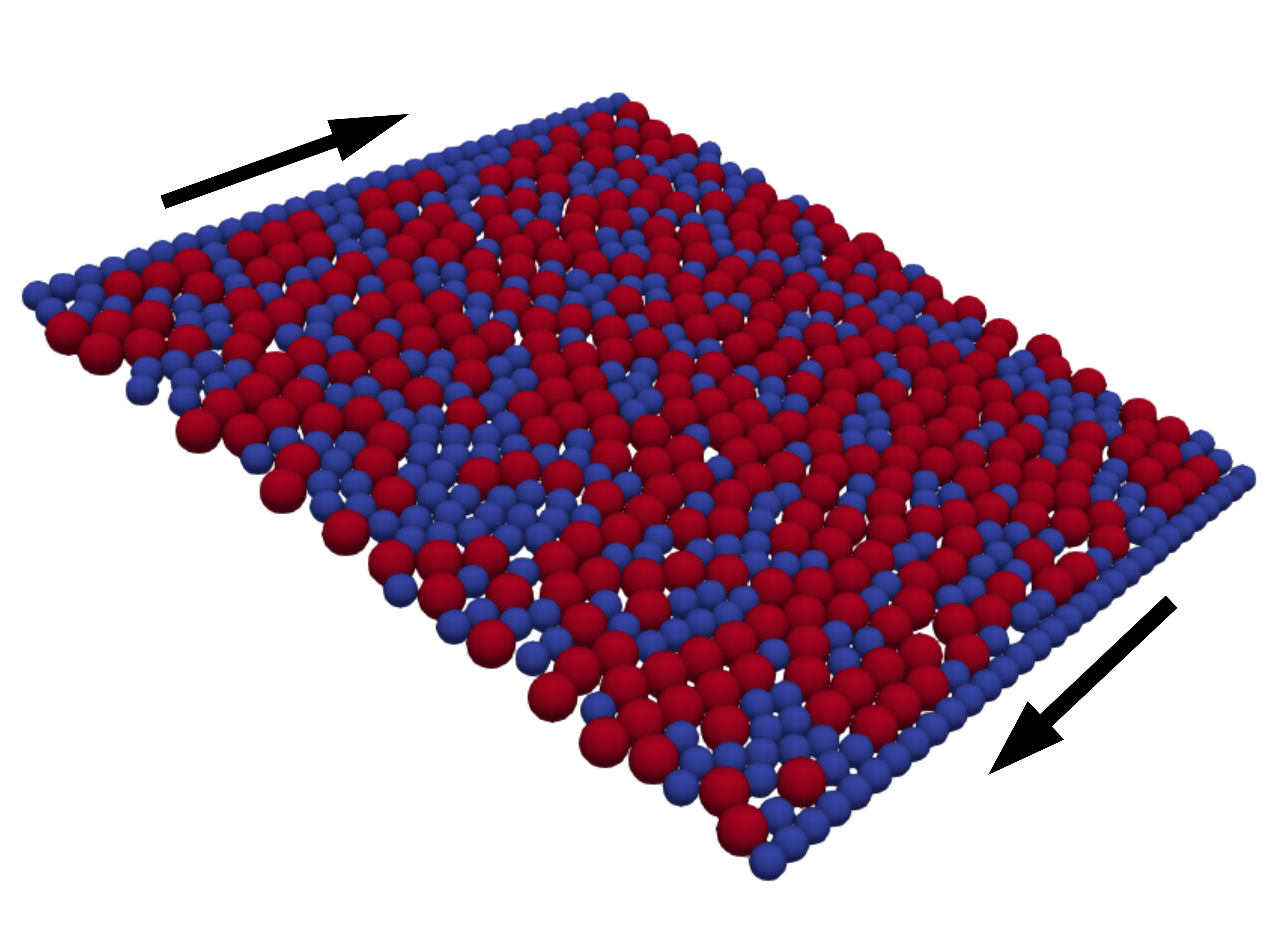}
	 \caption{A snapshot of the monolayer configuration in an oscillatory shear simulation. The arrows represent shear directions.}
	 \label{fgr:snapshot}
	\end{figure}
	
Let us consider a suspension consisting of spherical particles.
A set of equations of motion of the suspended particles is given by
	  \begin{equation}
	\frac{d}{dt} 
	  	\begin{pmatrix}
	  		m\bm{U} \\
	  		I\bm{\Omega}
	  		\end{pmatrix}
	  		=
	  		\begin{pmatrix}
	  			 \bm{F}^{h} + \bm{F}^{c} +  \bm{F}^{r}\\ 
	  			\bm{T}^{h} + \bm{T}^{c}
	  			\end{pmatrix}, \label{eq:1}
	  	\end{equation}
where $m$, $I$, $\bm{U}$, and $\bm{\Omega}$  are the diagonal matrix for mass of the particles, the diagonal matrix for the moment of inertia of particles, the translation velocity, and the angular velocity, respectively.
The forces are the sum of the hydrodynamic force $\bm{F}^{h} $, the contact force $\bm{F}^{c} $, and the electrostatic repulsive force $ \bm{F}^{r}$, while the torques are the sum of hydrodynamic torque $\bm{T}^{h}$ and contact torque $\bm{T}^{c}$.	
	
In the LBM, the hydrodynamic force $\bm{F}^{h} $ is calculated by computing the discrete distribution function on lattices.
Due to this lattice-based calculation, one needs to discretize the unit of length and time into the lattice units $\Delta x$.
This discretization is related to the stability and accuracy of the scheme.
This paper adopts $\Delta x = 0.24a_{\text{min}}$, where $a_{\rm min}$ is the radius of the smallest suspended particle.
Details of the LBM can be seen in Appendix A.
On the other hand, the LF-DEM includes only two-body lubrication interactions between particles of the Stokes flow.

Perfect spherical hard particles in the Stokes flow do not allow any contact since the lubrication force diverges if the gap between particles becomes zero.
Here, we introduce a cutoff length $\delta$ of the lubrication forces to allow the contact between rough particles.
Our simulations adopts $\delta = 0.01a_\text{min}$, which is used in Ref. \cite{mari2014} \footnote{the viscosity for $ \delta= 0.01 a_\text{min}$ is about $20 \%$ smaller than that for $\delta = 0.001 a_\text{min} $ under stress control simulation \cite{mari2014}.}

We adopt the linear spring model for the contact interaction between particles commonly used in discrete element method (DEM) \cite{luding2008} which involves both the normal and the tangential contact forces.
Note that we omit the dissipative contact force for the tangential part.
For particle $i$, the contact force $\bm{F}^{c}_{i}$ and torque $\bm{T}^{c}_{i}$ are, respectively, written as
\begin{equation}
	\bm{F}^{c}_{i} = \sum_{i \neq j} ( \bm{F}^{\text{nor}}_{ij} + \bm{F}^{\text{tan}}_{ij} ), \label{eq:11}
	\end{equation}	
	\begin{equation}
		\bm{T}^{c}_{i} = \sum_{i \neq j} a_i \bm{n}_{ij} \times \bm{F}^{\text{tan}}_{ij}, \label{eq:11t}
		\end{equation}			
where $a_i$ is the radius of the particle $i$.
The normal force is expressed as $\bm{F}^{\text{nor}}_{ij} =( k_n \delta_{ij}^{n} +  \eta^{(n)} u_{ij}^{(n)} ) \bm{n}_{ij}$ where $k_n$ is the spring constant, $\delta_{ij}^{n}$ is the normal overlap, $\bm{n}_{ij}$ is the normal unit vector between particles, $u_{ij}^{(n)}$ is the normal velocity of the contact point, and $ \eta^{(n)} = \sqrt{m_0 k_n}$ is the damping constant, where $m_0$ is the average mass of the particles.
The stick tangential counterpart is represented as $\tilde{\bm{F}}^{\text{tan}}_{ij} = k_t \delta_{ij}^{t} \bm{t}_{ij}$ where $k_t$ is tangential spring constant equals to $0.2k_n$, $\delta_{ij}^{t}$ is the tangential compression and $\bm{t}_{ij}$ is the tangential unit vector at the contact point between particles $i$ and $j$.
We also adopt the Coulomb friction rules, where $|\bm{F}_{ij}^{\text{tan}}|$ is replaced by $\mu |\bm{F}_{ij}^{\text{nor}}|$ if  $|\tilde{\bm{F}}^{\text{tan}}_{ij}| \geq \mu |\bm{F}^{\text{nor}}_{ij}|$ for slip contacts and use $|\tilde{\bm{F}}^{\text{tan}}_{ij}|$ if $|\tilde{\bm{F}}^{\text{tan}}_{ij}| \leq \mu |\bm{F}^{\text{nor}}_{ij}|$ for stick contacts, whereas $\delta_{ij}^{t}$ is updated each time with the relative tangential velocity\cite{luding2008}.
Here, $\mu$ is the friction coefficient for the particle interaction.
All results of our simulation presented in this paper are obtained for $\mu=1.0$.
Note that the normal spring constant $k_n$ determines the time scale $\tau$ as $\tau=\sqrt{m_0/k_n}$.
In our simulation, $\tau$ is related to the time step of the LBM $\Delta t$ as $\Delta t = 0.6 \tau$.  

The stress contribution from the contact forces is given by
		\begin{equation}
		 \sigma^{c}_{\alpha \beta} = - \frac{1}{2V} \sum_{i} \sum_{j \neq i} ( r_{ij, \alpha} F^{c}_{ij,\beta} + r_{ij, \beta} F^{c}_{ij,\alpha}), \label{eq:12}
		 \end{equation}			 
where $r_{ij, \alpha}$ is the distance between the center of masses of two contacting particles in $\alpha$ direction.
Although there are asymmetric contributions of the stress tensor for frictional particles~\cite{mitarai2002,goldhirsch2010}, we have ignored such contributions because the asymmetric stress tensor is much smaller than the symmetric part. 

To stabilize the suspensions, we introduce a double-layer (repulsive) electrostatic force in the Derjaguin-Landau-Verwey-Overbeek (DLVO) theory\cite{derjaguin1941,verwey1948,israelachvili2011}
	\begin{equation}
		\bm{F}^{r}_{ij} =  F_0\exp(-h/\lambda) \bm{n}_{ij}, \label{eq:13}
		\end{equation}		
where $F_0=k_B T \lambda_B \hat{Z}^2(e^{a_{\text{min}}/\lambda}/(1+a_{\text{min}}/\lambda))^2/h^2$ with the charge number $\hat{Z}$, the Bjerrum length $\lambda_B$ and the Debye-H\"{u}ckel length $\lambda$. Note that $\lambda_B$ can be expressed as $\lambda_B=e^2/(4\pi \epsilon_0 \epsilon_r k_B T)$ where $e$, $\epsilon_0$, $\epsilon_r$, and $k_B$ are the elementary charge, the vacuum permittivity, the dielectric constant, and the Boltzmann constant, respectively \cite{israelachvili2011}.
Here, we adopt the Debye length $\lambda = 0.02a_{\text{min}}$.
This stabilizing force is relevant to get quantitative agreements with the experiments as shown in Ref.\cite{mari2015}. 
The contribution to the stress from this double-layer force is given by
	\begin{equation}
	 \sigma^{r}_{\alpha \beta} = - \frac{1}{2V} \sum_{i} \sum_{j \neq i} ( r_{ij, \alpha} F^{r}_{ij,\beta}+r_{ij, \beta} F^{r}_{ij,\alpha}). \label{eq:14}
	 \end{equation} 		 
The total shear stress contains all contributions from the hydrodynamic stress, the contact stress, and the electrostatic stress as
\begin{equation}
	\sigma_{\alpha \beta} = \sigma^{h}_{\alpha \beta} + \sigma^{c}_{\alpha \beta} +\sigma^{r}_{\alpha \beta}. \label{eq:15}
	\end{equation}
	
We use bi-disperse particles which includes equal number of particles with ratio $a_{\text{max}}=1.4a_{\text{min}}$ to prevent them from crystallization.
In both the simple and the oscillatory cases, we apply shears by moving bumpy walls made of particles (see Figs. \ref{fgr:snapshot_std} and \ref{fgr:snapshot}), whereas we use the periodic boundary conditions for the other directions.  

\section{Simple shear simulation}

This section consists of four parts. In the first part, we explain the setup of our simulation. In the second part, we demonstrate that our LBM simulation successfully reproduces the DST. In the third part, we illustrate the role of anisotropy of the local contact force on the particles, which can capture the behavior of the macroscopic stress tensor including the DST. In the last part, we compare the results of the LBM with those of the LF-DEM.

\subsection{Setup}
Our simulation for simple shear flows contains $N=450$ particles in a three-dimensional box. 
We use the volume fraction $\phi_3 = 2 N\pi (a_{\text{max}}^3 + a_{\text{min}}^3)/(3V)$ to characterize the density of suspensions in 3D systems.
The viscosity $\eta=\sigma_{xy}/\dot{\gamma}$ with the shear rate $\dot\gamma$ is time averaged between the strains $\gamma := \dot{\gamma} t = 5$ and $\gamma=11$.
Then, we plot the viscosity against the dimensionless shear rate $\dot{\gamma}^{*}=6\pi \eta_0 a_{\text{min}}^2\dot\gamma/F_0$ (see Fig. \ref{fgr:snapshot_std}), where $\eta_0$ is the viscosity of the solvent fluid \cite{mari2014,seto2013}. 
Note that we do not have to specify the value of $F_0$ in the simulation of simple shears because we only use $\dot{\gamma}^{*}$ in our simulation.
This dimensionless shear rate $\dot{\gamma}^{*}$ is analogous to the Stokes number $St = m \dot{\gamma} / 6 \pi \eta_0 a_{\text{min}}$.
However, since the particles' inertia is irrelevant, using $\dot{\gamma}^{*}$ is more appropriate than $St$.
Using $\dot{\gamma}^{*}$ has been adopted in the previous numerical simulations \cite{seto2013,mari2014,mari2015}.

\subsection{Discontinuous shear thickening}

\begin{figure}[htbp]
 \centering
 \includegraphics[height=0.65\linewidth]{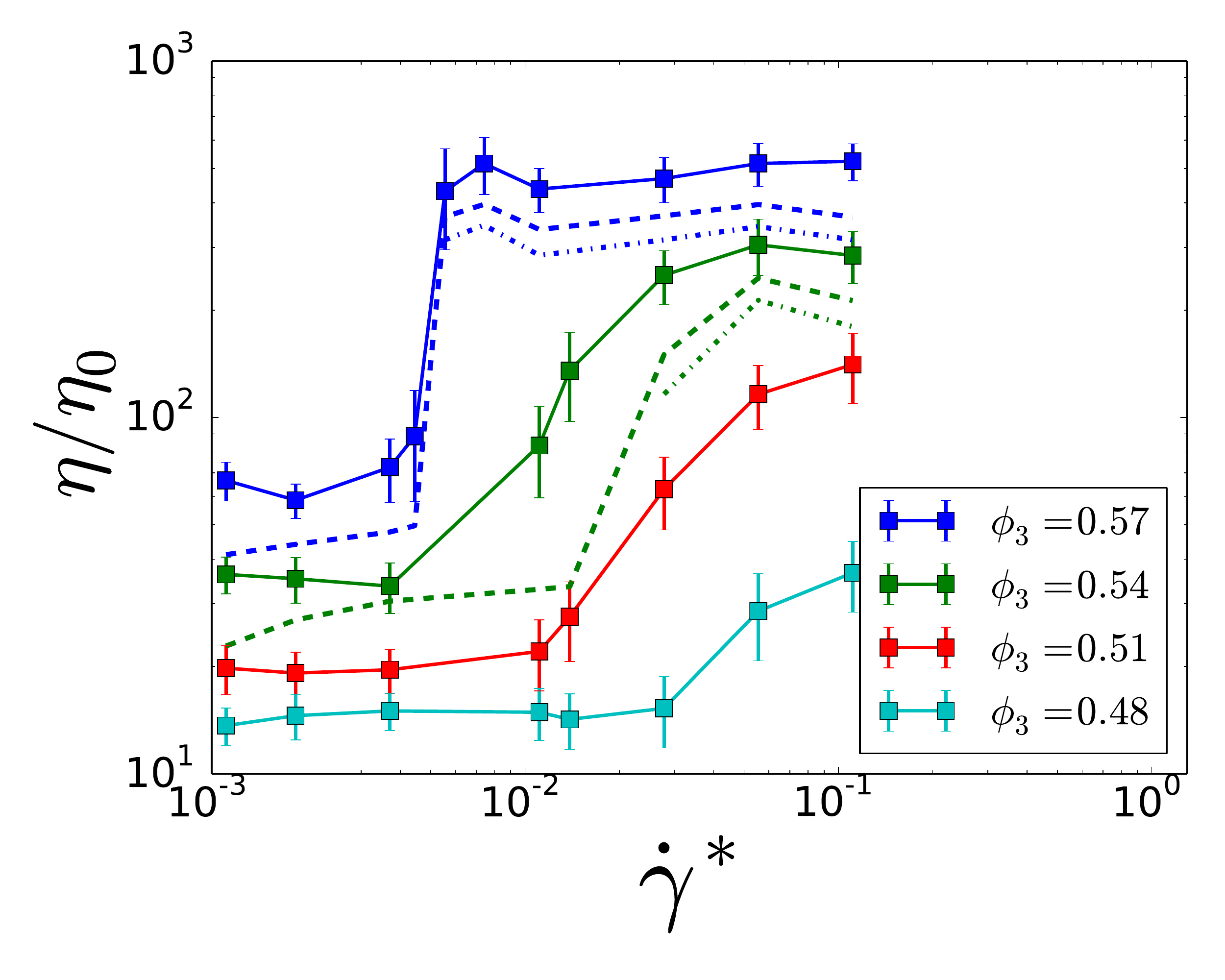}
 \caption{Plots of the apparent viscosity $\eta/\eta_0$ against the dimensionless shear rate $\dot{\gamma}^{*}$ for various $\phi_3$. Dashed lines are obtained from Eq. \eqref{eq:27} and the lubrication expressions in the Appendix A. The dotted-chain lines are obtained only from Eq. \eqref{eq:27}.}
 \label{fgr:steadyvisc}
\end{figure}

First, we present the results for the viscosity under a simple shear where we observe discontinuous jumps of the viscosity in dense situations (e.g., $\phi_3 = 0.57$) above critical shear rates as shown in Fig. \ref{fgr:steadyvisc}.
These results semi-quantitatively agree with the DST observed in the LF-DEM simulations \cite{seto2013,mari2014} and experiments \citep{frith1996,bender1996,maranzano2001a,maranzano2001b,lootens2005,fall2010,larsen2010,brown2012,brown2014}.
The onset of the DST corresponds to the beginning of the frictional contact between grains.
At low shear rate regime, the particles are prevented from contact by the hydrodynamic lubrication and the electrostatic repulsive forces.

\subsection{Anisotropies of contact forces and stress}
	\begin{figure}[htbp]
	 \centering
	 \begin{subfigure}{.5\textwidth}
	 \centering
	 \includegraphics[height=.65\linewidth]{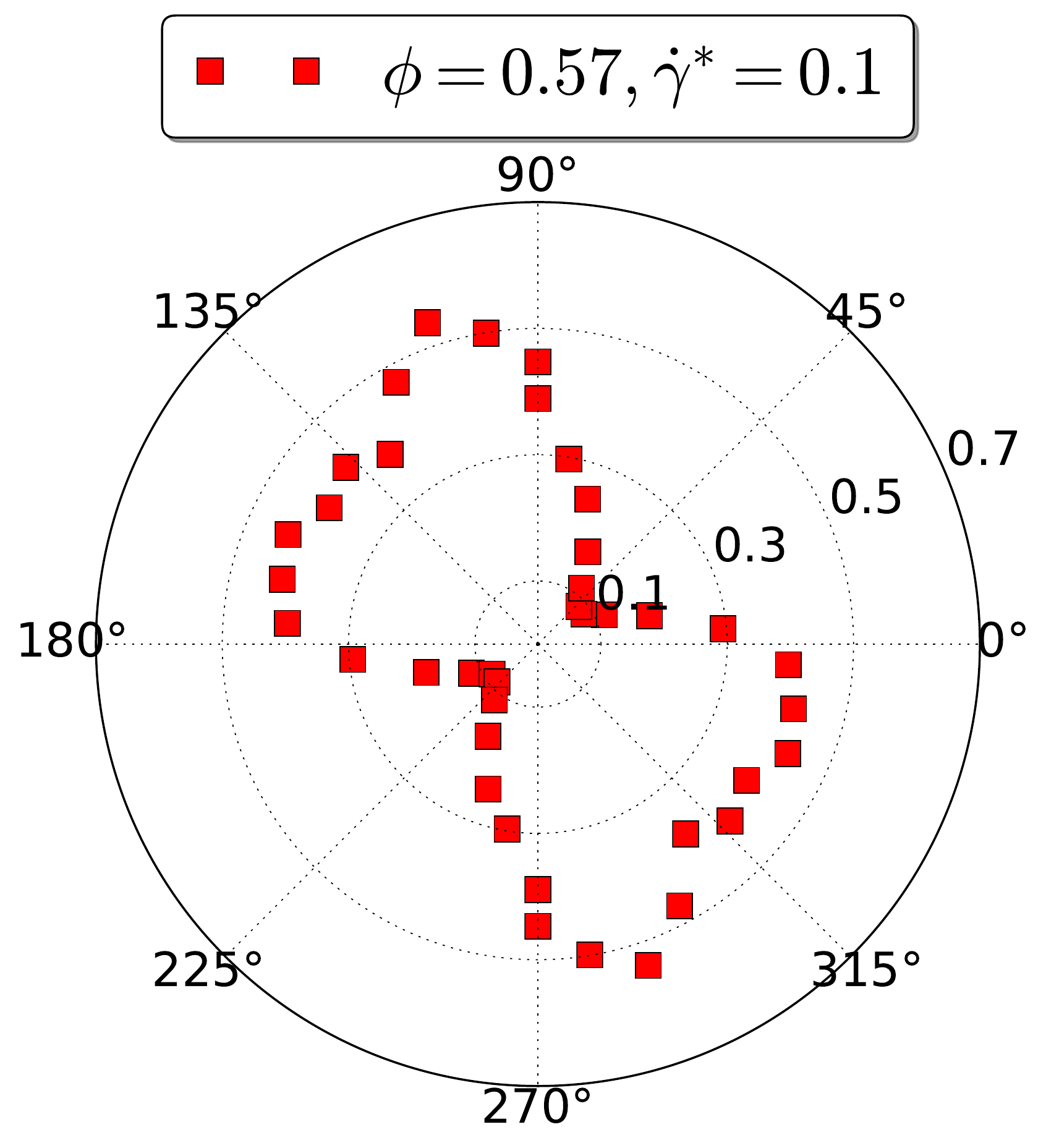}
	 \caption{}
	 \label{fgr:angular_std_a}
	 \end{subfigure}
	 \centering
	 \begin{subfigure}{.5\textwidth}
	 \centering
	 \includegraphics[height=.65\linewidth]{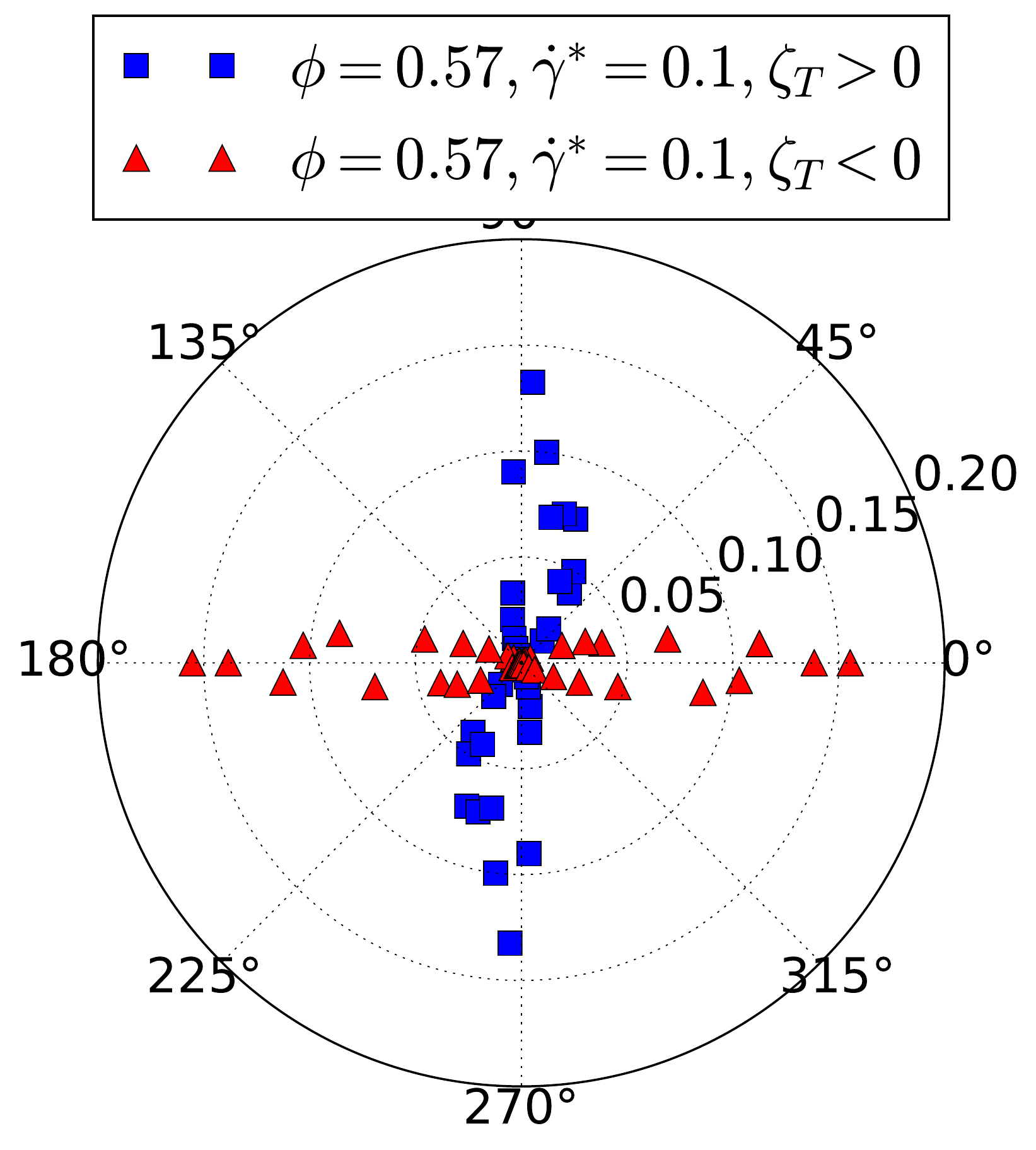}
	 \caption{}
	 \label{fgr:angular_std_b}
	 \end{subfigure}
	 \caption{(a) The angular distributions of the normal contact forces $\zeta_N$ and (b) of the tangential contact forces $\zeta_T$ under the simple shear where the shear rate is larger than the critical one for the DST. The data are obtained from the time average between $\gamma=10.5$ and $\gamma=11$. }
	 \label{fgr:angular_std}
	 \end{figure}
	 
	 	 \begin{figure}[htbp]
	  \centering
	  \includegraphics[height=0.7\linewidth]{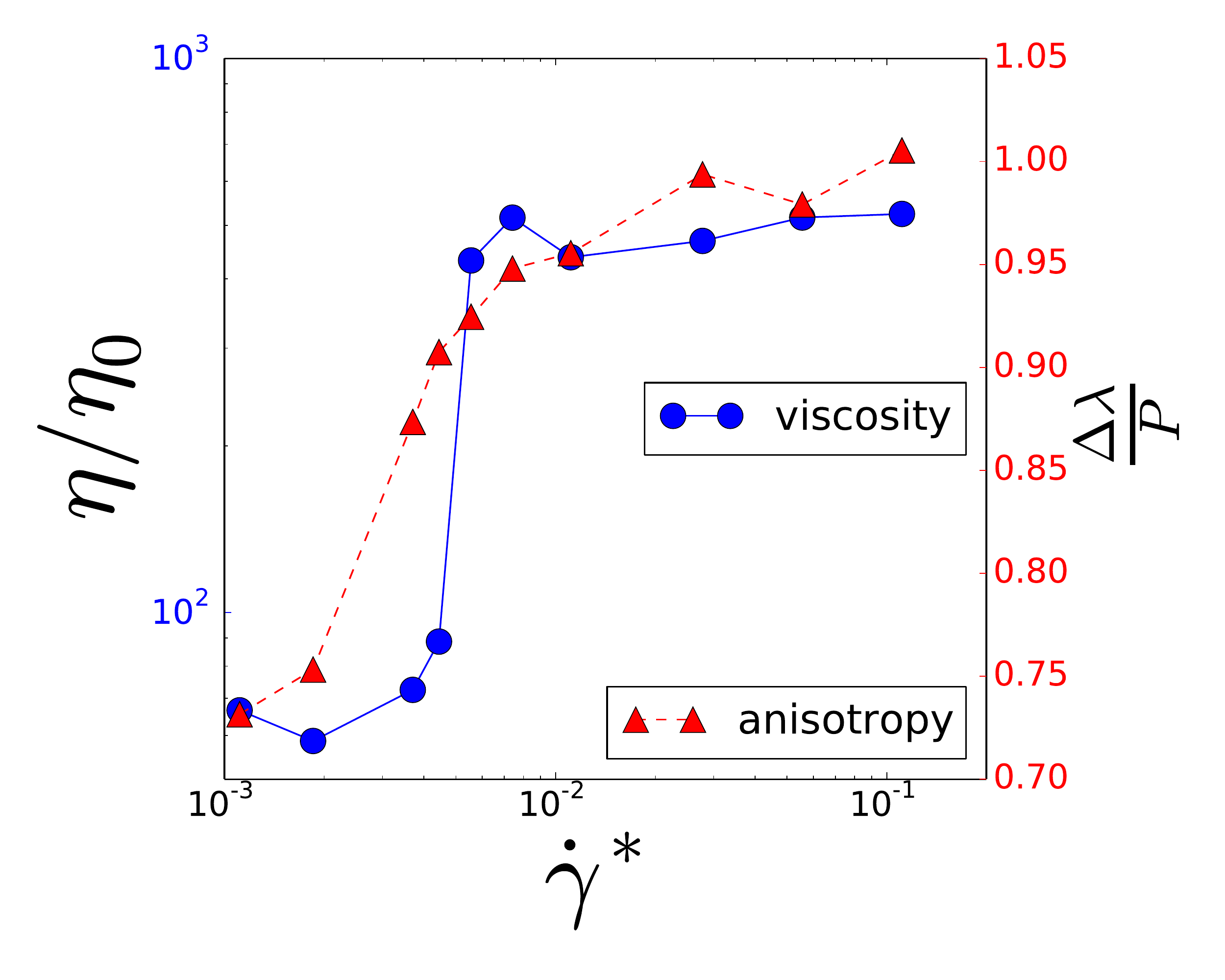}
	  \caption{Plots of anisotropy of the stress tensor $\Delta \lambda /P$ and the corresponding viscosity under a simple shear for $\phi_3=0.57$.}
	  \label{fgr:eigenvalues_std}
	 \end{figure}
	 
We quantify the role of the contact forces by analyzing the angular distributions of the contact forces\cite{dacruz2005} $\zeta _ { N }$ and $\zeta _ { T }$ for the normal and the tangential parts, respectively, defined as
\begin{align}
	\zeta _ { N } ( \theta ) &= \rho ( \theta ) \xi _ { N } ( \theta ), \label{eq:19} \\
	\zeta _ { T } ( \theta ) &= \rho ( \theta ) \xi _ { T } ( \theta ). \label{eq:20}
	\end{align}
The contact angle $\theta$ satisfies $\theta={\rm tan}^{-1}(r_{ij,y}/r_{ij,x})$ which is ranged between $0$ and $\pi$ and calculated counterclockwise from the $x$-axis.
Here, $\rho(\theta)$ is the angular distribution of the contact orientations. $\xi _ { N } ( \theta )$ and  $\xi _ { T } ( \theta )$ are the angular distributions of intensities of normal and tangential forces, respectively, defined as
\begin{align}
	\xi _ { N } ( \theta ) &=  F^{\text{nor}} ( \theta )  / \langle F^{\text{nor}} \rangle, \label{eq:21}\\ 
	 \xi _ { T } ( \theta ) &=  F^{\text{tan}} ( \theta )  / \langle F^{\text{nor}} \rangle. \label{eq:22}
	\end{align}  
Here, $  F^{\text{nor}} ( \theta ) $ and $  F^{\text{tan}} ( \theta ) $ are the normal and tangential forces in the direction of $\theta$ and $\langle F^{\text{nor}} \rangle$ is the average normal forces $\langle F^{\rm nor}\rangle=\int_0^{\pi}d\theta \rho(\theta)F^{\rm nor}(\theta)$.
The angular distribution $\rho(\theta)$ satisfies the following normalization:
\begin{align}
	 \int _ { 0 } ^ { \pi } \rho ( \theta ) d \theta &= 1.   \label{eq:23}
	\end{align}

Figures \ref{fgr:angular_std_a} and \ref{fgr:angular_std_b} present the angular distributions of the normal and tangential contact forces, respectively.
Our results show that the anisotropy exists in the normal part, where $\zeta_N(\theta)$ takes maxima in the directions of compression ($\theta \approx 3 \pi/4, 7 \pi /4$) and becomes minima in the expansion directions ($\theta\approx \pi/4, 5\pi/4$).
The tangential counterpart also shows anisotropy by exhibiting an 8-shaped figure, where the positive  $\zeta_{T} (\theta)$ (counterclockwise rotations) is most likely oriented at the directions of shear gradient $\theta = \pi/2$ while the negative $\zeta_{T} (\theta)$  (clockwise rotations) is at the directions of shear.
We can express the stress tensor as a function of these angular distributions as \cite{dacruz2005,kanatani1981}
\begin{equation}
	\hat{\sigma}^{(3)}_{\alpha \beta} = \frac {  3 \phi_3 Z \langle F^{\text{nor}} \rangle } { 2\pi a^2 } \int _ { 0 } ^ { \pi } \left[ \zeta _ { N } ( \theta )  n  _ { \theta, \alpha } - \zeta _ { T } ( \theta ) \ t  _ { \theta ,\alpha} \right]  n  _ { \theta, \beta } d \theta, \label{eq:27}
	\end{equation}	
where $Z$ is the average contact number, $\bm{n} _ { \theta } = ( \cos \theta , \sin \theta )$ and $\bm{t} _ { \theta } = (- \sin \theta , \cos \theta )$.
The derivation of this formula can be seen in Appendix C.
Note that three-dimensional effect only appears as the geometrical factor in Eq. \eqref{eq:27} because of the azimuthal symmetry.
This is acceptable since the dominant part of the shear stress comes from the $x$ (shear) and $y$ (shear gradient) directions.
In other words, we neglect the motion of particles in $z$-direction due to tangential forces. This formulation (Eq. \eqref{eq:27}), in addition to the calculation of the lubrication interactions in the Appendix A \cite{kimkarilla1991}, can reproduce the viscosity above the critical shear rate of the DST under the simple shear as shown in the dashed line in Fig. \ref{fgr:steadyvisc}. 
Note that the main contribution for the shear rate larger than the critical one comes from the contact stress in Eq. \eqref{eq:27}.
However, Eq. \eqref{eq:27} can only be used for large shear rates since almost no contact exists for low shear rates. 

We further explore the role of anisotropy as suggested from the angular distribution of the contact forces in Fig. \ref{fgr:angular_std} by analyzing the anisotropy of the stress tensor with the aid of $\Delta \lambda/P$.
Here we introduce the stress anisotropy $\Delta \lambda$ as $\Delta \lambda = \sigma_1 - \sigma_3$, where $\sigma_1$ and $\sigma_3$ are the maximum and the minimum eigenvalues of the stress tensor in 3D system.
On the other hand, the pressure $P$ is calculated as $P=-(\sigma_1+\sigma_2+\sigma_3)/3$ where $\sigma_2$ is the second largest eigenvalue of the stress.
We have confirmed that this stress anisotropy increases as $\dot{\gamma}^{*}$ increases as shown in Fig. \ref{fgr:eigenvalues_std}.
This behavior is similar to in a stress-controlled simulation under a simple shear \cite{thomas2018}.

\subsection{LBM vs LF-DEM}
\begin{figure}[htbp]
 \centering
 \includegraphics[height=0.65\linewidth]{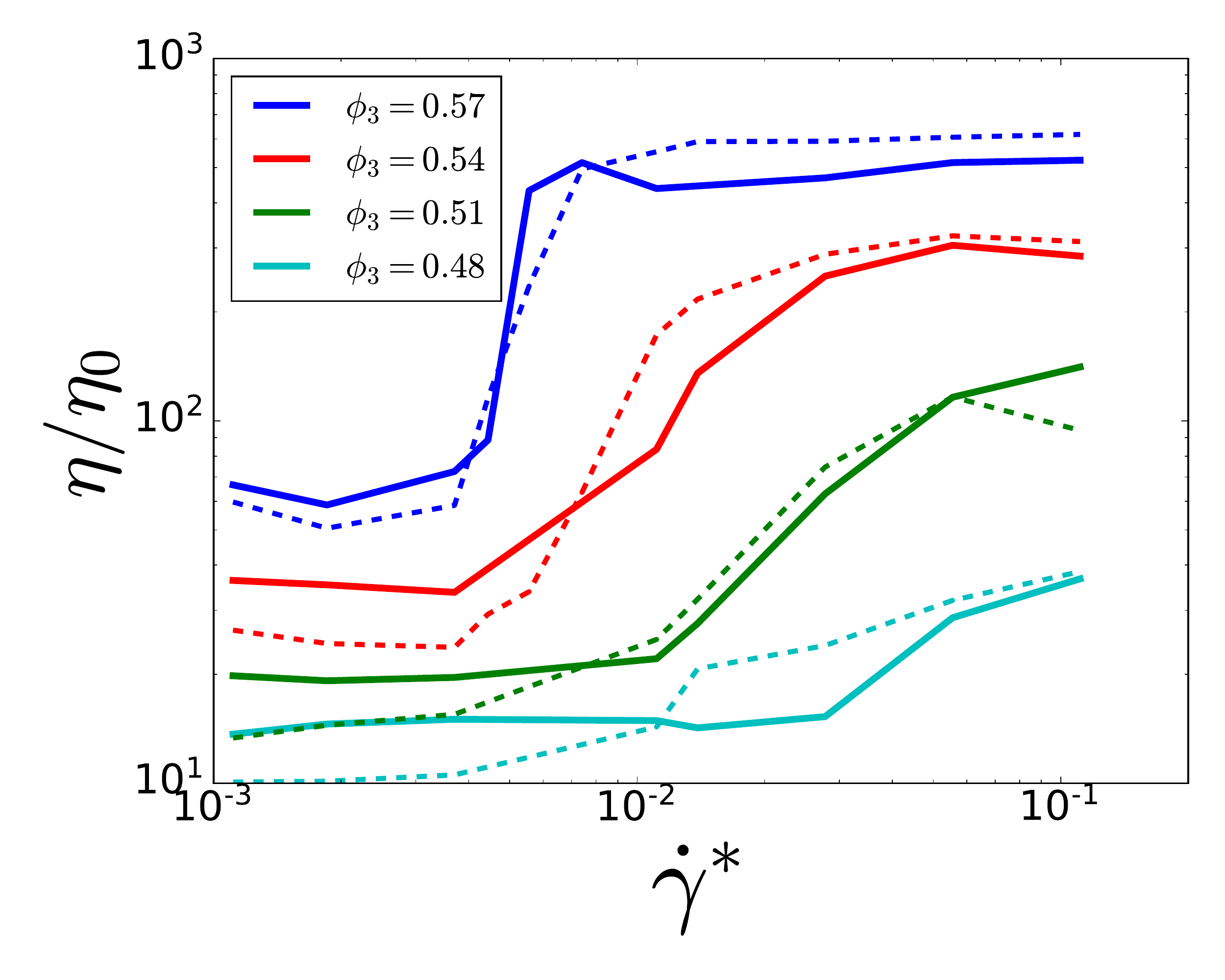}
 \caption{Comparison between the LBM (solid lines) and the LF-DEM (dashed lines) in simple shear for various densities.}
 \label{fgr:lblub_std}
\end{figure}
In this part, let us compare the results of our LBM simulation with those of the LF-DEM to clarify the role of long-range hydrodynamic interaction between particles.
As seen in Fig. \ref{fgr:lblub_std}, one can get qualitatively similar results without using the LBM for high shear regime, though the viscosity by the LBM in the low shear regime quite differs from that by the LF-DEM.  
We also find that the slope of the continuous shear thickening obtained from the LBM is smaller than that for the LF-DEM. This might suggest the shift of the critical density of the DST if we consider the full hydrodynamic interactions. 
Thus, the LF-DEM gives only qualitative results.
The long-range hydrodynamic interactions between particles play some roles if we are interested in a wide range of parameters.
This is one merit to introduce the LBM simulation even for the rheology of dense suspensions.

\section{Oscillatory shear simulation}
This section summarizes the results of our simulation under oscillatory shear flows, which consists of five parts.
In the first part, we explain the setup and the protocol of our simulation.
In the second part, we summarize the results of our simulation using the storage and the loss moduli to characterize the mechanical responses under our protocol.
In the third part, similar to the simple shear case, we illustrate the role of anisotropy of the local contact force to the macroscopic stress anisotropy.
In the fourth part, we discuss the existence of the fragile phase which depends on the initial phase of the applying the oscillatory shear. In the last part, we compare our results of the LBM with those of the LF-DEM.
\subsection{Setup and Protocol}
 	\begin{figure}[htbp]
 	 \centering
 	 \includegraphics[height=0.6\linewidth]{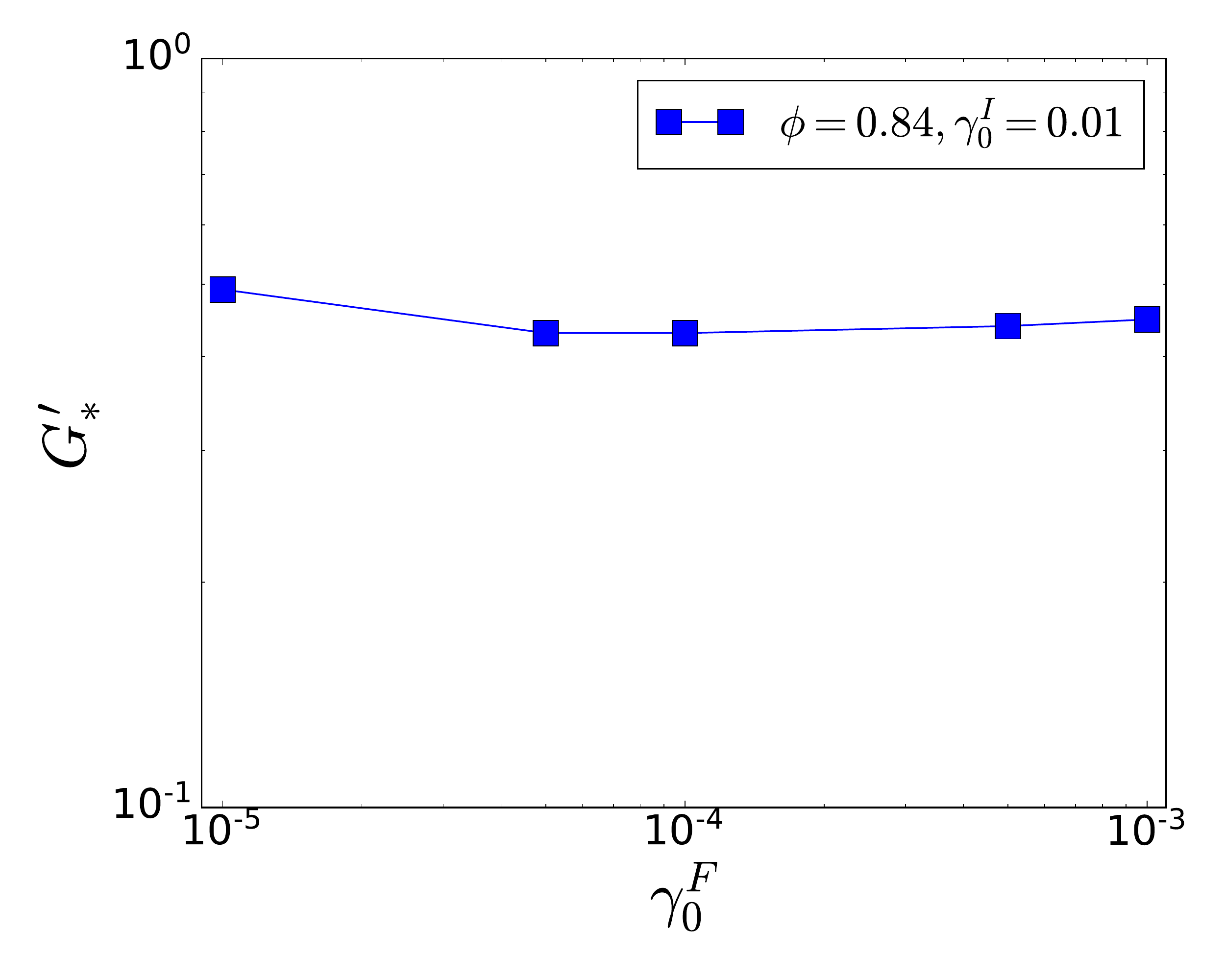}
 	 \caption{Plot of the dimensionless storage modulus $G^{\prime}_{*}$ against various strain amplitude $\gamma_0^F$ after the reduction for $\phi=0.84$ and $\gamma_0^I = 1 \times 10^{-2}$.}
 	 \label{gam}
 	\end{figure}
	 
	\begin{figure}[htbp]
	 \centering
	 \includegraphics[height=0.65\linewidth]{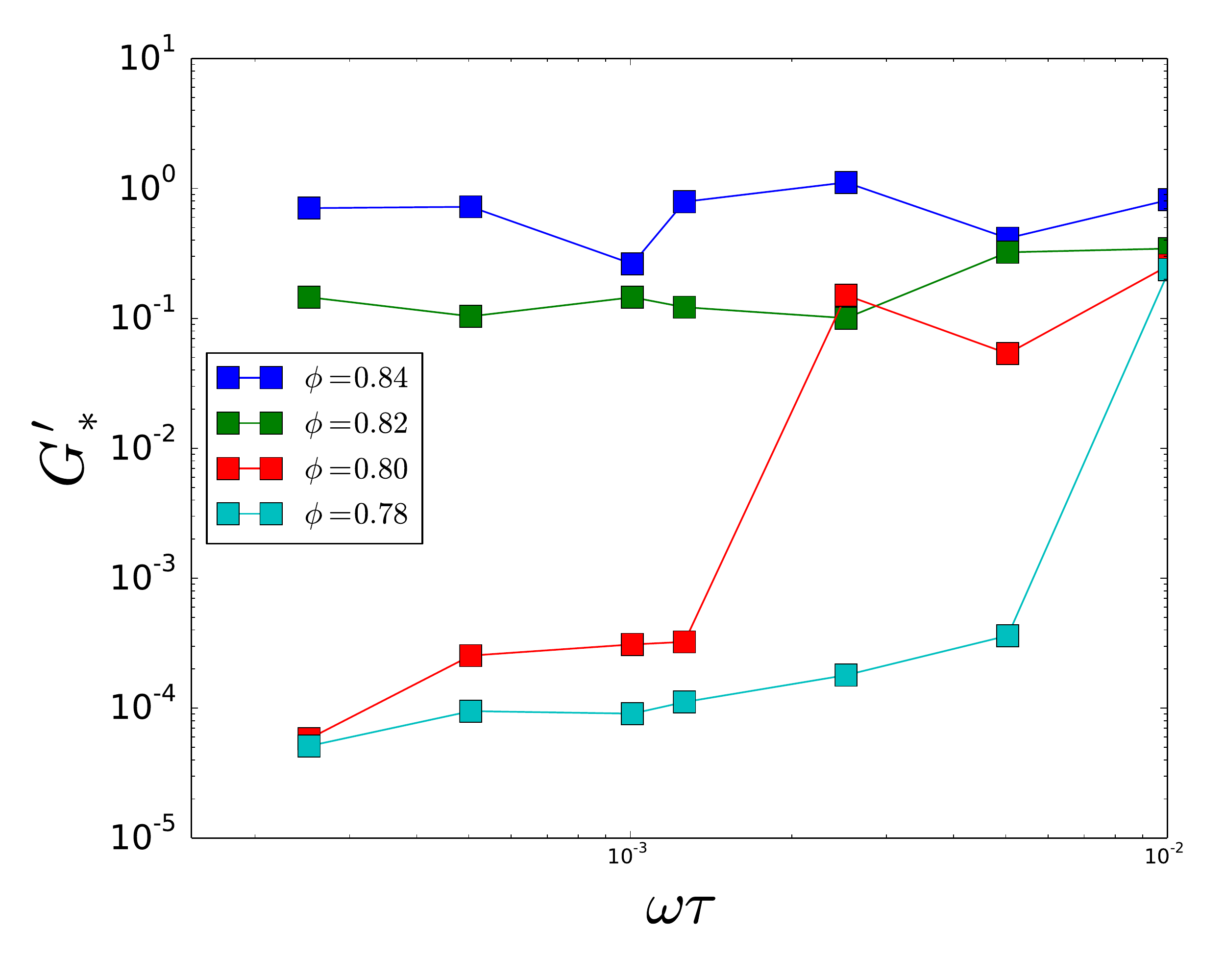}
	 \caption{Plots of the dimensionless storage modulus $G^{\prime}_{*}$ against $\omega \tau$ for $\gamma_0^{I}=1.0$.}
	 \label{fgr:omega}
	\end{figure}
Our simulation for oscillatory shears contains $N=450$ particles confined in a quasi-two-dimensional box with $L_z=4a_{\rm max}
$ with the radius $a_{\rm max}$ of the largest particle.
We discuss size dependence for simulations by the LF-DEM in the Appendix D.
The motion of particles is considered as complete two-dimensional one to keep the monolayer configuration.
Therefore, instead of using $\phi_3$ we use the area fraction  $\phi = N\pi (a_{\text{max}}^2 + a_{\text{min}}^2)/(2L_xL_y)$ to characterize the density in our monolayer system.
Note that we simulate 3D spheres instead of 2D disks since the hydrodynamic interactions are only well defined in three dimensional systems.
One of the reasons to use the monolayer configurations is to save computational time, which is commonly used by some previous simulations \cite{thomas2018,otsuki2018}. The second reason to use the configurations is the easiness of the visualization of the microstructure of particles.
Furthermore, monolayered suspensions have also been studied in some experiments on lipid domains \citep{klingler1993} and on fluid surfaces \cite{langevin2014,petkov2014,kralchevsky2015,higuera2016}.
The snapshot of the monolayer configuration is shown in Fig. \ref{fgr:snapshot}.

We implement a time-dependent oscillatory shear strain as
\begin{equation}
	 \gamma(t) = \gamma_0 [\cos \Theta - \cos(\omega t + \Theta)], \label{eq:16}
	\end{equation} 
where $\gamma_0$ is the strain amplitude, $\omega$ denotes the angular frequency, and $\Theta$ is the initial phase which characterizes the asymmetricity of the strain.
This is equivalent as implementing a strain rate $\dot{\gamma}(t) = \omega \gamma_0 \sin (\omega t + \Theta) $.
We measure the mechanical response in terms of \cite{doiedwards}
\begin{align}
	G^{\prime} &= - \frac{\omega}{\pi} \int_0^{2 \pi/ \omega} dt \frac{\sigma_{xy}(t) \cos(\omega t + \Theta)}{\gamma_0},  \label{eq:17} \\ 
	G^{\prime \prime} &= \frac{\omega}{\pi} \int_0^{2 \pi/ \omega} dt \frac{\sigma_{xy}(t) \sin(\omega t + \Theta)}{\gamma_0}, \label{eq:18}
	\end{align}		
where $G^{\prime}$ is the storage modulus corresponding to the rigidity and $G^{\prime \prime}$ is the loss modulus which is proportional to the dynamic viscosity $\eta := G^{\prime \prime}/\omega$.
For later discussion we have introduced the dimensionless storage and loss moduli $G^{\prime}_{*}$ and $G^{\prime \prime}_{*}$ as $G^{\prime }_*=G^{\prime}/k_n$ and $G^{\prime \prime}_{*}=G^{\prime \prime}/k_n$. As a result, the dimensionless viscosity $\eta^{*}$ is given by $\eta^{*} = G^{\prime \prime}_{*} / \omega \tau$. We have also introduced the dimensionless stress $\sigma^{*}_{\alpha  \beta}$ as $\sigma^{*}_{\alpha  \beta}=\sigma_{\alpha  \beta}/k_n$.  
As mentioned in the introduction, this formulation is only reliable in the linear response regime.
Some authors use different definitions of $G^{\prime}$ and $G^{\prime \prime}$ to handle the nonlinear responses\cite{marenne2017,ness2017,brader2010,ewoldt2008}.
Note that the stress-strain curve in the linear response regime is not expressed as a simple straight line because of the existence of $G^{\prime \prime}$.
The linear response means that $G'$ and $G"$ are independent of the strain amplitude.
To avoid uncertainty of the definition of the linear response functions $G^{\prime}$ and $G^{\prime \prime}$ in the nonlinear regime, we follow the protocol introduced by Otsuki and Hayakawa \cite{otsuki2018} in which we reduce the strain amplitude $\gamma_0$ to be in the linear response regime as $\gamma_0^{F}$ for observation after initial cycles with the large initial strain amplitude $\gamma_0^{I}$.

In other words, our protocol tries to extract the memory effect of the initial oscillation.
When $\gamma_0^{F} \leq 10^{-3}$, the storage modulus $G^{\prime}$ is almost independent of $\gamma_0^{F}$ (Fig. \ref{gam}), which confirms that our observation of the mechanical responses is in the linear response regime.
Therefore, all oscillatory shear simulations adopt $\gamma_0^{F}=10^{-4}$.

We have simulated range of $\omega$ as shown in Fig. \ref{fgr:omega} to check its dependencies and confirmed that the response is almost independent of $\omega$ for $\omega \tau \le 1.1 \times 10^{-3}$.
Therefore, we use $\omega \tau  = 1 \times 10^{-3}$ in our simulations.
For electrostatic repulsive force between particles in the oscillatory shear, we adopt $F_0=2\times 10^{-4}k_n$.
The control parameters of our simulation are the initial strain amplitude $\gamma_0^{I}$, the area fraction $\phi$, and the initial phase $\Theta$.
Our results below are obtained after ten initial cycles and averaged over nine final observation cycles after the reduction of the strain amplitude.
We average the results for three ensembles of different initial configurations.
The convergences for the number of initial cycles $N^I_c$ and observation cycles $N^F_c$ can be seen on the Appendix B. 

\subsection{Mechanical responses}

\begin{figure}[htbp]
 \centering
 \includegraphics[height=0.75\linewidth]{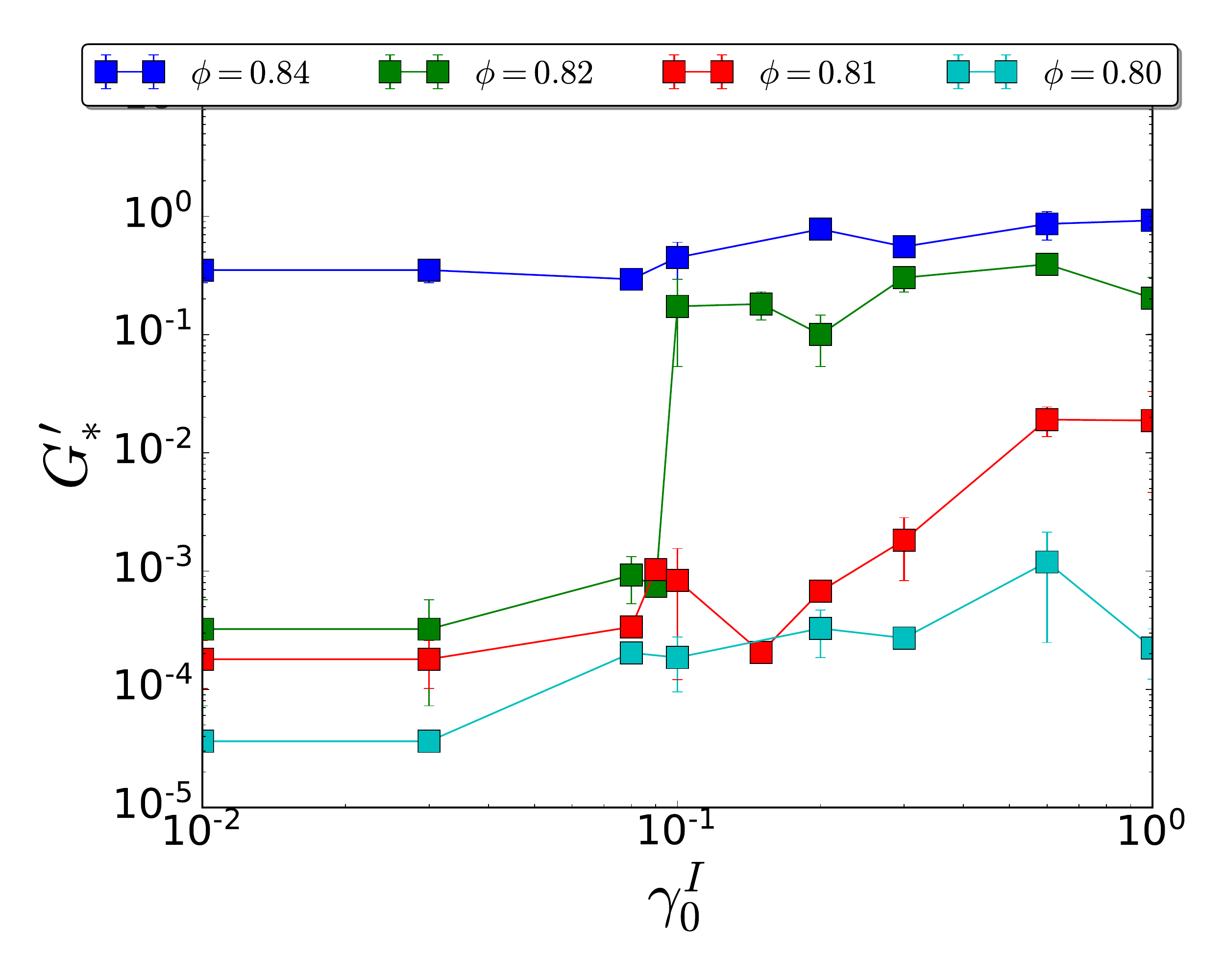}
 \caption{Plots of the dimensionless storage modulus against the initial strain amplitude $\gamma_0^{I}$ for various $\phi$.}
 \label{fgr:mechresponsestor}
\end{figure}
\begin{figure}[htbp]
 \centering
 \includegraphics[height=0.75\linewidth]{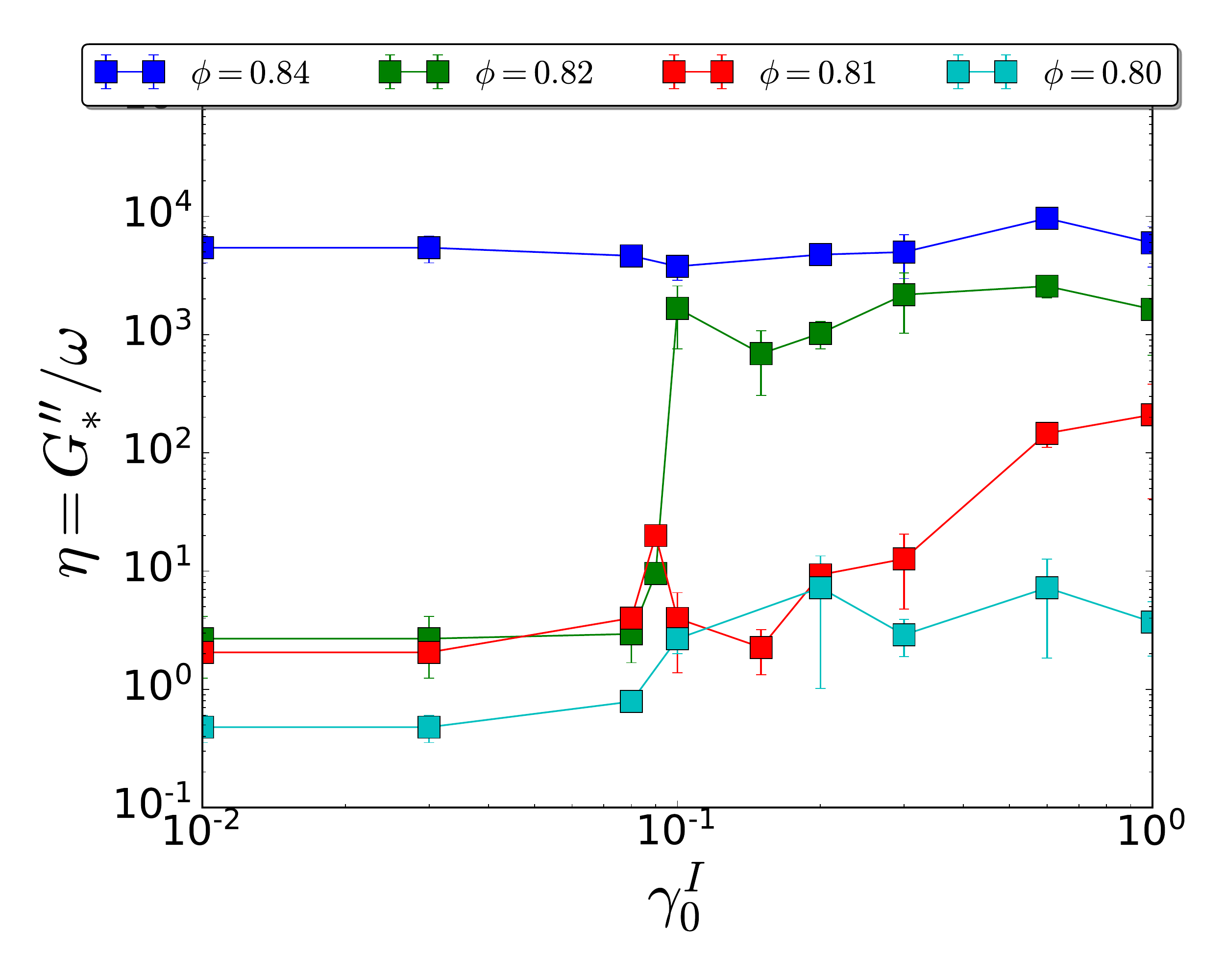}
 \caption{Plots of the dimensionless viscosity against the initial strain amplitude $\gamma_0^{I}$ for various $\phi$.}
 \label{fgr:mechresponsevisc}
\end{figure}

\begin{figure}[htbp]
 \centering
 \begin{subfigure}{.5\textwidth}
 \centering
 \includegraphics[height=.8\linewidth]{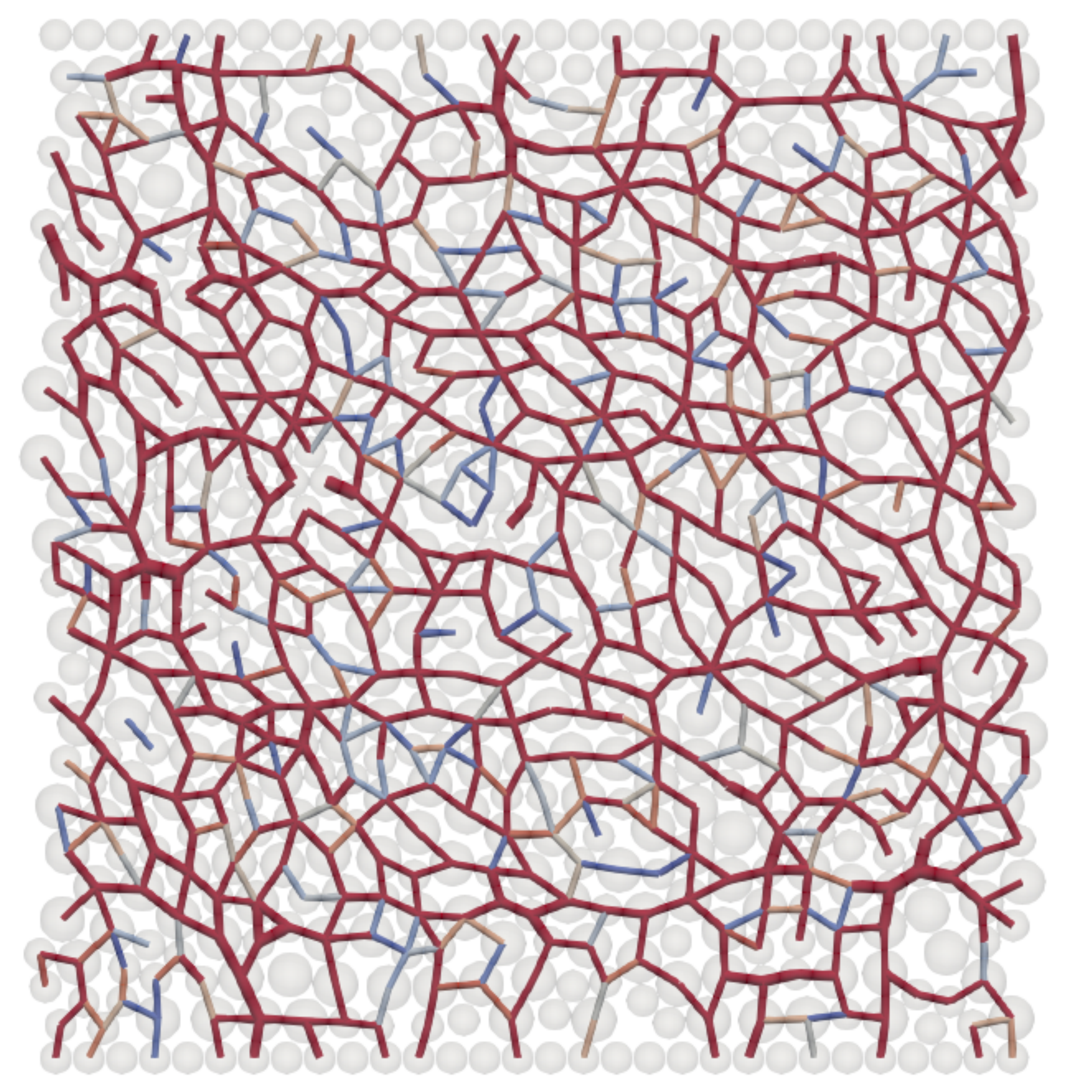}
 \caption{}
 \end{subfigure}
 \centering
 \begin{subfigure}{.5\textwidth}
 \centering
 \includegraphics[height=.8\linewidth]{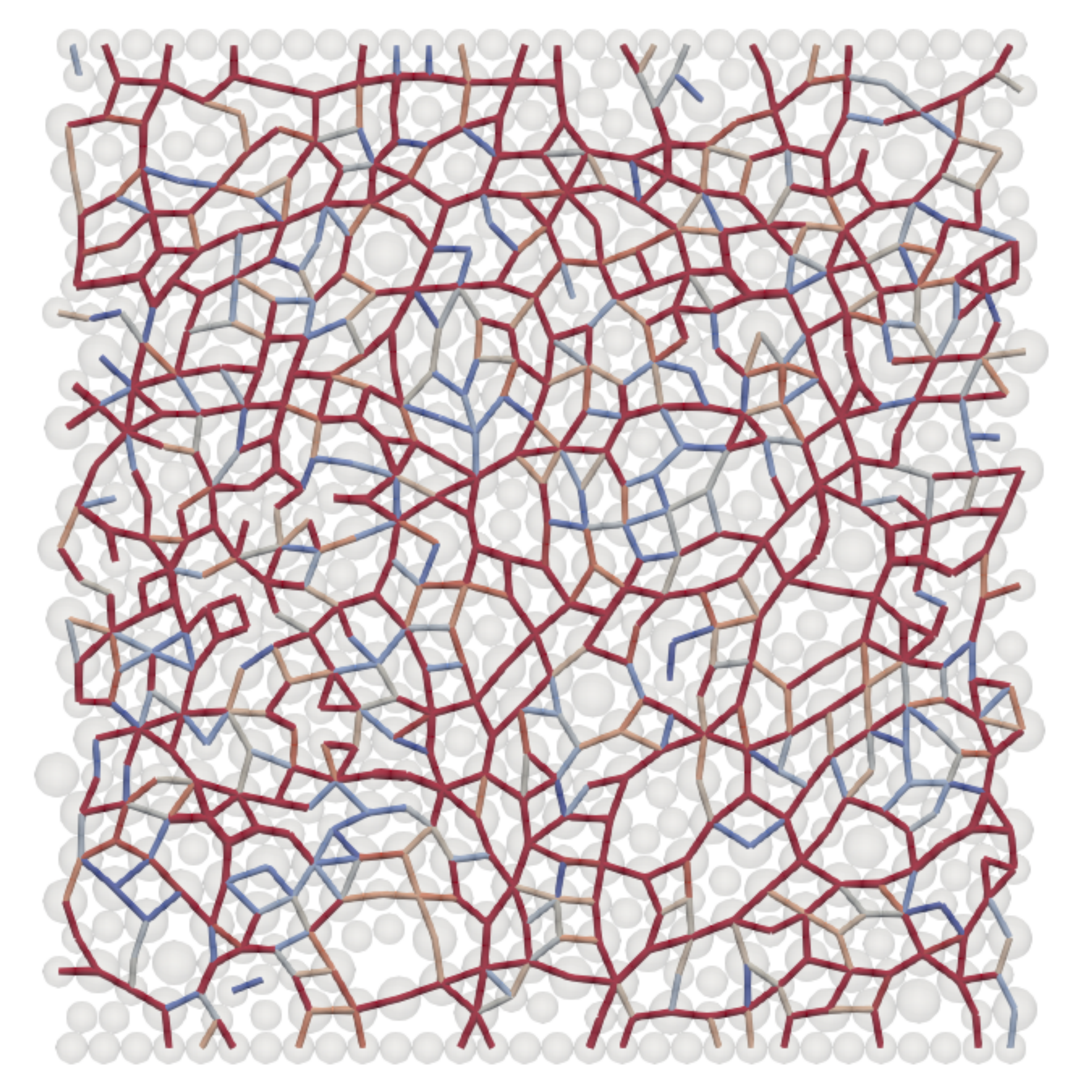}
 \caption{}
 \end{subfigure}
 \caption{Snapshots of force chains for $\phi=0.82$ and $\gamma_0^{I}=0.1$ (a) before the reduction ($\omega t = 17 \pi$) of the strain amplitude and (b) after the reduction ($\omega t = 37 \pi$) of the amplitude where the line thickness represents the magnitude of the normal forces $|\bm{F}^{\text{nor}}|$.}
 \label{fgr:fig}
 \end{figure}

\begin{figure}[htbp]
  	 \centering
  	 \begin{subfigure}{.5\textwidth}
  	 \centering
  	 \includegraphics[height=.8\linewidth]{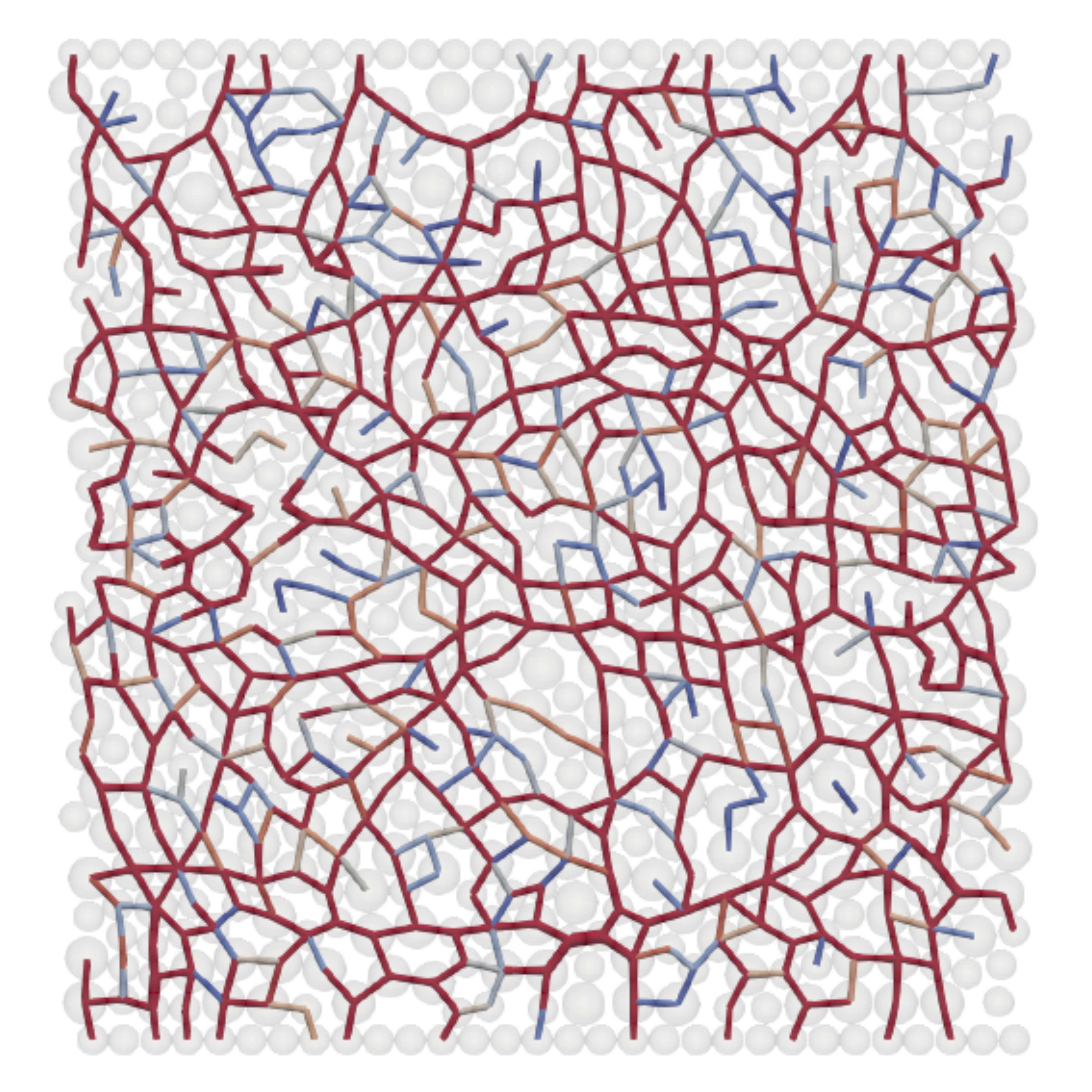}
  	 \caption{}
  	 \end{subfigure}
  	 \centering
  	 \begin{subfigure}{.5\textwidth}
  	 \centering
  	 \includegraphics[height=.8\linewidth]{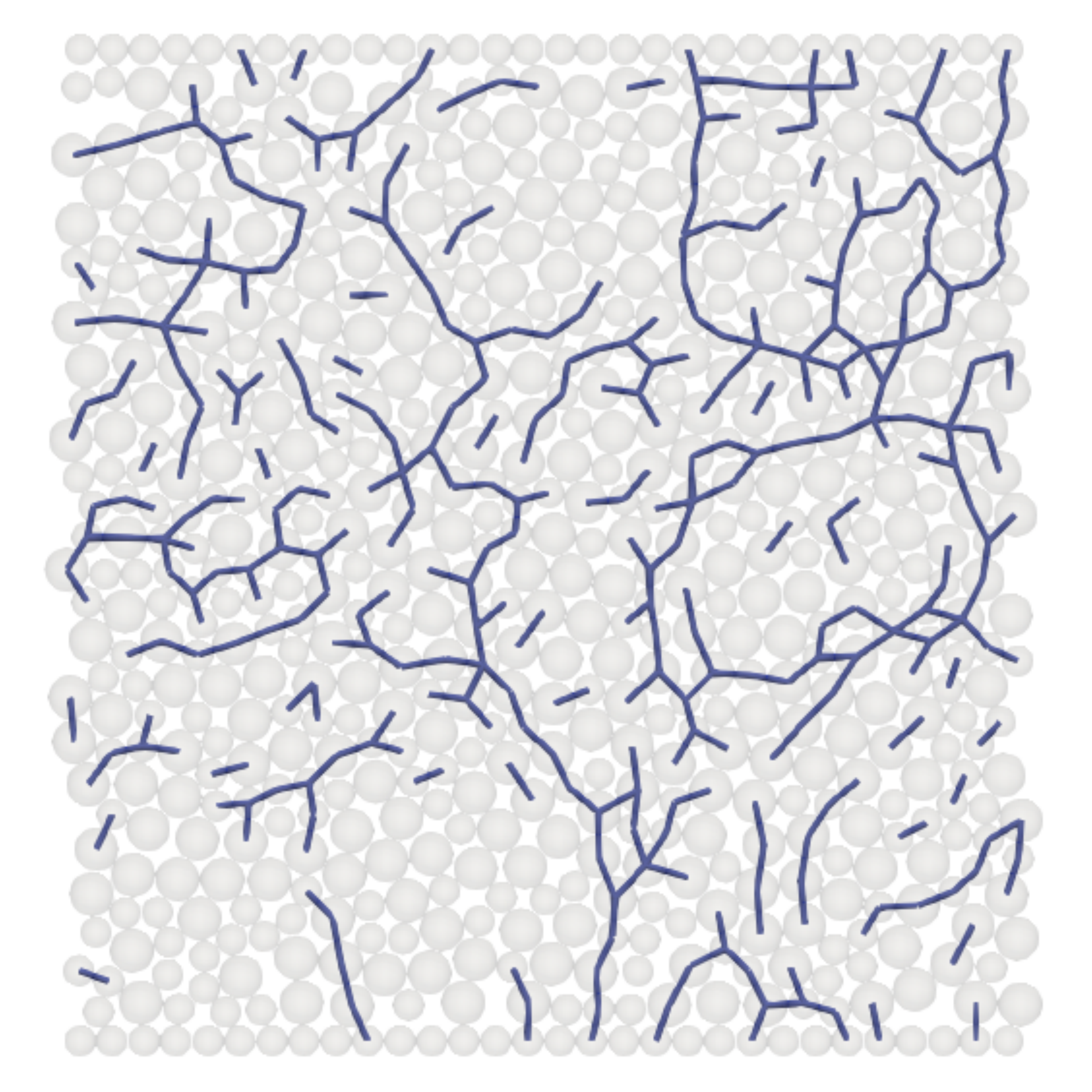}
  	 \caption{}
  	 \end{subfigure}
 	 \caption{Snapshots of force chains for $\phi=0.82$ and $\gamma_0^{I}=0.08$ (a) before the reduction ($\omega t = 17 \pi$) of the strain amplitude and (b) after the reduction ($\omega t = 37 \pi$) of the amplitude, where the line thickness stands for the magnitude of the normal forces $|\bm{F}^{\text{nor}}|$.}
 	 \label{fgr:fig2}
  	 \end{figure}

 \begin{figure}[htbp]
 	 \centering
 	 \begin{subfigure}{.5\textwidth}
 	 \centering
 	 \includegraphics[height=.7\linewidth]{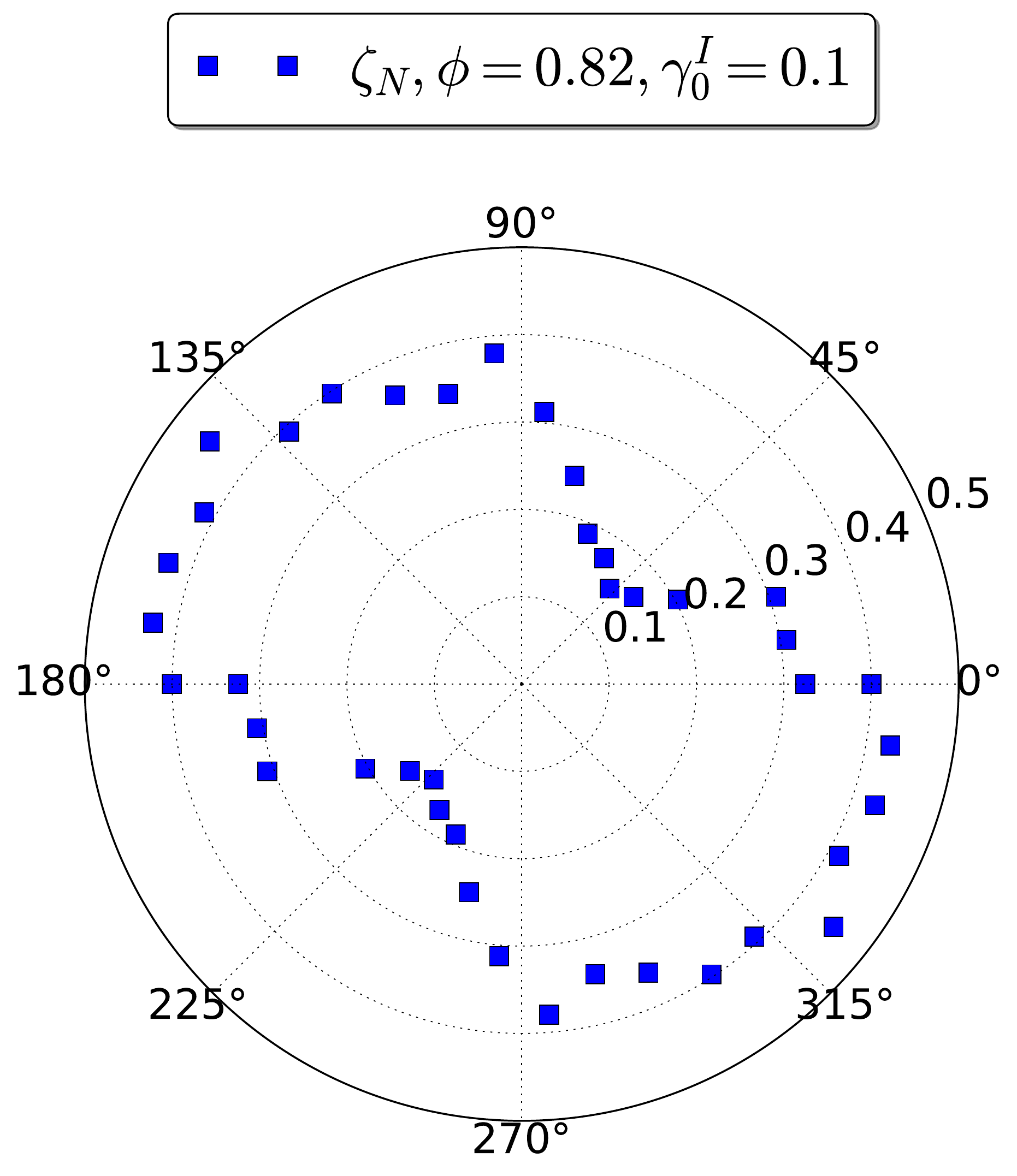}
 	 \caption{}
 	 \end{subfigure}
 	 \centering
 	 \begin{subfigure}{.5\textwidth}
 	 \centering
 	 \includegraphics[height=.7\linewidth]{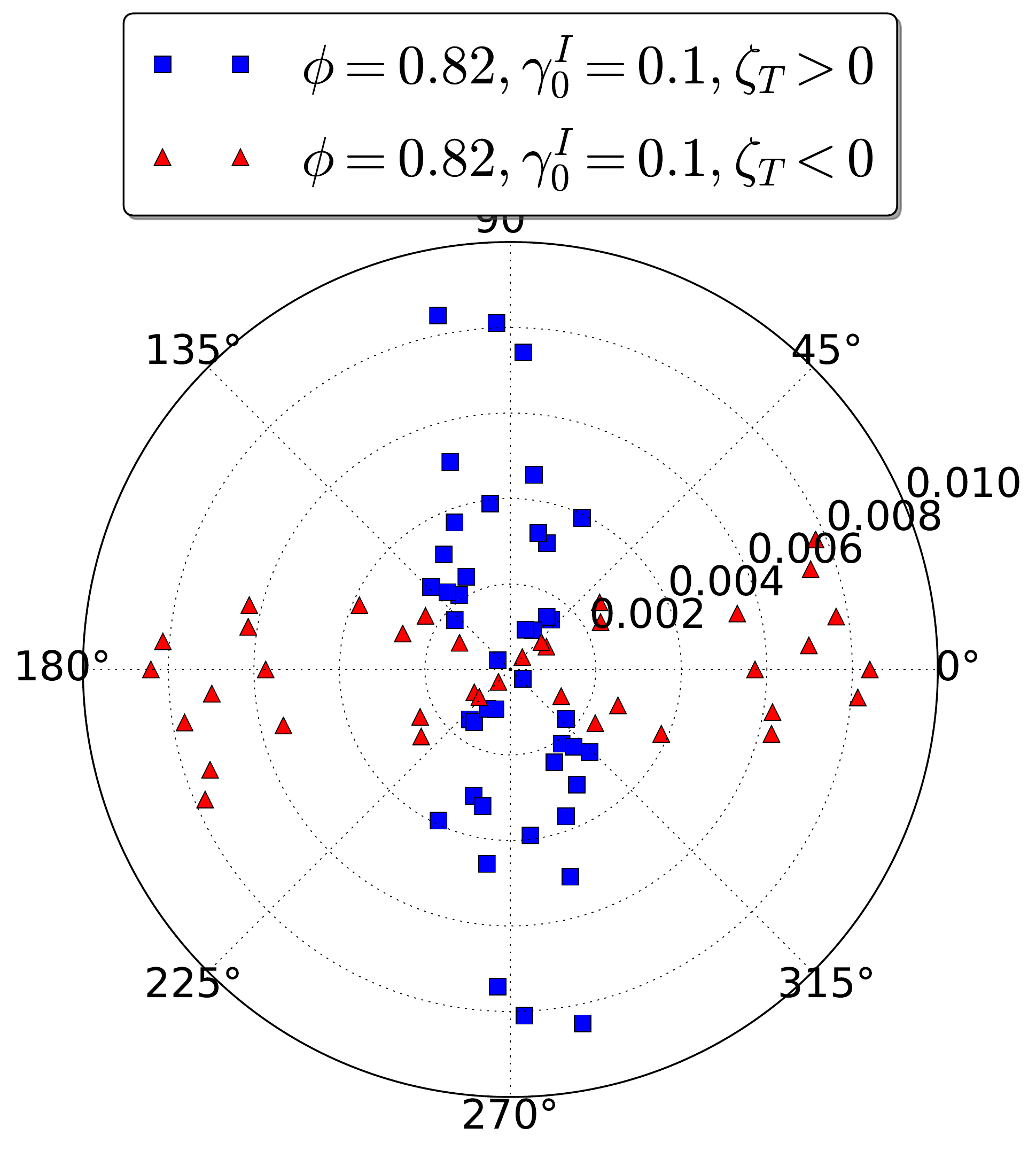}
 	 \caption{}
 	 \end{subfigure}
 	 \caption{The angular distributions of the normal (a) and the tangential (b) contact forces under oscillatory shear, respectively,  averaged from $\omega t =  11 \pi$, $\omega t = 13 \pi$, $\omega t = 15 \pi$, $\omega t = 17 \pi$, and $\omega t = 19 \pi$.}
 	 \label{fgr:angular}
 	 \end{figure}
	 
	\begin{figure}[htbp]
	 \centering
	 \begin{subfigure}{.5\textwidth}
	 \centering
	 \includegraphics[height=.65\linewidth]{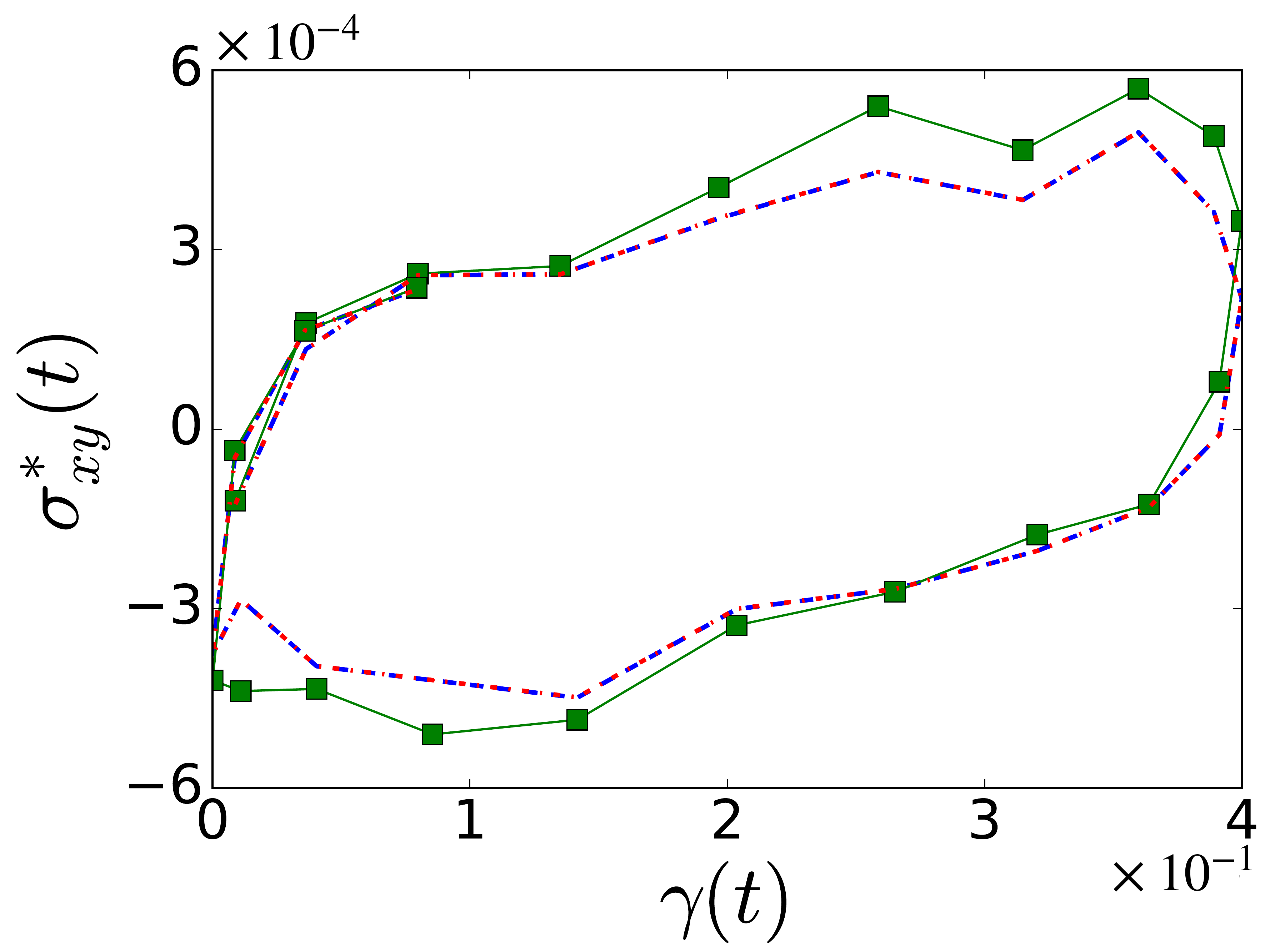}
	 \caption{}
	 \end{subfigure}
	 \centering
	 \begin{subfigure}{.5\textwidth}
	 \centering
	 \includegraphics[height=.65\linewidth]{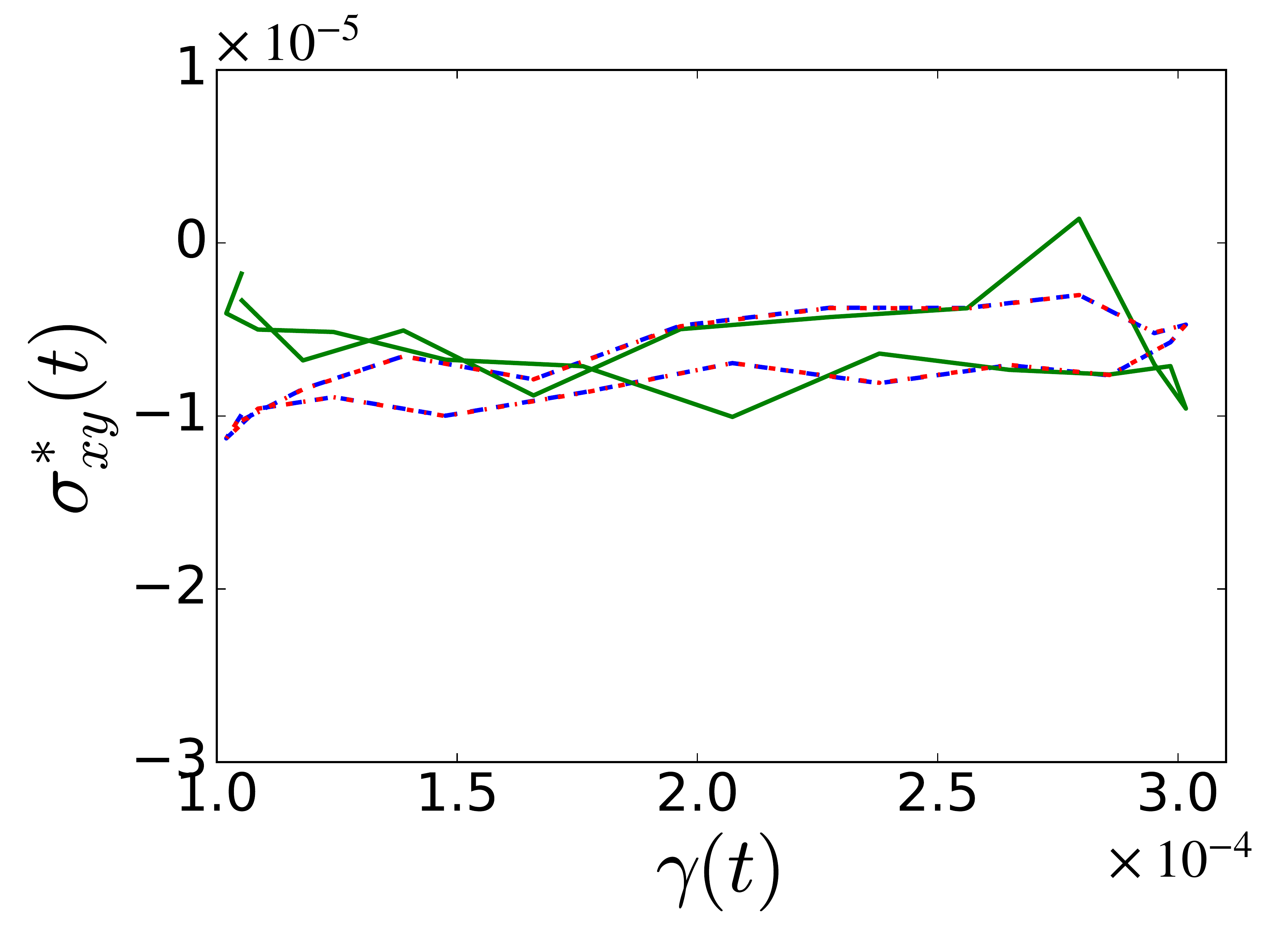}
	 \caption{}
	 \end{subfigure}
	 \caption{The stress-strain curve for $\phi = 0.82$ from Eq. \eqref{eq:26} (the dot-dashed lines), Eq. \eqref{eq:26} plus the lubrication forces (the dashed lines), compared with our simulation results (the solid lines) (a) before the reduction of the amplitude and (b) after the reduction.}
	 \label{fgr:theorstress}
	 \end{figure}

Let us look at the storage and the loss moduli in the linear response regime under the oscillatory shear.
Our obtained results of the storage modulus and the viscosity under oscillatory shear are shown in Figs. \ref{fgr:mechresponsestor} and \ref{fgr:mechresponsevisc}, respectively.
One can observe the existence of finite storage modulus for all $\gamma_0^{I}$ on $\phi= 0.84$.
For $\phi=0.82$, the storage and the loss moduli have discontinuous jumps at critical value of $\gamma_0^{I}$.
Note that the storage modulus below the critical $\gamma_{0}^{I}$ is almost zero while the loss modulus in this region is small but finite. 

\begin{figure}[htbp]
	 \centering
	 \includegraphics[height=0.65\linewidth]{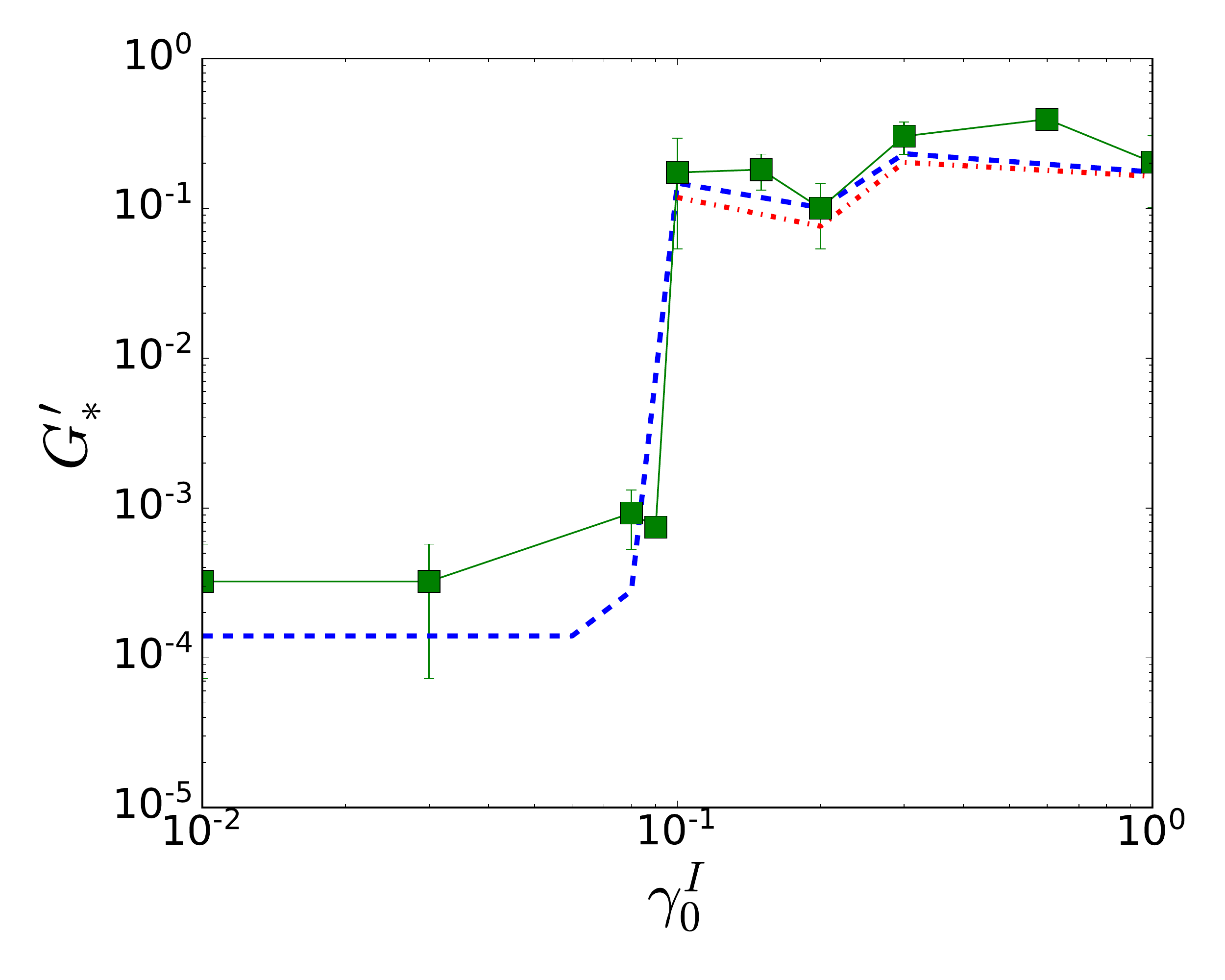}
	 \caption{Phenomenological storage modulus obtained from Eq. \eqref{eq:26} only (the dot-dashed line) and Eq. \eqref{eq:26} plus the lubrication contribution (the dashed line) compared with our simulation results (the solid line) for $\phi =  0.82$.}
	 \label{fgr:stortheor}
	\end{figure} 	

From these results, we can identify the jammed, unjammed, and shear-jammed states with the aid of $G^{\prime}$.
First we introduce a threshold value of the storage modulus, $G^{\prime, \text{th}}=10^{-3}k_n$. If $G^{\prime} > G^{\prime, \text{th}}$, we regard the system as having finite storage modulus.
The unjammed state is when $G^{\prime} < G^{\prime, \text{th}}$.
The isotropic jammed state is when $G^{\prime} > G^{\prime, \text{th}}$ for all $\gamma_0^{I}$ at a given $\phi$ or $G^{\prime} > G^{\prime, \text{th}}$ for low $\gamma_0^{I}$.
Meanwhile, we define the shear-jammed state as the state when $G^{\prime} > G^{\prime,\text{th}}$ for high $\gamma_0^{I}$ and unjammed for low $\gamma_0^{I}$. 
From the results of the viscosity in Fig. \ref{fgr:mechresponsevisc}, we observe finite jumps for $\phi=0.81$ and $\phi=0.82$, while the viscosity is almost independent of $\gamma_0^{I}$ on $\phi=0.84$.
The jumps of the viscosity observed here correspond to the DST in the simple shear\cite{williamson1931,barnes1989,egres2005,seto2013}.

Both of the abrupt increases on the storage modulus and the viscosity can be understood as the appearance of a percolating force chains after the reduction of the strain amplitude.
This clearly can be seen in Figs. \ref{fgr:fig} and \ref{fgr:fig2} where we visualize the force chains before and after the reduction for $\phi=0.82$, $\gamma_0^{I}=0.1$ in the shear-jammed state and $\gamma_0^{I}=0.08$ in the unjammed state in Figs. \ref{fgr:fig} and \ref{fgr:fig2}.
It is easy to find that the percolated force chains survive after the reduction of the strain amplitude in the shear-jammed state, while they disappear after the reduction in the unjammed state.

\subsection{Anisotropy of the contact forces}
Here, we explore the anisotropy of the stress tensor and the contact forces in our simulation by analyzing the angular distributions of the contact forces (Fig. \ref{fgr:angular}) as in the simple shear case (Eqs.\eqref{eq:19} and \eqref{eq:20}).
The angular distribution of the contact force for large $\gamma_0^{I}$ are almost equivalent to those observed in the simple shear.
Moreover, the angular distribution of the normal contact force is almost isotropic for isotropic jammed state, while the stress anisotropy is clearly visible in the shear jammed state.
For two-dimensional cases, we replace Eq. \eqref{eq:27} by \cite{dacruz2005}
\begin{equation}
	\hat{\sigma}_{\alpha \beta} = \frac {  \phi Z \langle F^{\text{nor}} \rangle } { \pi a } \int _ { 0 } ^ { \pi } \left[ \zeta _ { N } ( \theta )  n  _ { \theta, \alpha } - \zeta _ { T } ( \theta ) \ t  _ { \theta ,\alpha} \right]  n  _ { \theta, \beta } d \theta. \label{eq:26}
	\end{equation}

Figure \ref{fgr:theorstress} compares the theoretical shear stress in Eq.\eqref{eq:26} with the results of our simulation.
This agreement confirms that the contact contributions are dominant.
However, this analysis cannot be applied to the unjammed states since almost no contact exists in these regimes, whereas only the hydrodynamic contributions exist as shown in Fig. \ref{fgr:stortheor}. 
	 			
\subsection{Initial phase dependence and fragile state}

\begin{figure}[htbp]
 \centering
 \includegraphics[height=0.75\linewidth]{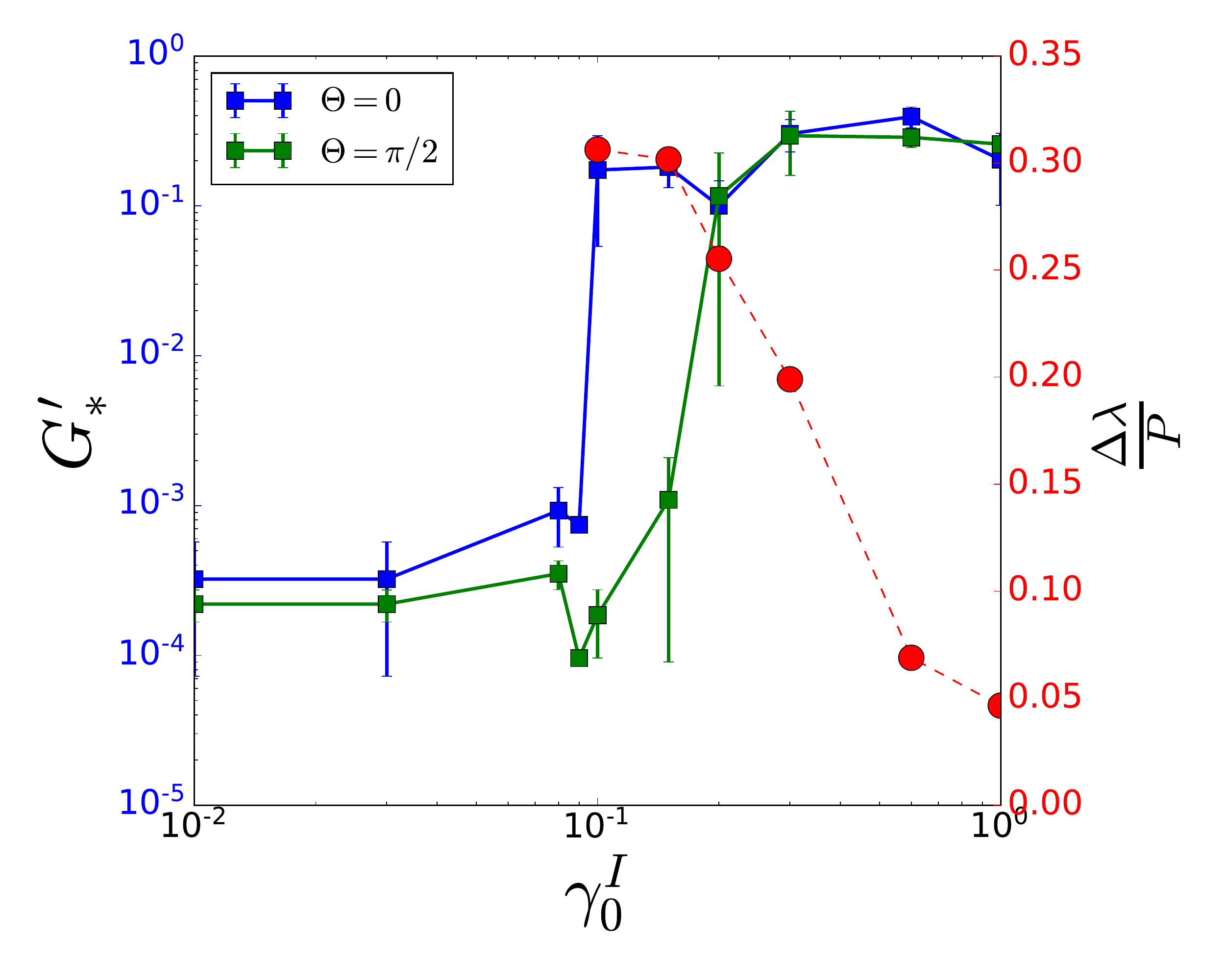}
 \caption{Plots of the dimensionless storage modulus $G^{\prime}_{*}$ against $\gamma_0^{I}$ for $\phi=0.82$ with $\Theta=0$ (blue filled squares) and $\Theta=\pi/2$ (green filled squares) alongside with the stress anisotropy $\Delta \lambda/P$ (red circles).} 
 \label{fgr:fragile82}
\end{figure}

\begin{figure}[htbp]
   \centering
   \includegraphics[height=0.7\linewidth]{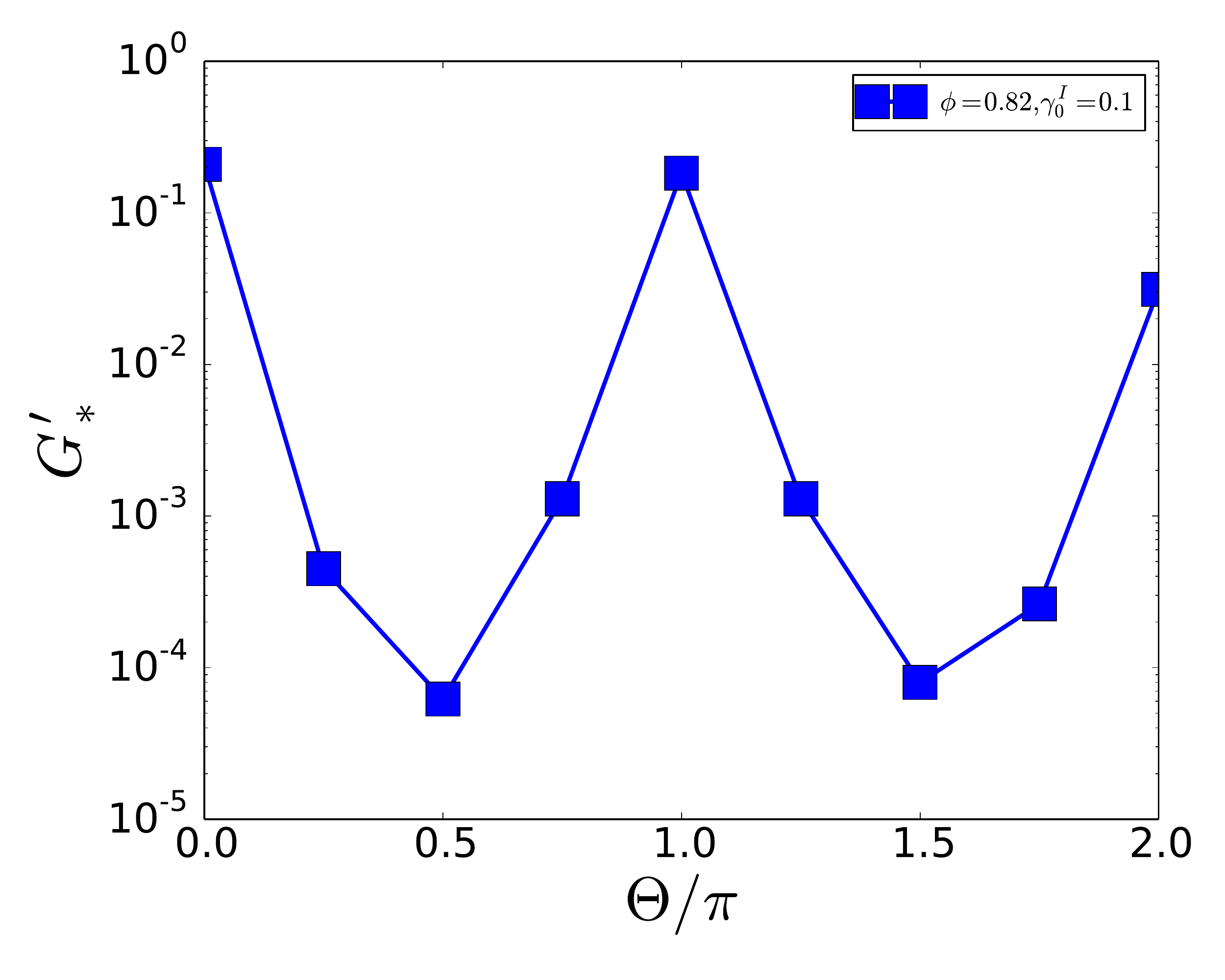}
   \caption{Plot of the dimensionless storage modulus $G^{\prime}_{*}$ against various initial phases $\Theta$ for $\phi = 0.82$ and $\gamma_0^{I} = 0.1$.}
   \label{fgr:phasedepend}
   \end{figure}

 \begin{figure}[htbp]
  \centering
  \begin{subfigure}{.5\textwidth}
  \centering
  \includegraphics[height=.6\linewidth]{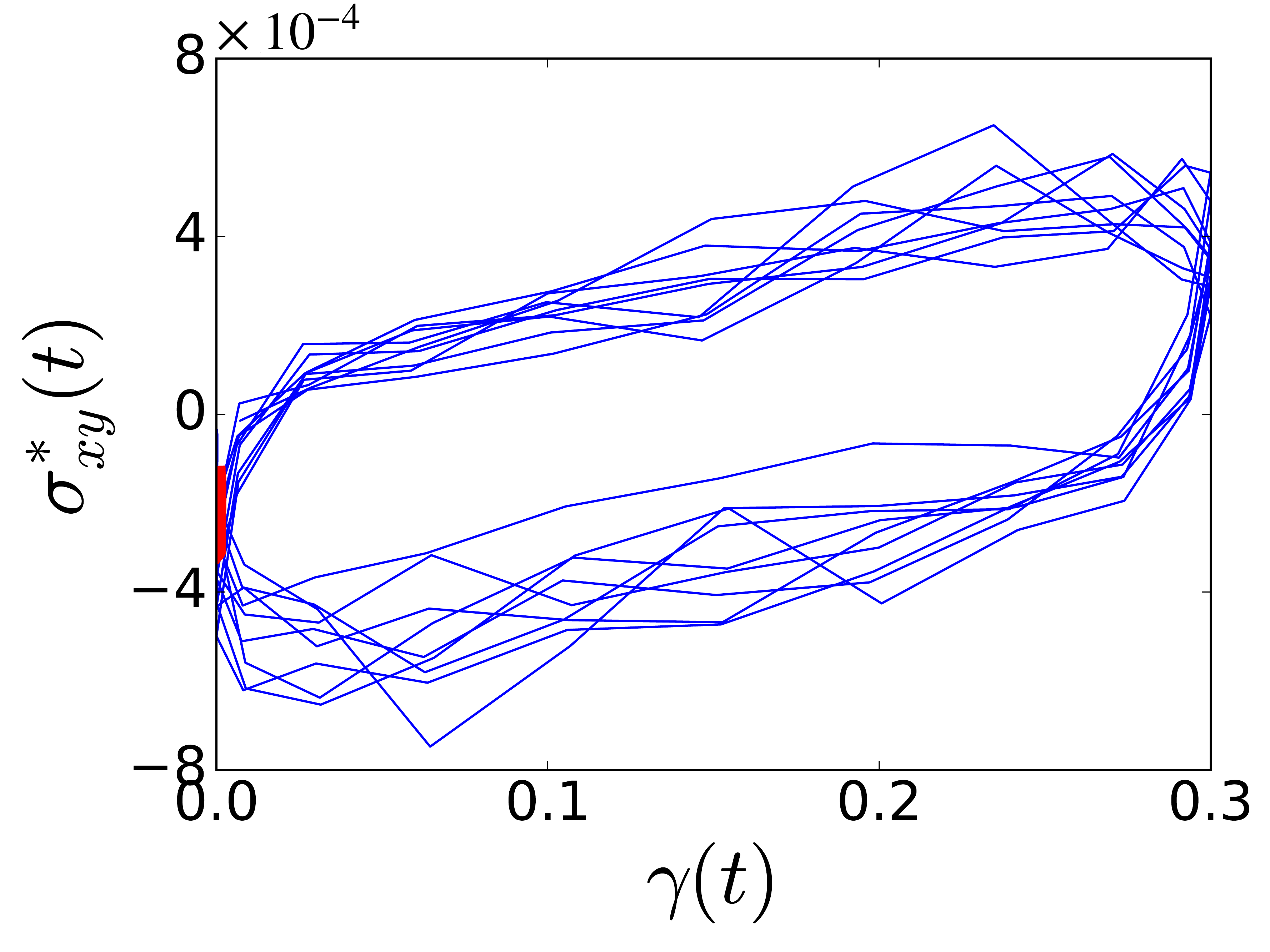}
  \caption{}
  \end{subfigure}
  \centering
  \begin{subfigure}{.5\textwidth}
  \centering
  \includegraphics[height=.6\linewidth]{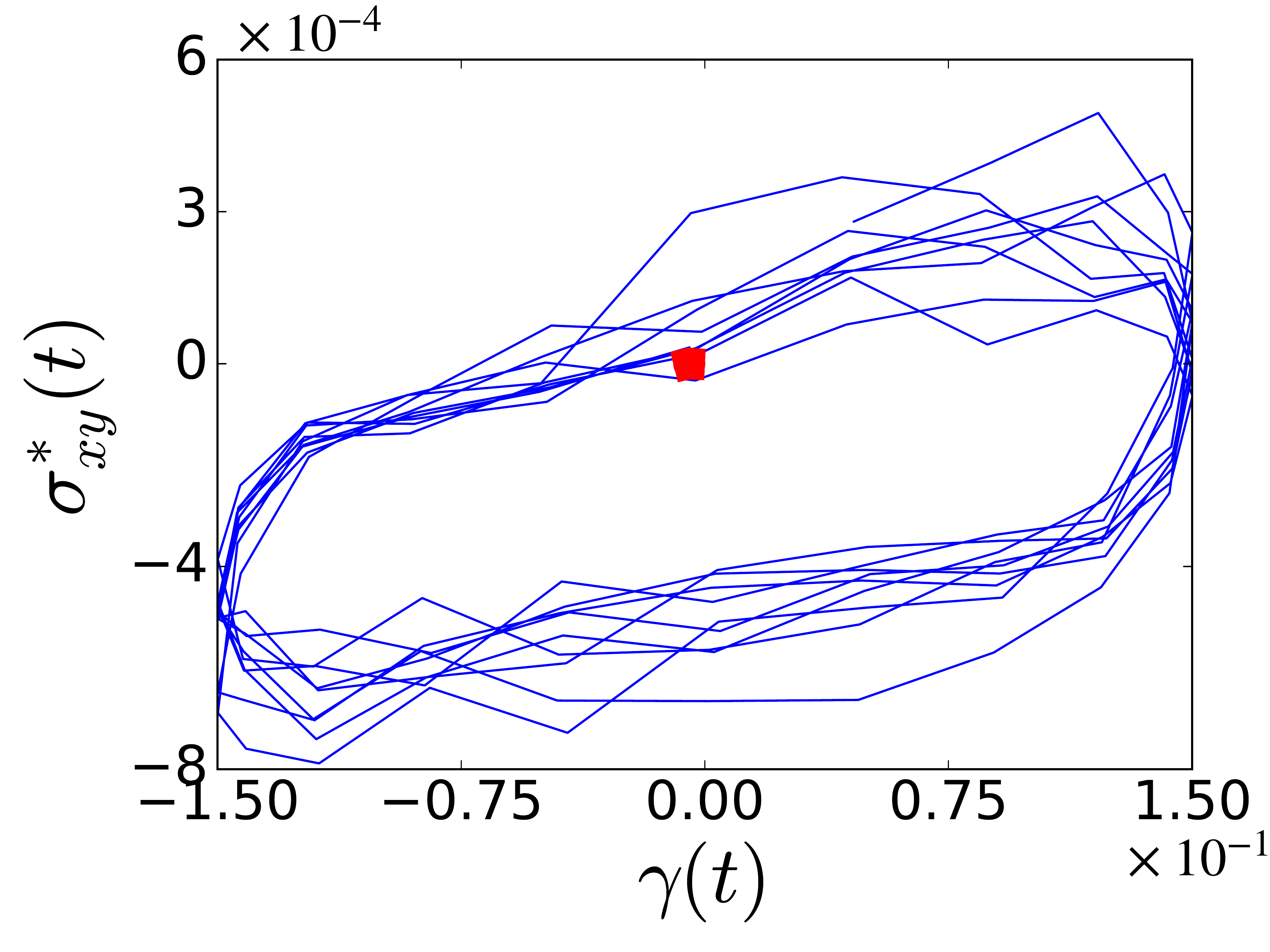}
  \caption{}
  \end{subfigure}
  \caption{Plots of the stress-strain curves for $\phi=0.82$ and $\gamma_0^{I} = 0.15$.  Plots of the data for (a) $\Theta = 0$ and (b) $\Theta = \pi/2$, respectively. The red lines show the stress-strain plot after the reduction of the strain amplitude.}
  \label{fgr:fragilestressstrain}
  \end{figure}

Now let us clarify properties of the fragile state within our simulations.
Originally, the fragile state is defined as the state where the system can only sustain load in a particular direction \cite{cates1998,bi2011}.
This suggests that both solid-like and fluid-like responses can be observed depending on the the initial phase $\Theta$ or the asymmetricity of the strain introduced in Eq. \eqref{eq:16} in the fragile state.
Therefore we try to explore how this duality can happen for given control parameters $\phi$, $\gamma_0^{I}$, and the initial phase $\Theta$.
In our protocol, % where we measure the response after the reduction of the strain amplitude,
the different initial phase $\Theta$ essentially corresponds to the different position of the observation in the stress-strain curve (see Fig. \ref{fgr:fragilestressstrain}).
If the state is fragile, we expect that the response depends on $\Theta$.

In Fig. \ref{fgr:fragile82}, we plot the storage modulus against $\gamma_0^{I}$ for $\phi = 0.82$ and $\Theta = 0$ and $\Theta = \pi/2$.
Here we can see that the points at $ \gamma_0^{I} = 0.1$ and $ \gamma_0^{I} = 0.15$ have $G^{\prime} > G^{\prime,\text{th}}$ for $\Theta = 0$ while they have $G^{\prime} < G^{\prime,\text{th}}$ for $\Theta = \pi/2$.
Hence, we confirm that this $\Theta$-dependence can be used for the definition of the fragile state as in the dry granular case \cite{otsuki2018}.
We also analyze the stress anisotropy as in the simple shear by using $\Delta \lambda = \sigma_1 - \sigma_2$ divided by the pressure $P = - \frac{1}{2}(\sigma_1 + \sigma_2)$.
Here, $\sigma_1$ and $\sigma_2$ are the maximum and the minimum eigenvalues of the stress tensor, respectively.
In Fig. \ref{fgr:fragile82}, we plot the stress anisotropy averaged in the last five cycles after the reduction.
We confirm that the stress anisotropy in this fragile state is much larger than that in the shear jammed state as expected in Refs. \cite{bi2011,otsuki2018}.
This observation agrees with the experiment of granular materials \cite{sarkar2016}, the simulations of frictionless dry granular particles\cite{chen2018} and frictional dry granular disks \cite{otsuki2018}. 
It is reasonable that anisotropy is large in the fragile state while it is small in the shear jammed state, because the shear jammed state has more rigid percolating force chains to sustain in all directions while the fragile state has weak force chains which are connected only in the compressive direction.

\begin{figure}[htbp]
  \centering
  \includegraphics[height=0.7\linewidth]{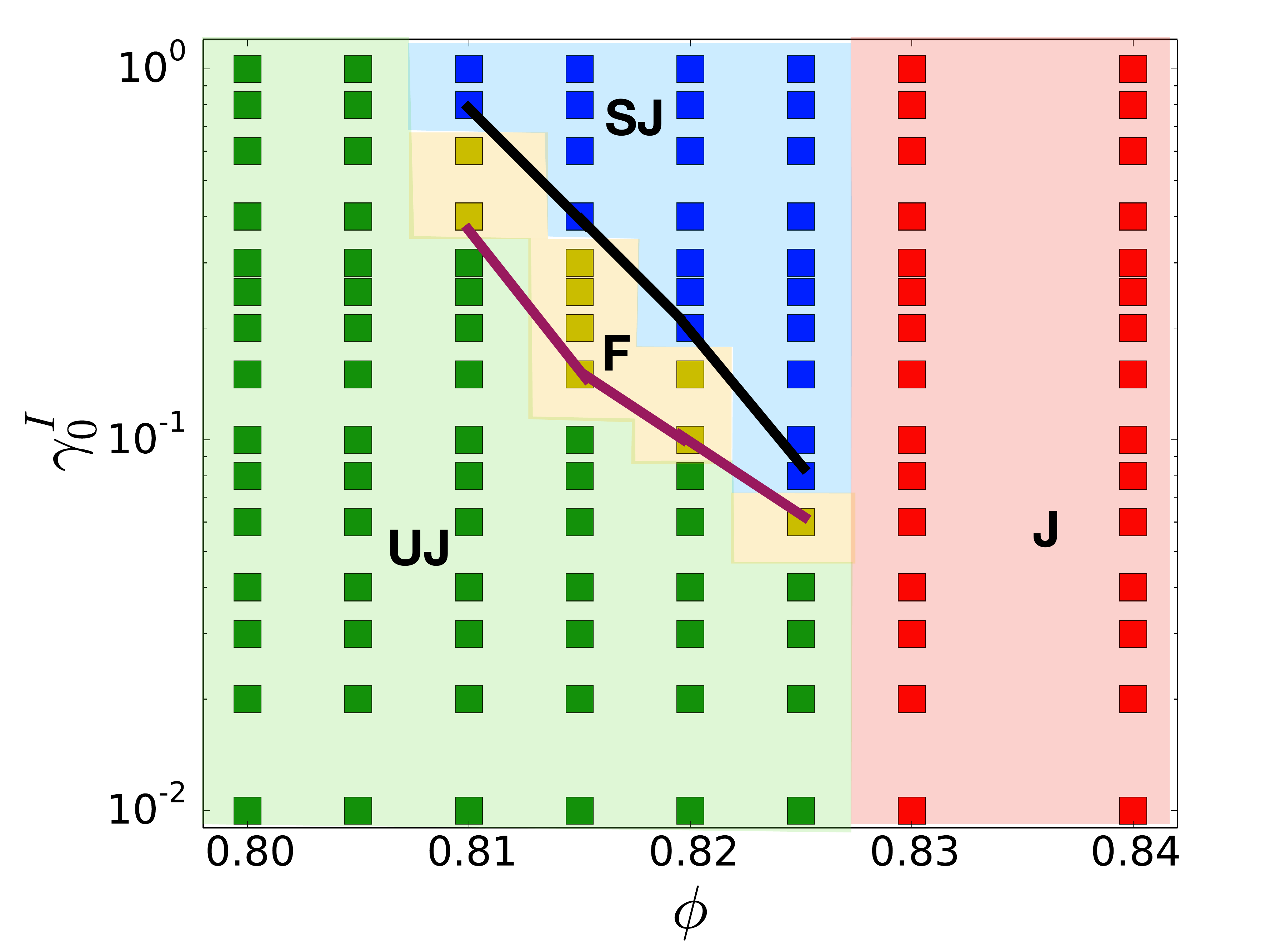}
  \caption{Phase diagram on the plane of the control parameters $\phi$ and $\gamma_0^{I}$ where red, blue, green, and yellow symbols represent the jammed (\textbf{J}), shear-jammed (\textbf{SJ}), unjammed (\textbf{UJ}), and fragile states (\textbf{F}), respectively. The purple and the black lines represent the DST-like phenomena for $\Theta=0$ and $\Theta=\pi/2$, respectively.}
  \label{fgr:phasediagram}
  \end{figure}

The $\Theta$-dependence of $G^{\prime}$ can be seen clearly in Fig. \ref{fgr:phasedepend} for $\phi = 0.82$ and $\gamma_0^{I} = 0.1$.
Note that Fig. \ref{fgr:phasedepend} is obtained from the LF-DEM simulation.
Furthermore, the change of the observation point depending on $\Theta$ can be seen from the stress-strain curve in Fig. \ref{fgr:fragilestressstrain} which is obtained from the LBM. This $\Theta-$dependent relationship between the initial stress-strain curve and the observation point is the origin of the $\Theta$-dependent response.

To summarize the results of our oscillatory shear simulation, we draw the phase diagram for the dense monolayer suspensions under oscillatory shears (with our control parameters, area fraction $\phi$ and initial strain amplitude $\gamma_0^{I}$) in Fig. \ref{fgr:phasediagram}.
Note that this phase diagram is drawn by using the LF-DEM because of the limitation of our computational resources.
Our phase diagram is quite similar to that for the dry granular materials in the corresponding protocol \cite{otsuki2018}.
Furthermore, our distinction within the shear-jammed and isotropic jammed states is also similar to the experimental results by controlling the shear rate for cornstarch \cite{fall2015} with additional fragile states on the onset of the shear jamming states. 

\subsection{LBM vs LF-DEM}

\begin{figure}[htbp]
 \centering
 \includegraphics[height=0.7\linewidth]{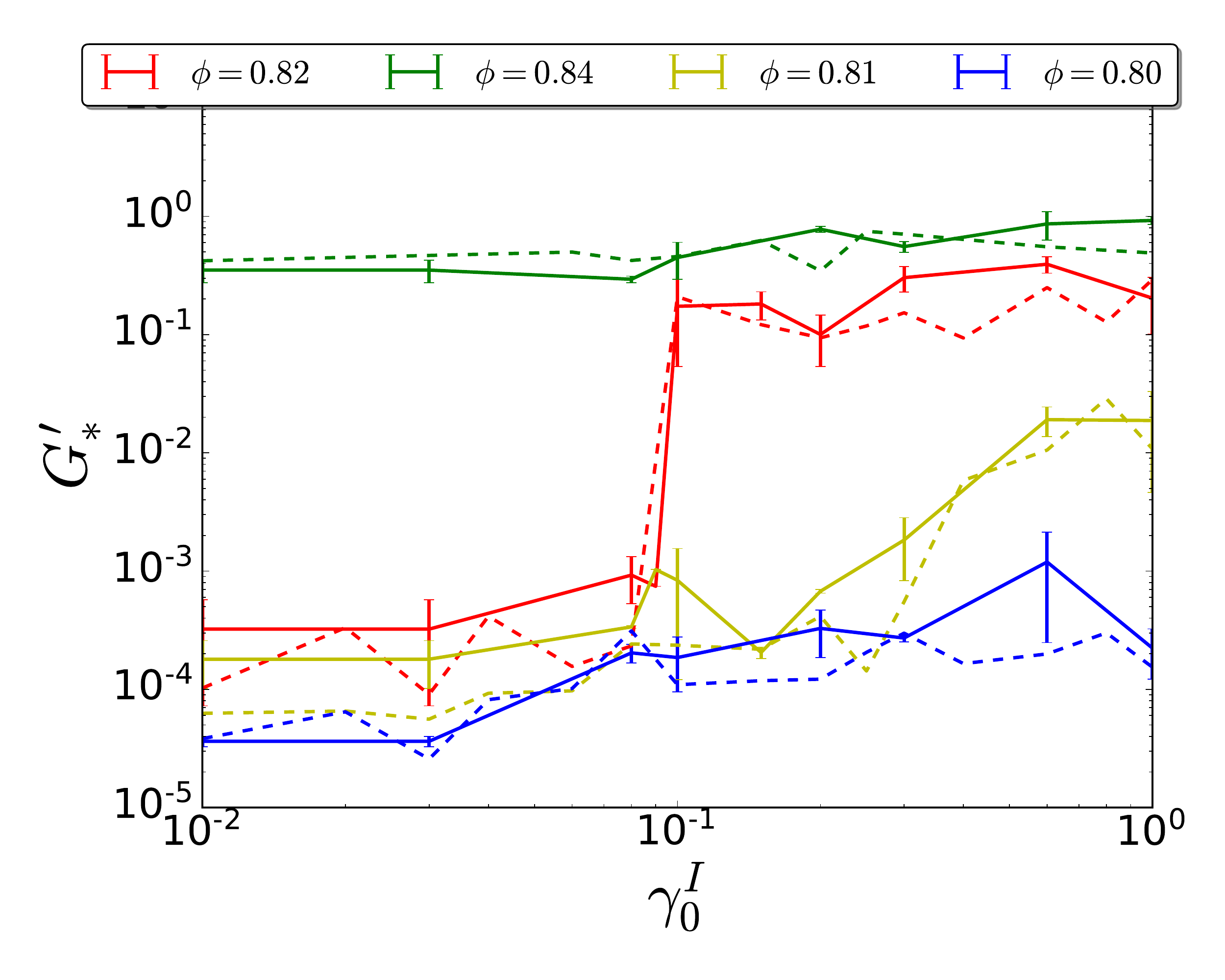}
 \caption{Comparison between the LBM (solid lines) and the LF-DEM (dashed lines) in oscillatory shear for various densities.}
 \label{fgr:lblub_osc}
\end{figure}

In this part, we clarify the difference between the LF-DEM and the LBM simulation.
From Fig. \ref{fgr:lblub_osc}, we can observe a qualitative agreement between both approaches for high shear regime.
It is natural that the hydrodynamic long range interactions play a minor role for the emergence of exotic states such as the shear jamming and the fragile states.
Nevertheless, the LF-DEM cannot be used for low density regime and the low shear regime because it does not contain the full hydrodynamic interactions.

\section{Discussions and Conclusions}

In conclusion, we have numerically studied dense non-Brownian suspensions under the simple shear and the oscillatory shear.
We have successfully reproduced the DST in our simple shear simulation.
For oscillatory shear, we have successfully characterized the unjammed, the jammed, the shear-jammed, and the fragile states as a memory effect of the initial shear by the reduction of the strain amplitude.
We have drawn the phase diagram for monolayered dense non-Brownian suspensions on the plane of the initial strain amplitude and the area fraction.
We have found that the viscosity under oscillatory shear also has an abrupt increase which corresponds to the DST.
Furthermore, we have defined the fragile state as a state which depends on the asymmetricity of the strain.
Moreover, we have also analyzed the angular distributions of the contact forces characterizing the DST and the shear jamming under the simple shear and the oscillatory shear, respectively.
We have expressed the shear stress, the storage modulus at high initial strain amplitude, and the viscosity for large shear rate as functions of these angular contact distributions.
The agreement between this phenomenology and the results of our simulation clarifies that the contact contributions are much larger than those of the hydrodynamics in the shear-jammed regime under the oscillatory shear and for large shear rate under the simple shear.
This agreement also indicates that one needs to construct a theory for the angular distribution\cite{suzuki2019} of contact forces, i.e., derive the $\zeta_N$ and $\zeta_T$ theoretically to recover the mechanical responses at high strain regime and the viscosity for large shear rate.
Finally, we have also illustrated that the stress anisotropy takes its maximum on the onset of the shear jamming, where it is corresponds to the fragile state and DST.

Let us discuss our future perspectives.
There are several unsolved problems which have not been addressed in this paper.
First, the effect of dilatancy is missing in our analysis because the system is incompressible.
However, the effect of dilatancy becomes significant in sand beaches, which are mixtures of air, water, and grains.
Recently, dilatancy and compaction of dry granular materials are studied under an oscillatory shear and in a pressure-control simulation\cite{daisuke2019arxiv}, where the anisotropy of the stress tensor also plays an important role.
Therefore, we need to perform the simulation of a mixture of air, water, and grains.
Second, one can also investigate the rheology of an intruder particle and measure the drag force and effective friction coefficient, which has been performed experimentally on the granular hydrogels immersed in waters\cite{panaitescu2017}.
Finally, we can also numerically study a tapping problem or impact-activated solidification which has been experimentally showed that it is related to the formation of the jamming front\cite{waitukaitis2012,peters2014,han2018}.
One may simulate this process by \textit{the coupled DEM-LBM for the free surface simulation}\cite{leonardi2014}.

\section*{Appendix A Details of the simulation method}
\subsection*{Lattice Boltzmann Method}
To calculate the hydrodynamic forces acting on the particles in the LBM, we introduce solid nodes inside the particles, fluid nodes on the surrounding fluid, and boundary nodes on the surface of the particles.
The hydrodynamic field is calculated from the time evolution of the discrete distribution function $n(\bm{r},\bm{c}_{\bm{q}},t)$ which has the dimension of mass density.
At each fluid node $\bm{r}$, $n(\bm{r},\bm{c}_{\bm{q}},t)$ is updated as			
  	\begin{equation}
  		n (\bm{r} + \bm{c}_{\bm{q}} \Delta t,\bm{c}_{\bm{q}}, t + \Delta t ) = n(\bm{r},\bm{c}_{\bm{q}},t) + \Delta_{\bm{q}} [n(\bm{r},\bm{c}_{\bm{q}},t)], \label{eq:3} 
  		\end{equation}
where $\bm{c}_{\bm{q}}$ is the lattice velocity with direction $\bm{q}$,  $\Delta t$ is the time step, and $\Delta_{\bm{q}}$ is the collision operator that depends on all $n(\bm{r},\bm{c}_{\bm{q}},t)$.
The left-hand side of Eq. \eqref{eq:3} expresses the change of the distribution function by the streaming effect, whereas the right-hand side expresses the change by the collision.
Our simulation uses nineteen directions of $\bm{c}_{\bm{q}}$ (D3Q19 lattice for 19 directions/quadratures in 3 dimensions) \cite{succi2001}.
Some moments of $n(\bm{c}_{\bm{q}})$ which is the abbreviation of $n(\bm{r},\bm{c}_{\bm{q}}, t)$ are related to the hydrodynamic variables as
	\begin{align}
		\rho = \sum_{\bm{q}} n(\bm{c}_{\bm{q}}), \qquad \bm{j} = \sum_{\bm{q}} n (\bm{c}_{\bm{q}})\bm{c}_{\bm{q}},  \qquad \bm{\Pi} = \sum_{\bm{q}} n(\bm{c}_{\bm{q}}) \bm{c}_{\bm{q}} \textbf{c}_{\bm{q}}. \label{eq:4}  
		\end{align}
In this paper, we use the linearized collision operator near the equilibrium distribution 
\begin{equation}
	\Delta_{\bm{q}} (n) = \sum_p \mathcal{L}_{\bm{qp}}n^{\rm neq}(\bm{c}_{\bm{p}}), \label{eq:5}  
	\end{equation}
where $n^{\rm neq}(\bm{c}_{\bm{p}})=n(\bm{c}_{\bm{p}})-n^{\rm eq}(\bm{c}_{\bm{p}})$ is the deviation from the equilibrium distribution and $\Delta_{\bm{q}}(n)$ satisfies $\Delta_{\bm{q}} (n^{\rm eq})=0$.
It is not necessary to specify an explicit collision operator $\mathcal{L}_{\bm{qp}}$ for the calculation of the hydrodynamic variables introduced in Eq. \eqref{eq:4}.
Instead, it is sufficient to consider the eigenvalues and eigenvectors of $\mathcal{L}_{\bm{qp}}$ for the collision process.
Details of this collision operator are not discussed in this paper and can be found in Refs. \cite{ladd1994a,ladd1994b,nguyen2002}.

The discrete distribution function on the boundary node is updated with a bounce-back rule on the surface of each particle.
At each boundary node, there exists virtual (post-collisional) distribution function  $n^{*}(\bm{r},\bm{c}_{\bm{q}},t)=n(\bm{r},\bm{c}_{\bm{q}},t) + \Delta_{\bm{q}} (\bm{r},t)$ instead of using Eq. \eqref{eq:3}.
Moreover, we need an additional term for the time updating of the boundary nodes
			\begin{equation}
				n (\bm{r},\bm{c}_{\bm{q}^{\prime}},t+\Delta t) = n^{*}(\bm{r},\bm{c}_{\bm{q}},t) - \frac{2 C^{\bm{q}} \rho_0 \bm{u}_b \cdot \bm{c}_{\bm{q}}}{c_{s}^{2}}, \label{eq:7}
				\end{equation}
where $\bm{c}_{\bm{q}^{\prime}} =- \bm{c}_{\bm{q}}$, i. e. $n(\bm{r},\bm{c}_{\bm{q}^{\prime}},t)=n_{\bm{q}}(\bm{r},-\bm{c}_{\bm{q}},t)$, and $c_{s}$ is the lattice sound speed, $c_s = \sqrt{c^2/3}$ with $c=\Delta x / \Delta t$.
Here, we choose the coefficient $C^{\bm{q}}$ equals to $1/3$ if the vector $\bm{q}$ is at rest, $1/18$ at the nearest neighbor site and $1/36$ at the second nearest neighbor site, respectively \cite{ladd1994a,ladd1994b,nguyen2002}.  
The velocity of the boundary node $\bm{u}_b$ satisfies the relation			
				\begin{equation}
					\bm{u}_b = \bm{U} + \bm{\Omega} \times (\bm{r}_b - \bm{R}), \label{eq:8}
					\end{equation}
where $\bm{r}_b$ is the position of the boundary nodes $\bm{r}_b = \bm{r} + \frac{1}{2} \bm{c}_{\bm{q}} \Delta t$ (halfway between the fluid and the solid nodes), where $\bm{R}$ is the center of mass of the particle.
As a result, the hydrodynamic force $\bm{f}^{h}(\bm{r}_b,\bm{c}_{\bm{q}},t)$ exerted on the boundary node is calculated from the momentum transferred in Eq. \eqref{eq:7} as
			\begin{equation}
				\bm{f}^{h}(\bm{r}_b,\bm{c}_{\bm{q}}, t) = \frac{2\Delta x^3}{\Delta t} \Bigg[2 n^{*} (\bm{r},\bm{c}_{\bm{q}},t-\Delta t) - \frac{2 C^{\bm{q}} \rho_0 \bm{u}_b \cdot \bm{c}_{\bm{q}}}{c_{s}^{2}} \Bigg] \bm{c}_{\bm{q}}. \label{eq:6} 
				\end{equation}
For simplicity, we abbreviate the force at each time $\bm{f}^{h}(\bm{r}_b, \bm{c}_{\bm{q}}, t)$ to be $\bm{f}^{h}(\bm{r}_b)$.
Then, the hydrodynamic force at each time $\bm{F}^{h}$ and torque $\bm{T}^{h}$ exerted on the particle are just sum of the forces and torques on all boundary nodes on the particle as
	\begin{align}
		\bm{F}^{h} &= \sum_b \bm{f}^{h}(\bm{r}_b), \label{eq:9} \\
		\bm{T}^{h} &= \sum_b \bm{r}_b \times \bm{f}^{h}(\bm{r}_b). \label{eq:10}
		\end{align}	

With the aid of the hydrodynamic force $\bm{f}^{h}(\bm{r}_b)$, the $\alpha, \beta$ element of the hydrodynamic stress exerted on $i-$th particle $\sigma_{i,\alpha\beta}^h$ is given by $\sigma^{h}_{i,\alpha \beta} = -\sum_b r_{b,\alpha} f^{h}_{\beta}(\bm{r}_b)/V$ where $V=L_xL_yL_z$ is the volume of the system with the linear dimension $L_\alpha$ of $\alpha$ direction.
Then, we sum up all the contributions from particles to obtain the hydrodynamic stress  $\sigma^{h}_{\alpha \beta} = \sum_i \sigma^{h}_{i,\alpha \beta}$.

Note that there are two technical difficulties to adopt this formulation with the aid of the boundary nodes in simulating dense suspensions.
(i) The formulation based on boundary nodes crashes when the particles are overlapping.
This forces us to introduce a contact radius larger than the mapped radius for the boundary nodes with ratio $a_{\text{map}} = 0.95  a_{\text{contact}}$.
The mentioned radius in this paper is the contact radius.
(ii) This scheme works well for the gap between particles $h > 0.5 \Delta x$ and is inaccurate for $h \leq 0.5 \Delta x$ due to shared nodes between two particles in the opposite case \cite{ladd1994b}.
Therefore, we need to incorporate the lubrication correction when the gap between particles is small.
This correction is calculated by the grand-resistance matrix formulation of pairwise lubrication interaction \cite{kimkarilla1991,nguyen2002,jeffrey1984}
\begin{equation}
	\begin{pmatrix}
		\bm{F}^{h}_{i} \\
		\bm{T}^{h}_{i} \\
		\bm{T}^{h}_{j} \\
		\bm{\sigma}^{h}_{i} \\
		\bm{\sigma}^{h}_{j}
		\end{pmatrix}
		= -
		\begin{pmatrix}
			\bm{A}_{ii} & - \bm{B}_{ii} & \bm{B}_{jj} \\
			\bm{B}_{ii} &  \bm{C}_{ii} & \bm{C}_{ij} \\
			-\bm{B}_{jj} &  \bm{C}_{ij} & \bm{C}_{jj} \\
			\bm{G}_{ii} &  \bm{H}_{ii} & \bm{H}_{ij} \\
			-\bm{G}_{jj} &  \bm{H}_{ji} & \bm{H}_{jj} 
			\end{pmatrix}
			\begin{pmatrix}
				\bm{U}_{ij} \\
				\bm{\Omega}_i \\
				\bm{\Omega}_j
				\end{pmatrix},	
	\end{equation}
where $\bm{F}^{h}_{i}$ is the hydrodynamic (lubrication) force on particle $i$ and for particle $j$, $\bm{F}^{h}_{j}=-\bm{F}^{h}_{i}$. $\bm{U}_{ij}=\bm{U}_i-\bm{U}_j$ is the relative velocity. We adopt the notation in Ref. \cite{kimkarilla1991,jeffrey1984} such as $\bm{A}_{ij} = (A^{(ij)}_{\alpha \beta})$ and due to the Lorentz reciprocal theorem, we have the symmetry relations such as
\begin{equation}
	A^{(ij)}_{\alpha \beta} = A^{(ji)}_{\beta \alpha}.
	\end{equation}	
For axisymmetric geometries, the coefficients can be expressed in terms of scalar functions as
\begin{align}
	A^{(ij)}_{\alpha \beta} &= X^{A}_{ij} n^{(ij)}_{\alpha}n^{(ij)}_{\beta}(\delta_{\alpha \beta} - n^{(ij)}_{\alpha}n^{(ij)}_{\beta}) ,\\
	B^{(ij)}_{\alpha \beta} &= Y^{B}_{ij} \epsilon_{\alpha \beta \gamma} n^{(ij)}_{\gamma}  ,\\
	C^{(ij)}_{\alpha \beta } &= X^{C}_{ij} n^{(ij)}_{\alpha} n^{(ij)}_{\beta} + Y^{C}_{ij} (\delta_{\alpha \beta} - n^{(ij)}_{\alpha} n^{(ij)}_{\beta}),\\
	G^{(ij)}_{\alpha \beta \gamma} &= X^{G}_{ij} (n^{(ij)}_{\alpha}n^{(ij)}_{\beta} - \frac{1}{3}\delta_{\alpha\beta})n^{(ij)}_{\gamma} \notag\\ &+ Y^{G}_{ij} (n^{(ij)}_{\alpha} \delta_{\beta \gamma} + n^{(ij)}_{\beta} \delta_{\alpha \gamma} -  2 n^{(ij)}_{\alpha}n^{(ij)}_{\beta}n^{(ij)}_{\gamma}),\\
	H^{(ij)}_{\alpha \beta \gamma} &= Y^{H}_{ij} (\epsilon_{\alpha \gamma \kappa} n^{(ij)}_{\kappa} n^{(ij)}_{\beta} + \epsilon_{\beta \gamma \kappa}n^{(ij)}_{\kappa}n^{(ij)}_{\alpha}),
	\end{align}	
where $n^{(ij)}_{\alpha}$ is the normal unit vector between particles $i$ and $j$ in the $\alpha$ direction and $\epsilon_{\alpha\beta\gamma}$ is Levi-Citiva symbol, i. e. $\epsilon_{\alpha\beta\gamma}=1$ for an even permutation of $(\alpha,\beta.\gamma)$, $\epsilon_{\alpha\beta\gamma}=-1$ for an odd permutation of $(\alpha,\beta.\gamma)$, and $\epsilon_{\alpha\beta\gamma}=0$ otherwise.
The scalar functions $X$ and $Y$ are functions of interparticle gap $h$. For two spheres of arbitrary size with the leading order only, the scalar functions are written as
\begin{align}
	X^{A}_{ii} &= 6 \pi a \bigg[\frac{2 \beta^2}{(1+ \beta)^{3}} \frac{1}{h+\delta} \bigg],\\
	X^{G}_{ii} &= 4 \pi a^2 \bigg[ \frac{3 \beta^2}{(1+\beta)^3} \frac{1}{h+\delta} \bigg] ,\\
	Y^{A}_{ii} &=6 \pi a \bigg[\frac{4 \beta (2 + \beta + 2 \beta^2)}{15(1+ \beta)^{3}} \ln\frac{1}{h+\delta} \bigg] , \\
	Y^{B}_{ii} &= 4 \pi a^2 \bigg[ \frac{\beta (4+ \beta)}{5(1+\beta)^2} \ln \frac{1}{h+\delta} \bigg] ,\\
	Y^{G}_{ii} &= 4 \pi a^2 \bigg[\frac{\beta (4 - \beta + 7 \beta^2)}{10 (1 + \beta)^3}  \ln \frac{1}{h+\delta} \bigg], \\
	Y^{C}_{ii} &= 8 \pi a^3 \bigg[ \frac{2 \beta}{5 (1 + \beta)} \ln \frac{1}{h+\delta} \bigg],\\
	Y^{C}_{ij} &= 8 \pi a^3 \bigg[ \frac{\beta^2}{10(1+\beta)}\ln \frac{1}{h+\delta}  \bigg],\\
	Y^{H}_{ii} &= 8 \pi a^3 \bigg[ \frac{\beta(2-\beta)}{10(1+\beta)^2}\ln \frac{1}{h+\delta}  \bigg] , \\
	Y^{H}_{ij} &= 8 \pi a^3 \bigg[\frac{\beta^2 (1+7 \beta)}{20(1+\beta)^2} \ln \frac{1}{h+\delta} \bigg],
	\end{align}
where $\beta$ is the ratio of the particle radius defined as $\beta = a_i/a_j$ and $\delta$ is the roughness length.

\section*{Appendix B Dependences on number of initial and observation cycles}
  \begin{figure}[htbp]
   \centering
   \includegraphics[height=0.7\linewidth]{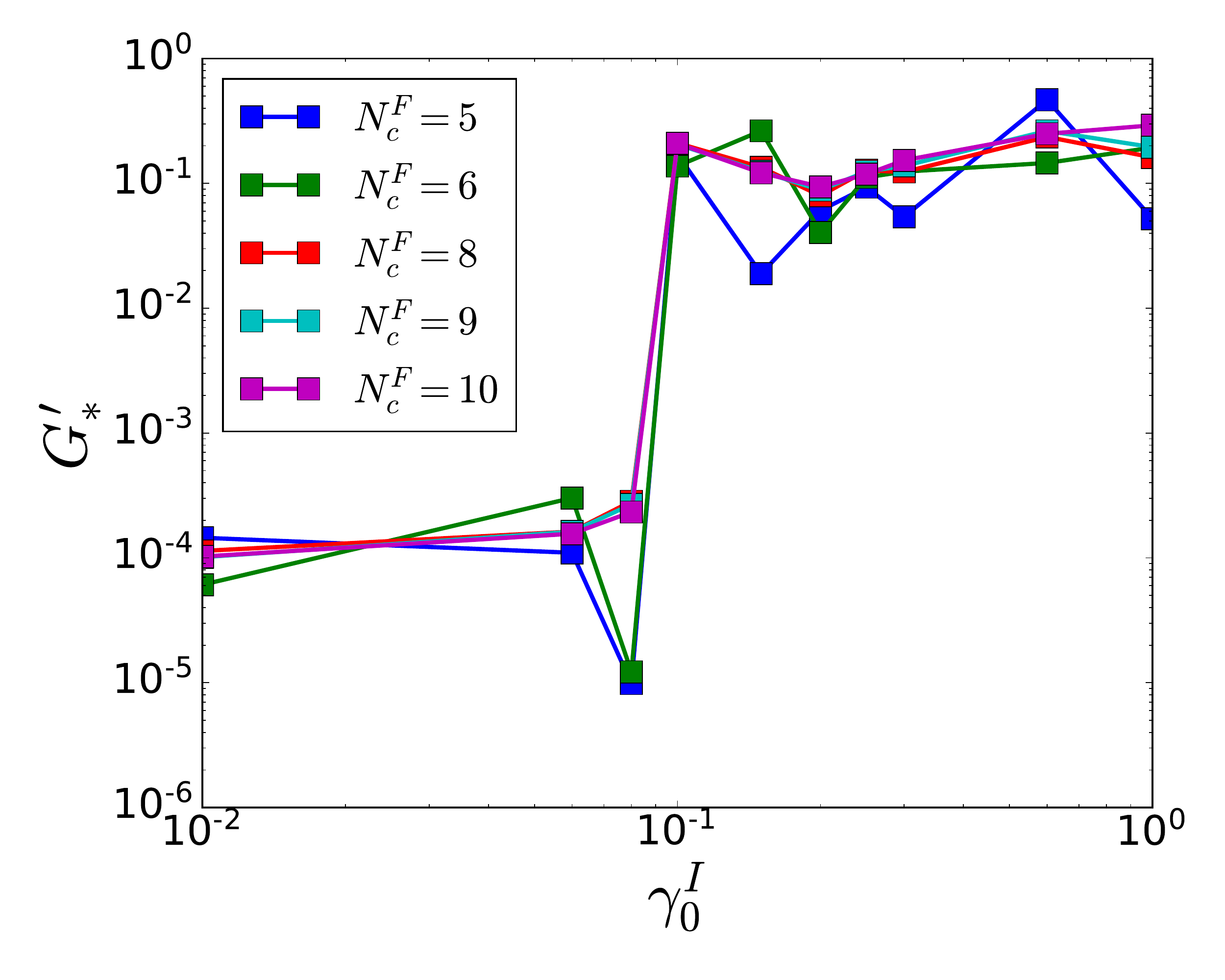}
   \caption{Plots of the dimensionless storage modulus $G^{\prime}_{*}$ against the initial strain amplitude $\gamma_0^{I}$ for several numbers of observation cycles $N_c^{F}$ for $\phi = 0.82$.}
   \label{fgr:ncf}
   \end{figure}
\begin{figure}[htbp]
   \centering
   \includegraphics[height=0.7\linewidth]{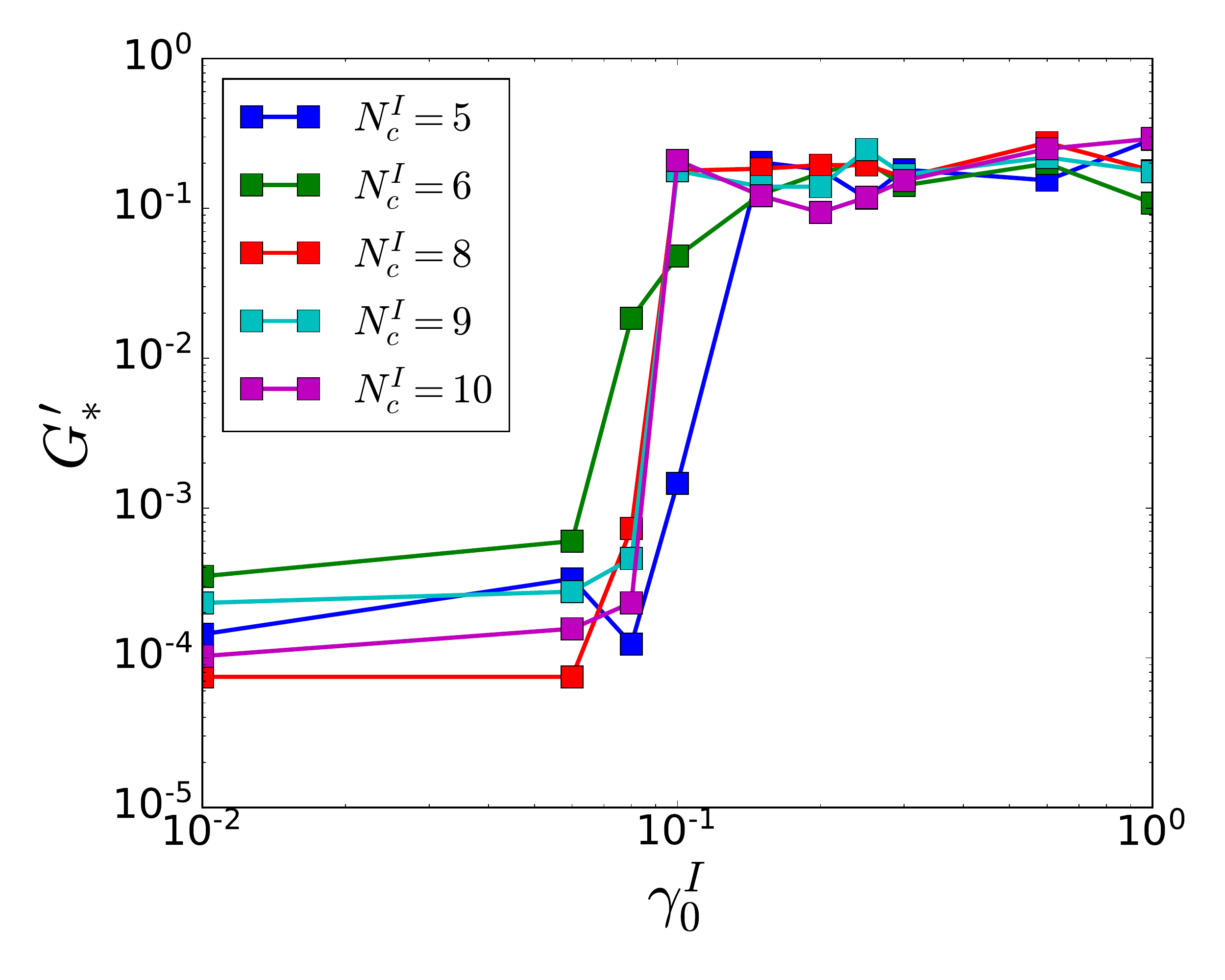}
   \caption{Plots of the dimensionless storage modulus $G^{\prime}_{*}$ against the initial strain amplitude $\gamma_0^{I}$ for several numbers of initial cycles $N_c^{I}$ for $\phi = 0.82$.}
   \label{fgr:nci}
   \end{figure}

In this section, we verify that the results for the mechanical responses in the main text is not merely transients.
First, we plot the storage modulus for $\phi=0.82$ against initial strain amplitude $\gamma_0^{I}$ to see how the results depend on number of observation cycles $N_c^{F}$. 
Our result in Fig. \ref{fgr:ncf} indicates that the results converges after 5 cycles.
Meanwhile, for the number of initial cycles, we also get convergence after eight initial cycles at least, for shear jammed states as seen in Fig. \ref{fgr:nci}.

\section*{Appendix C Brief derivation of the stress formulation}
\begin{figure}[htbp]
    \centering	
    \includegraphics[width=0.5\linewidth]{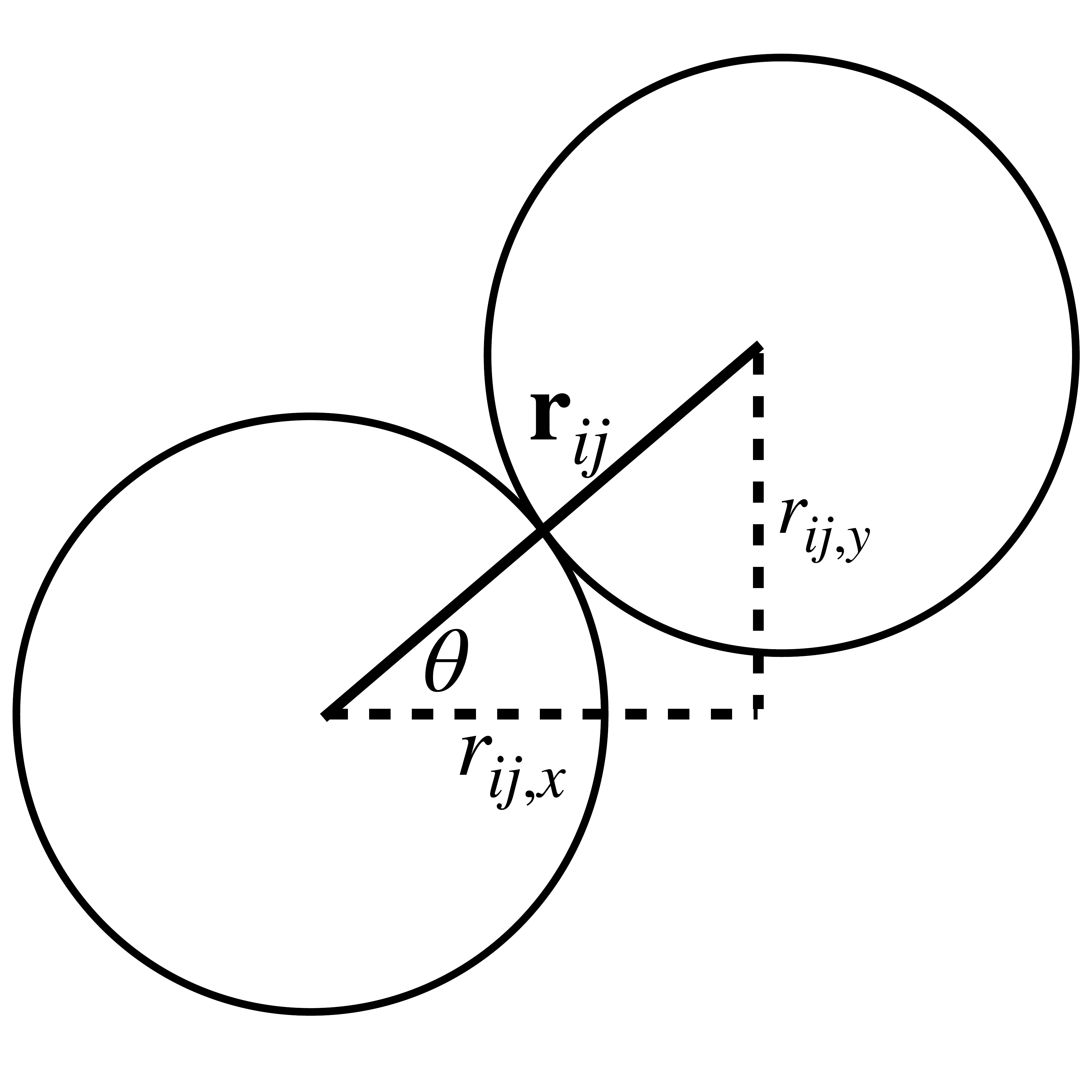}
    \caption{Illustration of contacting particles, $\theta$ is the contact angle.
    \label{app3}}
    \end{figure}

In this section we describe the derivation of the stress formulation by following Ref. \cite{kanatani1981}.
Let us consider a system consisting of $N$ particles with average radius $a$.
The particle radius is assumed to be uniform. 
From the contact angle $\theta={\rm tan}^{-1}(r_{ij,y}/r_{ij,x})$ (Fig. \ref{app3}) which is ranged between $0$ and $2 \pi$ and calculated counterclockwise from the $x$-axis, one can define $\rho_{\theta}$ as the angular distribution of the contact orientations.
$\rho_{\theta}$ satisfies the following normalization
\begin{align}
	 \int _ { 0 } ^ {2 \pi } \rho_ { \theta } d \theta = 1.   \label{eq:a14}
	\end{align}

Note that the azimuthal angle is ignored here.
Then, $F^{N} n_{\alpha}$ is the $\alpha-$ component of the normal contact forces vector on a contact point.
Suppose that the particles are in mechanical equilibrium, the force balance in $\theta$ directions
\begin{equation}
	Z \int _ { 0 } ^ {2 \pi }  F^N n_{\alpha} \rho_{\theta} d \theta = 0, \label{eq:a15}
	\end{equation}
is satisfied, where $Z$ is the coordination number or the average number of contact points on each particle.
Here, $n_{\alpha} \rho_{\theta}$ represents the expectation value of the contact unit vector.
The torque balance is written as
\begin{equation}
	Z \int _ { 0 } ^ {2 \pi }   F^N \epsilon_{\alpha\beta\gamma}n_\alpha n_\beta \rho_{\theta } d \theta = 0. \label{eq:a16}
	\end{equation}
Suppose that the material is subject to a macroscopically uniform stress.
Then, one can assume the first order deformation of the material
\begin{equation}
	x_{\alpha}^{\prime} = D_{\alpha \beta} x_{\beta},
	\end{equation}
under the deformation the point $x_{\alpha}$  to $x^{\prime}_{\alpha}$.
Thus, the displacement $u_{\alpha}=x^{\prime}_{\alpha}-x_{\alpha}$ is given by
\begin{align}
	u_{\alpha} = \Gamma_{\alpha \beta} x_{\beta}, \qquad \Gamma_{\alpha \beta} = D_{\alpha \beta} - \delta_{\alpha \beta},
	\end{align}
where $\delta_{\alpha \beta}$ is the Kronecker delta.
The distortion tensor $\Gamma_{\alpha \beta}$ contains the symmetric and anti-symmetric part as
\begin{equation}
	\Gamma_{\alpha \beta} = E_{\alpha \beta} + R_{\alpha \beta},	\label{eq:a17}
	\end{equation}
where $E_{\alpha \beta}$ is the strain (symmetric) tensor and $R_{\alpha \beta}$ is the rotation (anti-symmetric) tensor.

Here, we assume that the particles also experienced a virtual deformation.
This virtual deformation distorts a spherical particle into an ellipsoid due to the macroscopic deformation.
Therefore, the small displacement $\epsilon_{\alpha}$ in the contact direction $n_{\alpha}$ is given by
\begin{equation}
	\epsilon_{\alpha} = a \Gamma_{\alpha \beta} n_{\beta}.
	\end{equation}
Another assumption is that the contact forces are not changed during this virtual deformation.
Then, the virtual work done by the contact forces can be expressed in terms of angular integral (Eq. \eqref{eq:a15}) as
\begin{equation}
	Z \int _ { 0 } ^ { 2 \pi } \rho_{\theta}   F^N n_{\alpha} \epsilon_{\alpha}  d \theta = aZ \Gamma_{\alpha \beta}  \int _ { 0 } ^ {2 \pi } \rho_{\theta}  F^N n_{\alpha} n_{\beta} d \theta.  \label{eq:a50}
	\end{equation}
By the virtue of Eq. \eqref{eq:a17} and the torque balance equation (Eq. \eqref{eq:a16}), one can rewrite the virtual work Eq. \eqref{eq:a50} as	
\begin{equation}
	 a Z \Gamma_{\alpha \beta}  \int _ { 0 } ^ { 2 \pi } \rho_{\theta}  F^N n_{\alpha} n_{\beta} d \theta =  aZ E_{\alpha \beta}  \int _ { 0 } ^ {2 \pi } \rho_{\theta}  F^N n_{\alpha} n_{\beta} d \theta.
	\end{equation}		
Now, the number of particles in unit volume is $\phi_3 / (4/3) \pi a^3$, where $\phi_3$ is the solid volume fraction of the material in three dimensions.
Hence, the virtual work done in unit volume $W$ is
\begin{equation}
	W = \frac{3}{4} \frac{Z \phi_3}{\pi a^2} E_{\alpha \beta} \int _ { 0 } ^ {2 \pi } \rho_{\theta}  F^N n_{\alpha} n_{\beta} d \theta.
	\end{equation}
Meanwhile, the virtual work done per unit volume by the virtual strain $ E_{\alpha \beta}$ under stress $\sigma_{\alpha \beta}$ is
\begin{equation}
	W = \sigma_{\alpha \beta} E_{\alpha \beta}.
	\end{equation}
Therefore, the stress tensor can be expresssed as
\begin{equation}
	\sigma_{\alpha \beta} = \frac{3}{4} \frac{Z \phi_3}{\pi a^2}  \int _ { 0 } ^ {2 \pi } \rho_{\theta}  F^N n_{\alpha} n_{\beta} d \theta. 
	\end{equation}
Then, if we define the angular distributions of intensities of the normal contact forces $\xi_N = F^{N}/ \langle F^N \rangle $, and the angular distributions of the normal contact forces $\zeta_N = \rho_{\theta} \xi_N$, where $\langle F^N \rangle$ is the average normal forces, one could have
\begin{equation}
		\sigma_{\alpha \beta} = \frac{3}{4} \frac{Z \phi_3 \langle F^N \rangle}{\pi a^2}  \int _ { 0 } ^ {2 \pi } \zeta_N n_{\alpha} n_{\beta} d \theta. \label{eq:a18}
	\end{equation}		 		
Finally, one can also take into account the tangential forces by altering Eq. \eqref{eq:a18} as
\begin{equation}
		\sigma_{\alpha \beta} = \frac{3}{4} \frac{Z \phi_3 \langle F^N \rangle}{\pi a^2}  \int _ { 0 } ^ {2 \pi } [\zeta_N n_{\alpha} - \zeta_T t_{\alpha}]  n_{\beta} d \theta, \label{eq:a19}
	\end{equation}	
where $t_{\alpha}$ is the tangential unit vector.
Note that for a system of disks, the number of particles in the unit area is $\phi /  \pi a^2$, with $\phi$ is the area fraction.
Therefore, for 2-dimensional disks, Eq. \eqref{eq:a19} is rewritten as
	\begin{equation}
			\sigma_{\alpha \beta} =  \frac{Z \phi \langle F^N \rangle}{2 \pi a}  \int _ { 0 } ^ {2 \pi } [\zeta_N n_{\alpha} - \zeta_T t_{\alpha}]  n_{\beta} d \theta. \label{eq:a20}
		\end{equation}				

\section*{Appendix D System size dependence}
\begin{figure}[htbp]
   \centering
   \includegraphics[height=0.7\linewidth]{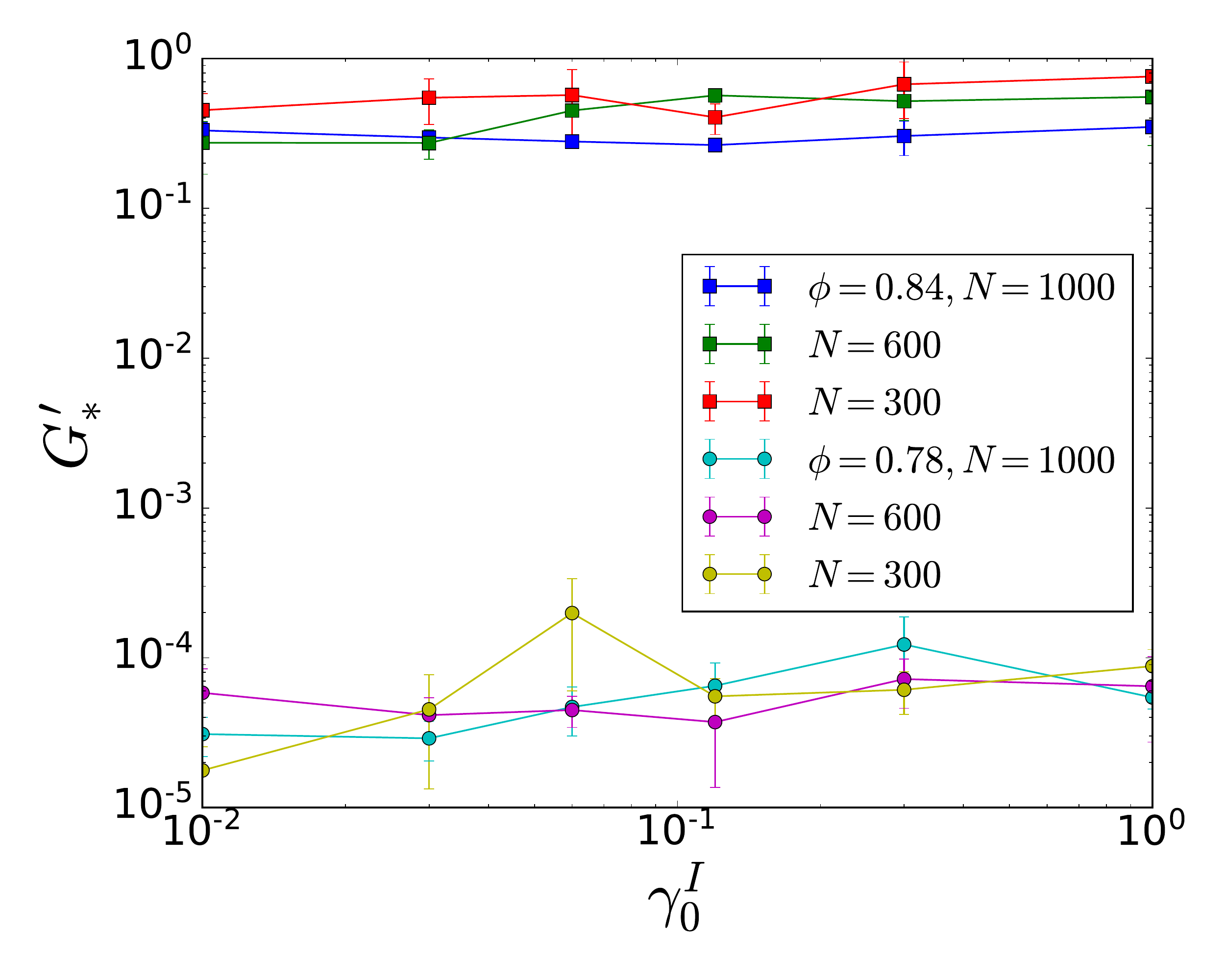}
   \caption{Plot of the dimensionless storage modulus $G^{\prime}_{*}$ against $\gamma_0^{I}$ for several numbers of particles $N$ for $\phi = 0.84$ and $\phi = 0.78$ .}
   \label{fgr:size} 
   \end{figure}

In this Appendix, we examine the system size dependence for the oscillatory shear simulations as shown in Fig. \ref{fgr:size} by using LF-DEM.
Here, one does not have to worry about serious finite size effects for $N \geq 300$.
Because of the limitation of our computational resources we have not examined the finite size effects for our LBM simulation.
Nevertheless, we believe that the finite size effect is not severe for $N\ge 300$ even for the LBM simulation  because the results of the LBM simulation are not quite different from those obtained from the LF-DEM.
\section*{Conflicts of interest}
There are no conflicts to declare.

\section*{Acknowledgements}
One of the authors (P) expresses his gratitude to A. J. C. Ladd for providing the Lattice Boltzmann susp3d code and S. Takada for his 3D-DEM code. The authors thank M. Otsuki, R. Seto, A. Santos, and D. Ishima for essential comments and fruitful discussions. This work is partially supported by the Grant-in-Aid of MEXT for Scientific Research (Grant No. 16H04025) and the Programs YITP-T-18-03 and YITP-W-18-17. All numerical calculations were carried out at the Yukawa Institute for Theoretical Physics (YITP) Computer Facilities, Kyoto University, Japan.
%%%END OF MAIN TEXT%%%

%The \balance command can be used to balance the columns on the final page if desired. It should be placed anywhere within the first column of the last page.

%\balance

%If notes are included in your references you can change the title from 'References' to 'Notes and references' using the following command:
%\renewcommand\refname{Notes and references}

%%%REFERENCES%%%
\bibliography{references} %You need to replace "rsc" on this line with the name of your .bib file

\providecommand*{\mcitethebibliography}{\thebibliography}
\csname @ifundefined\endcsname{endmcitethebibliography}
{\let\endmcitethebibliography\endthebibliography}{}
\begin{mcitethebibliography}{74}
\providecommand*{\natexlab}[1]{#1}
\providecommand*{\mciteSetBstSublistMode}[1]{}
\providecommand*{\mciteSetBstMaxWidthForm}[2]{}
\providecommand*{\mciteBstWouldAddEndPuncttrue}
  {\def\EndOfBibitem{\unskip.}}
\providecommand*{\mciteBstWouldAddEndPunctfalse}
  {\let\EndOfBibitem\relax}
\providecommand*{\mciteSetBstMidEndSepPunct}[3]{}
\providecommand*{\mciteSetBstSublistLabelBeginEnd}[3]{}
\providecommand*{\EndOfBibitem}{}
\mciteSetBstSublistMode{f}
\mciteSetBstMaxWidthForm{subitem}
{(\emph{\alph{mcitesubitemcount}})}
\mciteSetBstSublistLabelBeginEnd{\mcitemaxwidthsubitemform\space}
{\relax}{\relax}

\bibitem[Einstein(1905)]{einstein1905}
A.~Einstein, \emph{Ann. Phys. (Leipzig)}, 1905, \textbf{17}, 549--560\relax
\mciteBstWouldAddEndPuncttrue
\mciteSetBstMidEndSepPunct{\mcitedefaultmidpunct}
{\mcitedefaultendpunct}{\mcitedefaultseppunct}\relax
\EndOfBibitem
\bibitem[Williamson and Heckert(1931)]{williamson1931}
R.~V. Williamson and W.~W. Heckert, \emph{Ind. Eng. Chem}, 1931, \textbf{23},
  667--670\relax
\mciteBstWouldAddEndPuncttrue
\mciteSetBstMidEndSepPunct{\mcitedefaultmidpunct}
{\mcitedefaultendpunct}{\mcitedefaultseppunct}\relax
\EndOfBibitem
\bibitem[Barnes(1989)]{barnes1989}
H.~Barnes, \emph{J. Rheol}, 1989, \textbf{33}, 329--366\relax
\mciteBstWouldAddEndPuncttrue
\mciteSetBstMidEndSepPunct{\mcitedefaultmidpunct}
{\mcitedefaultendpunct}{\mcitedefaultseppunct}\relax
\EndOfBibitem
\bibitem[Egres and Wagner(2005)]{egres2005}
R.~G. Egres and N.~J. Wagner, \emph{J. Rheol}, 2005, \textbf{49},
  719--746\relax
\mciteBstWouldAddEndPuncttrue
\mciteSetBstMidEndSepPunct{\mcitedefaultmidpunct}
{\mcitedefaultendpunct}{\mcitedefaultseppunct}\relax
\EndOfBibitem
\bibitem[Brown and Jaeger(2012)]{brown2012}
E.~Brown and H.~M. Jaeger, \emph{J. Rheol}, 2012, \textbf{56}, 875--923\relax
\mciteBstWouldAddEndPuncttrue
\mciteSetBstMidEndSepPunct{\mcitedefaultmidpunct}
{\mcitedefaultendpunct}{\mcitedefaultseppunct}\relax
\EndOfBibitem
\bibitem[Brown \emph{et~al.}(2011)Brown, Zhang, Forman, Maynor, Betts,
  DeSimone, and Jaeger]{brown2011}
E.~Brown, H.~Zhang, N.~A. Forman, B.~W. Maynor, D.~E. Betts, J.~M. DeSimone and
  H.~M. Jaeger, \emph{Phys. Rev. E: Stat. Nonlinear Soft Matter Phys.}, 2011,
  \textbf{84}, 031408\relax
\mciteBstWouldAddEndPuncttrue
\mciteSetBstMidEndSepPunct{\mcitedefaultmidpunct}
{\mcitedefaultendpunct}{\mcitedefaultseppunct}\relax
\EndOfBibitem
\bibitem[Cwalina \emph{et~al.}(2016)Cwalina, Harrison, and Wagner]{cwalina2016}
C.~D. Cwalina, K.~J. Harrison and N.~J. Wagner, \emph{Soft Matter}, 2016,
  \textbf{12}, 4654\relax
\mciteBstWouldAddEndPuncttrue
\mciteSetBstMidEndSepPunct{\mcitedefaultmidpunct}
{\mcitedefaultendpunct}{\mcitedefaultseppunct}\relax
\EndOfBibitem
\bibitem[Allen \emph{et~al.}(2018)Allen, Sokol, Mukhopadhyay, Maharjan, and
  Brown]{allen2018}
B.~Allen, B.~Sokol, S.~Mukhopadhyay, R.~Maharjan and E.~Brown, \emph{Phys. Rev.
  E: Stat. Nonlinear Soft Matter Phys.}, 2018, \textbf{97}, 052603\relax
\mciteBstWouldAddEndPuncttrue
\mciteSetBstMidEndSepPunct{\mcitedefaultmidpunct}
{\mcitedefaultendpunct}{\mcitedefaultseppunct}\relax
\EndOfBibitem
\bibitem[Maharjan \emph{et~al.}(2018)Maharjan, Mukhopadhyay, Allen, Storz, and
  Brown]{maharjan2018}
R.~Maharjan, S.~Mukhopadhyay, B.~Allen, T.~Storz and E.~Brown, \emph{Phys. Rev.
  E: Stat. Nonlinear Soft Matter Phys.}, 2018, \textbf{97}, 052602\relax
\mciteBstWouldAddEndPuncttrue
\mciteSetBstMidEndSepPunct{\mcitedefaultmidpunct}
{\mcitedefaultendpunct}{\mcitedefaultseppunct}\relax
\EndOfBibitem
\bibitem[Seto \emph{et~al.}(2013)Seto, Mari, Morris, and Denn]{seto2013}
R.~Seto, R.~Mari, J.~F. Morris and M.~M. Denn, \emph{Phys. Rev. Lett.}, 2013,
  \textbf{111}, 218301\relax
\mciteBstWouldAddEndPuncttrue
\mciteSetBstMidEndSepPunct{\mcitedefaultmidpunct}
{\mcitedefaultendpunct}{\mcitedefaultseppunct}\relax
\EndOfBibitem
\bibitem[Mari and Seto(2014)]{mari2014}
R.~Mari and R.~Seto, \emph{J. Rheol}, 2014, \textbf{58}, 1693--1724\relax
\mciteBstWouldAddEndPuncttrue
\mciteSetBstMidEndSepPunct{\mcitedefaultmidpunct}
{\mcitedefaultendpunct}{\mcitedefaultseppunct}\relax
\EndOfBibitem
\bibitem[Mari \emph{et~al.}(2015)Mari, Seto, Morris, and Denn]{mari2015}
R.~Mari, R.~Seto, J.~F. Morris and M.~M. Denn, \emph{Proc. Natl. Acad. Sci. U.
  S. A.}, 2015, \textbf{112}, 15326--15330\relax
\mciteBstWouldAddEndPuncttrue
\mciteSetBstMidEndSepPunct{\mcitedefaultmidpunct}
{\mcitedefaultendpunct}{\mcitedefaultseppunct}\relax
\EndOfBibitem
\bibitem[Brady and Bossis(1988)]{brady1988}
J.~F. Brady and G.~Bossis, \emph{Annu. Rev. Fluid Mech.}, 1988, \textbf{20},
  111--157\relax
\mciteBstWouldAddEndPuncttrue
\mciteSetBstMidEndSepPunct{\mcitedefaultmidpunct}
{\mcitedefaultendpunct}{\mcitedefaultseppunct}\relax
\EndOfBibitem
\bibitem[Denn \emph{et~al.}(2018)Denn, Morris, and Bonn]{denn2018}
M.~M. Denn, J.~F. Morris and D.~Bonn, \emph{Soft Matter}, 2018, \textbf{14},
  170\relax
\mciteBstWouldAddEndPuncttrue
\mciteSetBstMidEndSepPunct{\mcitedefaultmidpunct}
{\mcitedefaultendpunct}{\mcitedefaultseppunct}\relax
\EndOfBibitem
\bibitem[Wyart and Cates(2014)]{wyart2014}
M.~Wyart and M.~E. Cates, \emph{Phys. Rev. Lett.}, 2014, \textbf{112},
  098302\relax
\mciteBstWouldAddEndPuncttrue
\mciteSetBstMidEndSepPunct{\mcitedefaultmidpunct}
{\mcitedefaultendpunct}{\mcitedefaultseppunct}\relax
\EndOfBibitem
\bibitem[Thomas \emph{et~al.}(2018)Thomas, Ramola, Singh, Mari, Morris, and
  Chakraborty]{thomas2018}
J.~E. Thomas, K.~Ramola, A.~Singh, R.~Mari, J.~F. Morris and B.~Chakraborty,
  \emph{Phys. Rev. Lett.}, 2018, \textbf{121}, 128002\relax
\mciteBstWouldAddEndPuncttrue
\mciteSetBstMidEndSepPunct{\mcitedefaultmidpunct}
{\mcitedefaultendpunct}{\mcitedefaultseppunct}\relax
\EndOfBibitem
\bibitem[Otsuki and Hayakawa(2011)]{otsuki2011}
M.~Otsuki and H.~Hayakawa, \emph{Phys. Rev. E: Stat. Nonlinear Soft Matter
  Phys.}, 2011, \textbf{83}, 051301\relax
\mciteBstWouldAddEndPuncttrue
\mciteSetBstMidEndSepPunct{\mcitedefaultmidpunct}
{\mcitedefaultendpunct}{\mcitedefaultseppunct}\relax
\EndOfBibitem
\bibitem[Liu and Nagel(1998)]{liu1998}
A.~J. Liu and S.~R. Nagel, \emph{Nature}, 1998, \textbf{396}, 21--22\relax
\mciteBstWouldAddEndPuncttrue
\mciteSetBstMidEndSepPunct{\mcitedefaultmidpunct}
{\mcitedefaultendpunct}{\mcitedefaultseppunct}\relax
\EndOfBibitem
\bibitem[Liu and Nagel(2010)]{liu2010}
A.~J. Liu and S.~R. Nagel, \emph{Soft Matter}, 2010, \textbf{6}, 2869\relax
\mciteBstWouldAddEndPuncttrue
\mciteSetBstMidEndSepPunct{\mcitedefaultmidpunct}
{\mcitedefaultendpunct}{\mcitedefaultseppunct}\relax
\EndOfBibitem
\bibitem[Bi \emph{et~al.}(2011)Bi, Zhang, Chakraborty, and Behringer]{bi2011}
D.~Bi, J.~Zhang, B.~Chakraborty and R.~P. Behringer, \emph{Nature}, 2011,
  \textbf{480}, 355--358\relax
\mciteBstWouldAddEndPuncttrue
\mciteSetBstMidEndSepPunct{\mcitedefaultmidpunct}
{\mcitedefaultendpunct}{\mcitedefaultseppunct}\relax
\EndOfBibitem
\bibitem[Otsuki and Hayakawa(2018)]{otsuki2018}
M.~Otsuki and H.~Hayakawa, arXiv:1810.03846, 2018\relax
\mciteBstWouldAddEndPuncttrue
\mciteSetBstMidEndSepPunct{\mcitedefaultmidpunct}
{\mcitedefaultendpunct}{\mcitedefaultseppunct}\relax
\EndOfBibitem
\bibitem[Lee \emph{et~al.}(2015)Lee, Nam, Hyun, Ahn, and Lee]{lee2015}
Y.~K. Lee, J.~Nam, K.~Hyun, K.~H. Ahn and S.~J. Lee, \emph{Soft Matter}, 2015,
  \textbf{11}, 4061\relax
\mciteBstWouldAddEndPuncttrue
\mciteSetBstMidEndSepPunct{\mcitedefaultmidpunct}
{\mcitedefaultendpunct}{\mcitedefaultseppunct}\relax
\EndOfBibitem
\bibitem[Marenne and Morris(2017)]{marenne2017}
S.~Marenne and J.~F. Morris, \emph{J. Rheol}, 2017, \textbf{61}, 797--815\relax
\mciteBstWouldAddEndPuncttrue
\mciteSetBstMidEndSepPunct{\mcitedefaultmidpunct}
{\mcitedefaultendpunct}{\mcitedefaultseppunct}\relax
\EndOfBibitem
\bibitem[Moghimi \emph{et~al.}(2017)Moghimi, Jacob, Koumakis, and
  Petekidis]{moghimi2017}
E.~Moghimi, A.~R. Jacob, N.~Koumakis and G.~Petekidis, \emph{Soft Matter},
  2017, \textbf{13}, 2371\relax
\mciteBstWouldAddEndPuncttrue
\mciteSetBstMidEndSepPunct{\mcitedefaultmidpunct}
{\mcitedefaultendpunct}{\mcitedefaultseppunct}\relax
\EndOfBibitem
\bibitem[Park \emph{et~al.}(2015)Park, Ahn, and Lee]{park2015}
J.~D. Park, K.~H. Ahn and S.~J. Lee, \emph{Soft Matter}, 2015, \textbf{11},
  9262\relax
\mciteBstWouldAddEndPuncttrue
\mciteSetBstMidEndSepPunct{\mcitedefaultmidpunct}
{\mcitedefaultendpunct}{\mcitedefaultseppunct}\relax
\EndOfBibitem
\bibitem[Ness \emph{et~al.}(2017)Ness, Xing, and Eiser]{ness2017}
C.~Ness, Z.~Xing and E.~Eiser, \emph{Soft Matter}, 2017, \textbf{13},
  3664\relax
\mciteBstWouldAddEndPuncttrue
\mciteSetBstMidEndSepPunct{\mcitedefaultmidpunct}
{\mcitedefaultendpunct}{\mcitedefaultseppunct}\relax
\EndOfBibitem
\bibitem[Doi and Edwards(1986)]{doiedwards}
M.~Doi and S.~F. Edwards, \emph{The Theory of Polymer Dynamics}, Oxford
  University Press, 1986\relax
\mciteBstWouldAddEndPuncttrue
\mciteSetBstMidEndSepPunct{\mcitedefaultmidpunct}
{\mcitedefaultendpunct}{\mcitedefaultseppunct}\relax
\EndOfBibitem
\bibitem[Ganeriwala and Rotz(1987)]{ganeriwala1987}
S.~N. Ganeriwala and C.~A. Rotz, \emph{Polym. Eng. Sci}, 1987, \textbf{27},
  165--178\relax
\mciteBstWouldAddEndPuncttrue
\mciteSetBstMidEndSepPunct{\mcitedefaultmidpunct}
{\mcitedefaultendpunct}{\mcitedefaultseppunct}\relax
\EndOfBibitem
\bibitem[Brader \emph{et~al.}(2010)Brader, Siebenb{\"u}rger, Ballauff,
  Reinheimer, Wilhelm, Frey, Weysserand, and Fuchs]{brader2010}
J.~M. Brader, M.~Siebenb{\"u}rger, M.~Ballauff, K.~Reinheimer, M.~Wilhelm,
  S.~J. Frey, F.~Weysserand and M.~Fuchs, \emph{Phys. Rev. E: Stat. Nonlinear
  Soft Matter Phys.}, 2010, \textbf{82}, 061401\relax
\mciteBstWouldAddEndPuncttrue
\mciteSetBstMidEndSepPunct{\mcitedefaultmidpunct}
{\mcitedefaultendpunct}{\mcitedefaultseppunct}\relax
\EndOfBibitem
\bibitem[Ewoldt \emph{et~al.}(2008)Ewoldt, Hosoi, and McKinley]{ewoldt2008}
R.~H. Ewoldt, A.~E. Hosoi and G.~H. McKinley, \emph{J. Rheol}, 2008,
  \textbf{52}, 1427--1458\relax
\mciteBstWouldAddEndPuncttrue
\mciteSetBstMidEndSepPunct{\mcitedefaultmidpunct}
{\mcitedefaultendpunct}{\mcitedefaultseppunct}\relax
\EndOfBibitem
\bibitem[Peters \emph{et~al.}(2016)Peters, Majumdar, and Jaeger]{peters2016}
I.~R. Peters, S.~Majumdar and H.~M. Jaeger, \emph{Nature}, 2016, \textbf{532},
  214--217\relax
\mciteBstWouldAddEndPuncttrue
\mciteSetBstMidEndSepPunct{\mcitedefaultmidpunct}
{\mcitedefaultendpunct}{\mcitedefaultseppunct}\relax
\EndOfBibitem
\bibitem[Fall \emph{et~al.}(2015)Fall, Bertrand, Hautemayou, Mezi{\`e}re,
  Moucheront, Lema{\^\i}tre, and Ovarlez]{fall2015}
A.~Fall, F.~Bertrand, D.~Hautemayou, C.~Mezi{\`e}re, P.~Moucheront,
  A.~Lema{\^\i}tre and G.~Ovarlez, \emph{Phys. Rev. Lett.}, 2015, \textbf{114},
  098301\relax
\mciteBstWouldAddEndPuncttrue
\mciteSetBstMidEndSepPunct{\mcitedefaultmidpunct}
{\mcitedefaultendpunct}{\mcitedefaultseppunct}\relax
\EndOfBibitem
\bibitem[Ladd(1994)]{ladd1994a}
A.~J.~C. Ladd, \emph{J. Fluid Mech}, 1994, \textbf{271}, 285--309\relax
\mciteBstWouldAddEndPuncttrue
\mciteSetBstMidEndSepPunct{\mcitedefaultmidpunct}
{\mcitedefaultendpunct}{\mcitedefaultseppunct}\relax
\EndOfBibitem
\bibitem[Ladd(1994)]{ladd1994b}
A.~J.~C. Ladd, \emph{J. Fluid Mech}, 1994, \textbf{271}, 311--339\relax
\mciteBstWouldAddEndPuncttrue
\mciteSetBstMidEndSepPunct{\mcitedefaultmidpunct}
{\mcitedefaultendpunct}{\mcitedefaultseppunct}\relax
\EndOfBibitem
\bibitem[Nguyen and Ladd(2002)]{nguyen2002}
N.~Q. Nguyen and A.~J.~C. Ladd, \emph{Phys. Rev. E: Stat. Nonlinear Soft Matter
  Phys.}, 2002, \textbf{66}, 046708\relax
\mciteBstWouldAddEndPuncttrue
\mciteSetBstMidEndSepPunct{\mcitedefaultmidpunct}
{\mcitedefaultendpunct}{\mcitedefaultseppunct}\relax
\EndOfBibitem
\bibitem[Shakib-Manesh \emph{et~al.}(2002)Shakib-Manesh, Raiskinm{\"a}ki,
  Koponen, Kataja, and Timonen]{shakib-manesh2002}
A.~Shakib-Manesh, P.~Raiskinm{\"a}ki, A.~Koponen, M.~Kataja and J.~Timonen,
  \emph{J. Stat. Phys.}, 2002, \textbf{107}, 67--84\relax
\mciteBstWouldAddEndPuncttrue
\mciteSetBstMidEndSepPunct{\mcitedefaultmidpunct}
{\mcitedefaultendpunct}{\mcitedefaultseppunct}\relax
\EndOfBibitem
\bibitem[Raiskinmaki \emph{et~al.}(2003)Raiskinmaki, {\AA}strom, Kataja,
  Latva-Kokko, Koponen, Jasberg, and Shakib-Manesh]{raiskinmaki2003}
P.~Raiskinmaki, J.~A. {\AA}strom, M.~Kataja, M.~Latva-Kokko, A.~Koponen,
  A.~Jasberg and A.~Shakib-Manesh, \emph{Phys. Rev. E: Stat. Nonlinear Soft
  Matter Phys.}, 2003, \textbf{68}, 061403\relax
\mciteBstWouldAddEndPuncttrue
\mciteSetBstMidEndSepPunct{\mcitedefaultmidpunct}
{\mcitedefaultendpunct}{\mcitedefaultseppunct}\relax
\EndOfBibitem
\bibitem[Kulkarni and Morris(2008)]{kulkarni2008}
P.~M. Kulkarni and J.~F. Morris, \emph{Phys. Fluids}, 2008, \textbf{20},
  040602\relax
\mciteBstWouldAddEndPuncttrue
\mciteSetBstMidEndSepPunct{\mcitedefaultmidpunct}
{\mcitedefaultendpunct}{\mcitedefaultseppunct}\relax
\EndOfBibitem
\bibitem[Lee \emph{et~al.}(2014)Lee, Ahn, and Lee]{lee2014}
Y.~K. Lee, K.~H. Ahn and S.~J. Lee, \emph{Phys. Rev. E: Stat. Nonlinear Soft
  Matter Phys.}, 2014, \textbf{90}, 062317\relax
\mciteBstWouldAddEndPuncttrue
\mciteSetBstMidEndSepPunct{\mcitedefaultmidpunct}
{\mcitedefaultendpunct}{\mcitedefaultseppunct}\relax
\EndOfBibitem
\bibitem[Nguyen and Ladd(2005)]{nguyen2004}
N.~Q. Nguyen and A.~J.~C. Ladd, \emph{J. Fluid Mech}, 2005, \textbf{255},
  73--104\relax
\mciteBstWouldAddEndPuncttrue
\mciteSetBstMidEndSepPunct{\mcitedefaultmidpunct}
{\mcitedefaultendpunct}{\mcitedefaultseppunct}\relax
\EndOfBibitem
\bibitem[Luding(2008)]{luding2008}
S.~Luding, \emph{Granul. Matter.}, 2008, \textbf{10}, 235--246\relax
\mciteBstWouldAddEndPuncttrue
\mciteSetBstMidEndSepPunct{\mcitedefaultmidpunct}
{\mcitedefaultendpunct}{\mcitedefaultseppunct}\relax
\EndOfBibitem
\bibitem[Mitarai \emph{et~al.}(2002)Mitarai, Hayakawa, and
  Nakanishi]{mitarai2002}
N.~Mitarai, H.~Hayakawa and H.~Nakanishi, \emph{Phys. Rev. Lett.}, 2002,
  \textbf{88}, \relax
\mciteBstWouldAddEndPuncttrue
\mciteSetBstMidEndSepPunct{\mcitedefaultmidpunct}
{\mcitedefaultendpunct}{\mcitedefaultseppunct}\relax
\EndOfBibitem
\bibitem[Goldhirsch(2010)]{goldhirsch2010}
I.~Goldhirsch, \emph{Granul. Matter.}, 2010, \textbf{12}, 239--252\relax
\mciteBstWouldAddEndPuncttrue
\mciteSetBstMidEndSepPunct{\mcitedefaultmidpunct}
{\mcitedefaultendpunct}{\mcitedefaultseppunct}\relax
\EndOfBibitem
\bibitem[Derjaguin and Landau(1941)]{derjaguin1941}
B.~Derjaguin and L.~D. Landau, \emph{Acta Physicochim. U.R.S.S}, 1941,
  \textbf{14}, 633--662\relax
\mciteBstWouldAddEndPuncttrue
\mciteSetBstMidEndSepPunct{\mcitedefaultmidpunct}
{\mcitedefaultendpunct}{\mcitedefaultseppunct}\relax
\EndOfBibitem
\bibitem[Verwey and Overbeek(1948)]{verwey1948}
E.~J.~W. Verwey and J.~T.~G. Overbeek, \emph{Theory of the Stability of
  Lyophobic Colloids: The Interaction of Sol Particles Having an Electric
  Double Layer}, Elsevier, 1948\relax
\mciteBstWouldAddEndPuncttrue
\mciteSetBstMidEndSepPunct{\mcitedefaultmidpunct}
{\mcitedefaultendpunct}{\mcitedefaultseppunct}\relax
\EndOfBibitem
\bibitem[Israelachvili(2011)]{israelachvili2011}
J.~Israelachvili, \emph{Intermolecular and Surface Forces}, Academic Press, 3rd
  edn, 2011\relax
\mciteBstWouldAddEndPuncttrue
\mciteSetBstMidEndSepPunct{\mcitedefaultmidpunct}
{\mcitedefaultendpunct}{\mcitedefaultseppunct}\relax
\EndOfBibitem
\bibitem[Frith \emph{et~al.}(1996)Frith, d'Haene, Buscall, and
  Mewis]{frith1996}
W.~J. Frith, P.~d'Haene, R.~Buscall and J.~Mewis, \emph{J. Rheol}, 1996,
  \textbf{40}, 531--548\relax
\mciteBstWouldAddEndPuncttrue
\mciteSetBstMidEndSepPunct{\mcitedefaultmidpunct}
{\mcitedefaultendpunct}{\mcitedefaultseppunct}\relax
\EndOfBibitem
\bibitem[Bender and Wagner(1996)]{bender1996}
J.~Bender and N.~J. Wagner, \emph{J. Rheol}, 1996, \textbf{40}, 899--916\relax
\mciteBstWouldAddEndPuncttrue
\mciteSetBstMidEndSepPunct{\mcitedefaultmidpunct}
{\mcitedefaultendpunct}{\mcitedefaultseppunct}\relax
\EndOfBibitem
\bibitem[Maranzano and Wagner(2001)]{maranzano2001a}
B.~J. Maranzano and N.~J. Wagner, \emph{J. Rheol}, 2001, \textbf{45},
  1205--1222\relax
\mciteBstWouldAddEndPuncttrue
\mciteSetBstMidEndSepPunct{\mcitedefaultmidpunct}
{\mcitedefaultendpunct}{\mcitedefaultseppunct}\relax
\EndOfBibitem
\bibitem[Maranzano and Wagner(2001)]{maranzano2001b}
B.~J. Maranzano and N.~J. Wagner, \emph{J. Chem. Phys}, 2001, \textbf{114},
  10514--10527\relax
\mciteBstWouldAddEndPuncttrue
\mciteSetBstMidEndSepPunct{\mcitedefaultmidpunct}
{\mcitedefaultendpunct}{\mcitedefaultseppunct}\relax
\EndOfBibitem
\bibitem[Lootens \emph{et~al.}(2005)Lootens, van Damme, Hemar, and
  Hebraud]{lootens2005}
D.~Lootens, H.~van Damme, Y.~Hemar and P.~Hebraud, \emph{Phys. Rev. Lett.},
  2005, \textbf{95}, year\relax
\mciteBstWouldAddEndPuncttrue
\mciteSetBstMidEndSepPunct{\mcitedefaultmidpunct}
{\mcitedefaultendpunct}{\mcitedefaultseppunct}\relax
\EndOfBibitem
\bibitem[Fall \emph{et~al.}(2010)Fall, Lema{\^\i}tre, Bertrand, Bonn, and
  Ovarlez]{fall2010}
A.~Fall, A.~Lema{\^\i}tre, F.~Bertrand, D.~Bonn and G.~Ovarlez, \emph{Phys.
  Rev. Lett.}, 2010, \textbf{105}, year\relax
\mciteBstWouldAddEndPuncttrue
\mciteSetBstMidEndSepPunct{\mcitedefaultmidpunct}
{\mcitedefaultendpunct}{\mcitedefaultseppunct}\relax
\EndOfBibitem
\bibitem[Larsen \emph{et~al.}(2010)Larsen, Kim, Zukoski, and Weitz]{larsen2010}
R.~J. Larsen, J.~W. Kim, C.~F. Zukoski and D.~A. Weitz, \emph{Phys. Rev. E:
  Stat. Nonlinear Soft Matter Phys.}, 2010, \textbf{81}, year\relax
\mciteBstWouldAddEndPuncttrue
\mciteSetBstMidEndSepPunct{\mcitedefaultmidpunct}
{\mcitedefaultendpunct}{\mcitedefaultseppunct}\relax
\EndOfBibitem
\bibitem[Brown and Jaeger(2014)]{brown2014}
E.~Brown and H.~M. Jaeger, \emph{Rep. Prog. Phys.}, 2014, \textbf{77},
  046602\relax
\mciteBstWouldAddEndPuncttrue
\mciteSetBstMidEndSepPunct{\mcitedefaultmidpunct}
{\mcitedefaultendpunct}{\mcitedefaultseppunct}\relax
\EndOfBibitem
\bibitem[da~Cruz \emph{et~al.}(2005)da~Cruz, Emam, Prochnow, Roux, and
  Chevoir]{dacruz2005}
F.~da~Cruz, S.~Emam, M.~Prochnow, J.~Roux and F.~Chevoir, \emph{Phys. Rev. E:
  Stat. Nonlinear Soft Matter Phys.}, 2005, \textbf{72}, 021309\relax
\mciteBstWouldAddEndPuncttrue
\mciteSetBstMidEndSepPunct{\mcitedefaultmidpunct}
{\mcitedefaultendpunct}{\mcitedefaultseppunct}\relax
\EndOfBibitem
\bibitem[Kanatani(1981)]{kanatani1981}
K.~Kanatani, \emph{Powder Technol.}, 1981, \textbf{28}, 167--172\relax
\mciteBstWouldAddEndPuncttrue
\mciteSetBstMidEndSepPunct{\mcitedefaultmidpunct}
{\mcitedefaultendpunct}{\mcitedefaultseppunct}\relax
\EndOfBibitem
\bibitem[Kim and Karrila(1991)]{kimkarilla1991}
S.~Kim and S.~J. Karrila, \emph{Microhydrodynamics: Principles and Selected
  Applications}, Butterworth-Heinemann, 1991\relax
\mciteBstWouldAddEndPuncttrue
\mciteSetBstMidEndSepPunct{\mcitedefaultmidpunct}
{\mcitedefaultendpunct}{\mcitedefaultseppunct}\relax
\EndOfBibitem
\bibitem[Klingler and McConnell(1993)]{klingler1993}
J.~F. Klingler and H.~M. McConnell, \emph{J. Phys. Chem.}, 1993, \textbf{97},
  6096--6100\relax
\mciteBstWouldAddEndPuncttrue
\mciteSetBstMidEndSepPunct{\mcitedefaultmidpunct}
{\mcitedefaultendpunct}{\mcitedefaultseppunct}\relax
\EndOfBibitem
\bibitem[Langevin(2014)]{langevin2014}
D.~Langevin, \emph{Annu. Rev. Fluid Mech.}, 2014, \textbf{46}, 47--65\relax
\mciteBstWouldAddEndPuncttrue
\mciteSetBstMidEndSepPunct{\mcitedefaultmidpunct}
{\mcitedefaultendpunct}{\mcitedefaultseppunct}\relax
\EndOfBibitem
\bibitem[Petkov \emph{et~al.}(2014)Petkov, Danov, and Kralchevsky]{petkov2014}
P.~V. Petkov, K.~D. Danov and P.~A. Kralchevsky, \emph{Langmuir}, 2014,
  \textbf{30}, 2768--2778\relax
\mciteBstWouldAddEndPuncttrue
\mciteSetBstMidEndSepPunct{\mcitedefaultmidpunct}
{\mcitedefaultendpunct}{\mcitedefaultseppunct}\relax
\EndOfBibitem
\bibitem[Kralchevsky \emph{et~al.}(2015)Kralchevsky, Danov, and
  Petkov]{kralchevsky2015}
P.~A. Kralchevsky, K.~D. Danov and P.~V. Petkov, \emph{Phil. Trans. R. Soc. A},
  2015, \textbf{374}, year\relax
\mciteBstWouldAddEndPuncttrue
\mciteSetBstMidEndSepPunct{\mcitedefaultmidpunct}
{\mcitedefaultendpunct}{\mcitedefaultseppunct}\relax
\EndOfBibitem
\bibitem[Higuera \emph{et~al.}(2016)Higuera, Perales, and Vega]{higuera2016}
M.~Higuera, J.~M. Perales and J.~M. Vega, \emph{Phys. Fluids}, 2016,
  \textbf{28}, year\relax
\mciteBstWouldAddEndPuncttrue
\mciteSetBstMidEndSepPunct{\mcitedefaultmidpunct}
{\mcitedefaultendpunct}{\mcitedefaultseppunct}\relax
\EndOfBibitem
\bibitem[Cates \emph{et~al.}(1998)Cates, Wittmer, Bouchaud, and
  Claudin]{cates1998}
M.~E. Cates, J.~P. Wittmer, J.-P. Bouchaud and P.~Claudin, \emph{Phys. Rev.
  Lett.}, 1998, \textbf{81}, 1841--1844\relax
\mciteBstWouldAddEndPuncttrue
\mciteSetBstMidEndSepPunct{\mcitedefaultmidpunct}
{\mcitedefaultendpunct}{\mcitedefaultseppunct}\relax
\EndOfBibitem
\bibitem[Sarkar \emph{et~al.}(2016)Sarkar, Bi, Zhang, Ren, Behringer, and
  Chakraborty]{sarkar2016}
S.~Sarkar, D.~Bi, J.~Zhang, J.~Ren, R.~P. Behringer and B.~Chakraborty,
  \emph{Phys. Rev. E: Stat. Nonlinear Soft Matter Phys.}, 2016, \textbf{93},
  042901\relax
\mciteBstWouldAddEndPuncttrue
\mciteSetBstMidEndSepPunct{\mcitedefaultmidpunct}
{\mcitedefaultendpunct}{\mcitedefaultseppunct}\relax
\EndOfBibitem
\bibitem[Chen \emph{et~al.}(2018)Chen, Bertrand, Jin, Shattuck, and
  O'Hern]{chen2018}
S.~Chen, T.~Bertrand, W.~Jin, M.~D. Shattuck and C.~S. O'Hern, \emph{Phys. Rev.
  E: Stat. Nonlinear Soft Matter Phys.}, 2018, \textbf{98}, 042906\relax
\mciteBstWouldAddEndPuncttrue
\mciteSetBstMidEndSepPunct{\mcitedefaultmidpunct}
{\mcitedefaultendpunct}{\mcitedefaultseppunct}\relax
\EndOfBibitem
\bibitem[Suzuki and Hayakawa(2019)]{suzuki2019}
K.~Suzuki and H.~Hayakawa, \emph{J. Fluid Mech}, 2019, \textbf{864},
  1125--1176\relax
\mciteBstWouldAddEndPuncttrue
\mciteSetBstMidEndSepPunct{\mcitedefaultmidpunct}
{\mcitedefaultendpunct}{\mcitedefaultseppunct}\relax
\EndOfBibitem
\bibitem[Ishima and Hayakawa(2019)]{daisuke2019arxiv}
D.~Ishima and H.~Hayakawa, arXiv:1902.04759, 2019\relax
\mciteBstWouldAddEndPuncttrue
\mciteSetBstMidEndSepPunct{\mcitedefaultmidpunct}
{\mcitedefaultendpunct}{\mcitedefaultseppunct}\relax
\EndOfBibitem
\bibitem[Panaitescu \emph{et~al.}(2017)Panaitescu, Clotet, and
  Kudrolli]{panaitescu2017}
A.~Panaitescu, X.~Clotet and A.~Kudrolli, \emph{Phys. Rev. E: Stat. Nonlinear
  Soft Matter Phys.}, 2017, \textbf{95}, 032901\relax
\mciteBstWouldAddEndPuncttrue
\mciteSetBstMidEndSepPunct{\mcitedefaultmidpunct}
{\mcitedefaultendpunct}{\mcitedefaultseppunct}\relax
\EndOfBibitem
\bibitem[Waitukaitis and Jaeger(2012)]{waitukaitis2012}
S.~R. Waitukaitis and H.~M. Jaeger, \emph{Nature}, 2012, \textbf{487},
  205--209\relax
\mciteBstWouldAddEndPuncttrue
\mciteSetBstMidEndSepPunct{\mcitedefaultmidpunct}
{\mcitedefaultendpunct}{\mcitedefaultseppunct}\relax
\EndOfBibitem
\bibitem[Peters and Jaeger(2014)]{peters2014}
I.~R. Peters and H.~M. Jaeger, \emph{Soft Matter}, 2014, \textbf{10},
  6564\relax
\mciteBstWouldAddEndPuncttrue
\mciteSetBstMidEndSepPunct{\mcitedefaultmidpunct}
{\mcitedefaultendpunct}{\mcitedefaultseppunct}\relax
\EndOfBibitem
\bibitem[Han \emph{et~al.}(2018)Han, Wyart, Peters, and Jaeger]{han2018}
E.~Han, M.~Wyart, I.~R. Peters and H.~M. Jaeger, \emph{Phys. Rev. Fluids},
  2018, \textbf{3}, 073301\relax
\mciteBstWouldAddEndPuncttrue
\mciteSetBstMidEndSepPunct{\mcitedefaultmidpunct}
{\mcitedefaultendpunct}{\mcitedefaultseppunct}\relax
\EndOfBibitem
\bibitem[Leonardi \emph{et~al.}(2014)Leonardi, Wittel, Mendoza, and
  Hermann]{leonardi2014}
A.~Leonardi, F.~K. Wittel, M.~Mendoza and H.~J. Hermann, \emph{Comp. Part.
  Mech.}, 2014, \textbf{1}, 3--13\relax
\mciteBstWouldAddEndPuncttrue
\mciteSetBstMidEndSepPunct{\mcitedefaultmidpunct}
{\mcitedefaultendpunct}{\mcitedefaultseppunct}\relax
\EndOfBibitem
\bibitem[Succi(2001)]{succi2001}
S.~Succi, \emph{The lattice Boltzmann equation: for fluid dynamics and beyond},
  Oxford University Press, 2001\relax
\mciteBstWouldAddEndPuncttrue
\mciteSetBstMidEndSepPunct{\mcitedefaultmidpunct}
{\mcitedefaultendpunct}{\mcitedefaultseppunct}\relax
\EndOfBibitem
\bibitem[Jeffrey and Onishi(1984)]{jeffrey1984}
D.~J. Jeffrey and Y.~Onishi, \emph{J. Fluid Mech}, 1984, \textbf{139},
  261--290\relax
\mciteBstWouldAddEndPuncttrue
\mciteSetBstMidEndSepPunct{\mcitedefaultmidpunct}
{\mcitedefaultendpunct}{\mcitedefaultseppunct}\relax
\EndOfBibitem
\end{mcitethebibliography}
\bibliographystyle{rsc} %the RSC's .bst file

\end{document}